\documentclass[12pt]{article}

\pdfoutput=1
	\usepackage[colorlinks,citecolor=blue,urlcolor=blue,linkcolor = blue]{hyperref}
	\newcommand{\red}[1]{{\textcolor{black}{#1}}}	
	\usepackage[comma,sort&compress]{natbib}
	\usepackage{amssymb, amsmath,bm,graphicx}
	\usepackage{mathtools}
	\usepackage{setspace}
	\usepackage{mathrsfs,amsthm}
	\usepackage[normalem]{ulem}
	\usepackage{subcaption}
	\usepackage{comment}
	\usepackage{times,bm}
	\usepackage{multirow}
	\usepackage{anyfontsize}
	\usepackage{graphicx,psfrag,epsf}
	\usepackage{float}
	\usepackage[font=small,labelfont=bf]{caption}
	\usepackage{tabularx}
	\usepackage{url}
	\usepackage{booktabs}
	\usepackage{hyperref}
	\usepackage{color}
	\usepackage{xcolor}
	\usepackage{caption}
	\usepackage{bm}
	\usepackage{bbm}
	\usepackage[mathscr]{euscript}
	\usepackage{bigstrut,enumerate}
	\usepackage[utf8]{inputenc}
	\usepackage[ruled,vlined]{algorithm2e}
	\usepackage{multirow}
	\renewcommand{\liminf}{\varliminf}

	\newtheorem{theorem}{Theorem}
	
	\newtheorem{example}{Example}
	\newtheorem{lemma}{Lemma}
	\newtheorem{proposition}{Proposition}	
	
	\newtheorem{remark}{Remark}
	\newtheorem{corollary}{Corollary}

	\newcommand{\PP}{\ensuremath{\tilde{P}}}

	\newcommand{\fclass}{\mathbb{W}}
	\DeclareMathOperator*{\argmin}{arg\,min}

	\newtheorem{definition}{Definition}

\usepackage{color,soul}	
		
	\newcommand{\blind}{1}
	
\theoremstyle{definition}

\newtheorem{assumption}{Assumption}

\SetKwInOut{Parameter}{Parameter}

\begin{document}
	\def\spacingset#1{\renewcommand{\baselinestretch}%
		{#1}\small\normalsize} \spacingset{1}
	
\if1\blind
{
\title{\vspace{-2.5cm}Nonparametric Empirical Bayes Estimation on Heterogeneous Data}
\author{$\text{Trambak Banerjee}^1, \text{Luella Fu},\text{Gareth M. James},$\\ $\text{Gourab Mukherjee}^{2}, \text{ and Wenguang Sun}^{2,3}$\\\\
University of Kansas, San Francisco State University, Emory University,\\ University of Southern California and Zhejiang University}
\date{}
\footnotetext[1]{T. Banerjee was partially supported by the University of Kansas General Research Fund allocation \#2302216.}
\footnotetext[2]{G.M. and W.S. thank Dr. Zeyu Yao at Zhejiang University for helpful discussions on the theory. }
\footnotetext[3]{W. Sun was supported in part by NSF grant DMS-2015339.}
\maketitle
\vspace{-1.0cm}
} \fi

\if0\blind
{
\title{\vspace{-2.5cm}{Nonparametric Empirical Bayes Estimation On Heterogeneous Data}}
\date{}
\maketitle
\vspace{-1.0cm}
%\medskip
} \fi
\begin{abstract}
The simultaneous estimation of many parameters based on data collected from corresponding studies is a key research problem that has received renewed attention in the high-dimensional setting. Many practical situations involve heterogeneous data where heterogeneity is captured by a nuisance parameter. Effectively pooling information across samples while correctly accounting for heterogeneity presents a significant challenge in large-scale estimation problems. We address this issue by introducing the ``Nonparametric Empirical Bayes Structural Tweedie" (NEST) estimator, which efficiently estimates the unknown effect sizes and properly adjusts for heterogeneity via a generalized version of Tweedie's formula. For the normal means problem, NEST simultaneously handles the two main selection biases introduced by heterogeneity: one, the selection bias in the mean, which cannot be effectively corrected without also correcting for, two, selection bias in the variance. 
We develop theory to show that NEST is asymptotically as good as the optimal Bayes rule that uniquely minimizes a weighted squared error loss.
In our simulation studies NEST outperforms competing methods, with much efficiency gains in many settings. The proposed method is demonstrated on estimating the batting averages of baseball players and Sharpe ratios of mutual fund returns. Extensions to other members of the two-parameter exponential family are discussed.
\end{abstract}
\noindent%
\spacingset{1.5} % DON'T change the spacing!
\section{Introduction}

Suppose that we are interested in %the classic problem of 
estimating a vector of parameters $\bm\mu=(\mu_1,\ldots, \mu_n)$ based on the summary statistics $Y_1, \ldots, Y_n$ from $n$ study units.
%, where the $ith$ unit has $j = 1, \ldots, m_i$ observations.  
%Let $\pmb y_i=(y_{i1}, \cdots, y_{im_i})$ be the vector of observations collected for unit $i$. 
%Let $Y_i=\sum_{j=1}^{m_i} Y_{ij}/m_i$. 
The setting where $Y_{i}\mid\mu_i\sim N(\mu_i,\sigma^2)$ is the most well--known example, %but 
%this parameter-estimation problem occurs in a wide range of applications % with many possible distributions. 
%Other common examples include: the Poisson distribution with $\eta_i=\lambda_i$ and the Binomial distribution with $\eta_i=p_i$.
but the broader scope includes
%including 
the compound estimation of %Normal means $\mu_i$, 
Poisson parameters $\lambda_i$, Binomial parameters $p_i$, and other members of the exponential family.

%self note: do we follow up on Poisson or Binomial examples? Should we instead sub in our Gamma, Beta examples?

%This fundamental problem %largely reso{\color{red}LV}ed in the  low-dimensional setting, (note by Wen: the problem has been extensively studied but not reso{\color{red}LV}ed)
%has recently become a key research question in the high-dimensional setting. 
In modern large-scale applications it is often of interest to perform simultaneous and selective inference \citep{BenjYeku2011, Berk2013, Weinetal2013}, which has called for solving the compound estimation problem in new ways and for new purposes. For example, there has been recent work on how to construct valid post-selection intervals for regression coefficients conditional on the selected model \citep{Leeetal2016}. %For example, there has been recent work on how to construct valid simultaneous confidence intervals of $\mu_i$'s after a selection procedure is applied \citep{Leeetal2016}. 
% first select interesting values of $X^i$ and then carry out inferential analysis for the associated $\eta_i$'s; see , which all
In multiple testing, as well as related ranking and selection problems, it is often desirable to incorporate estimates of the effect sizes $\mu_i$ in the decision process to prioritize the selection of more scientifically meaningful hypotheses \citep{BenHoc97, SunMcL12, Heetal2015, HenNew16, Basu2017}. 

However, the simultaneous inference of thousands of means, or other parameters, is challenging because, as described in \citet{Efr11}, the large scale of the problem introduces selection bias, wherein some data points are large merely by chance, causing traditional estimators to overestimate the corresponding means. 
Shrinkage estimation, exemplified by the seminal work of \citet{JamSte61}, has been widely used in  %provided an effective 
%bias reduction (in low-dimensional setting shrinkage idea is originally introduced to reduce the MSE at the price of having a biased estimator)
simultaneous inference. There are several popular classes of methods, including linear shrinkage estimators \citep{JamSte61, EfrMor75, Ber76}, non--linear thresholding--based estimators motivated by sparse priors \citep{DonohoJohnstone94, JohnstoneSilverman04, Abretal06}, and both Bayes or empirical Bayes 
%% (EB) 
estimators with unspecified priors  \citep{BroGre09, JiangZhang09, CastillovanderVaart12}. This article focuses on a class of estimators based on Tweedie's formula \citep{Dyson1926, Eddington1940,Rob1956}. 
% Mention Efron coined this name, but Dyson credited Eddington.
The formula is an elegant shrinkage estimator, for distributions from the exponential family, that has recently received renewed interest \citep{BroGre09, Efr11, KoeMiz14}. Tweedie's formula is simple and intuitive, and its implementation %under the $f$-modeling strategy for empirical Bayes estimation 
only requires estimating the marginal distribution of $Y_i$. This property is particularly appealing for large--scale estimation problems where such estimates can be easily constructed from the observed data. The resultant {empirical Bayes} 
%%EB 
estimator enjoys optimality properties \citep{BroGre09} and %delivers superior numerical performance. The 
the work of \citet{Efr11} further convincingly demonstrates that Tweedie's formula provides an effective bias correction tool  when estimating thousands of parameters simultaneously. 

\subsection{Issues with heterogeneous data}
Most of the research in this area has been restricted to models %of the form $f(y_i\mid\eta_i)$ 
where the distribution of $Y_i$ is solely a function of the parameter of interest $\mu_i$. In situations involving a nuisance parameter $\tau_i$ it is generally assumed to be known and identical for all $Y_i$. For example, homoskedastic Gaussian models of the form $Y_i\mid\mu_i, \sigma \overset{ind}{\sim} N(\mu_i, \sigma^2)$ involve a common nuisance parameter $\tau_i=1/\sigma^2$ for all $i$. However, in large-scale studies when the data are collected from heterogeneous sources, the nuisance parameters may vary over the $n$ study units. Perhaps the most common example, and the setting we concentrate most on, involves heteroskedastic errors, where $\sigma^2$ varies over $Y_i$. Microarray data  \citep{EricksonSabatti05, Chiarettietal04}, returns on mutual funds \citep{BrownGoetzmann92}, and the state-wide school performance gaps \citep{SunMcL12} are all instances of large-scale data where genes, funds, or schools have heterogeneous variances. Heteroskedastic errors also arise in analysis of variance \citep{Weinsteinetal18} and linear regression settings \citep{KouYang15}. Moreover, in compound binomial problems, heterogeneity arises through unequal sample sizes across different study units. Unfortunately, the conventional Tweedie's formula assumes identical nuisance parameters across study units and so cannot eliminate selection bias for heterogeneous data. Moreover, various works show that
failing to account for heterogeneity leads to inefficient shrinkage estimators \citep{Weinsteinetal18}, methods with invalid false discovery rates \citep{Efr08, CaiSun09}, unstable multiple testing procedures \citep{Tushetal2001} and suboptimal ranking and selection algorithms \citep{HenNew16}, exacerbating the replicability crisis in large-scale studies. Few methodologies are available to address this issue.

For Gaussian data, %\red{$Y_{ij}\stackrel{i.i.d}{\sim}N(\mu_i,1/\tau_i),~j=1,\ldots,m_i$},
a common goal is to find the estimator of the means $\mu_i$ that minimizes the expected squared error loss. %Consider the heteroscedastic normal means problem where the sample mean and sample variance for study unit $i$ are respectively denoted by $Y_i$ and $S_i^2$. 
A plausible--seeming solution might be to scale each sample mean $Y_i$ by its estimated standard deviation $S_i/\sqrt{m_i}$ so that a homoskedastic method could be applied to $X_i=\sqrt{m_i}Y_i/S_i$, before undoing the scaling on the final estimate of $\mu_i$. Indeed, this is essentially the approach taken whenever we compute standardized test statistics, such as $t$--values and $z$--values. %A similar standardization is performed in the Binomial setting when we compute $\hat p_i = X_i/m_i$, where $X_i$ is the number of successes and $m_i$ the number of trials. 
However, this approach, which disregards important structural information, can be highly inefficient. More advanced methods have been developed, but all suffer from various limitations. For instance, the methods proposed by \citet{Xieetal12},  \citet{Tan15}, \citet{jing2016sure}, \citet{KouYang15}, and \citet{ZhangBhattacharya17} are designed for heteroscedastic data but assume a parametric Gaussian prior or semi-parametric Gaussian mixture prior, which leads to loss of efficiency when the prior is misspecified. Moreover, existing methods, such as \citet{Xieetal12} and \cite{Weinsteinetal18}, often assume that the nuisance parameters, the variances $1/\tau_i$, are known and use a consistent estimator for implementation. However, when a large number of units are investigated simultaneously, traditional sample variance estimators may similarly suffer from selection bias, which often leads to severe deterioration in the MSE for estimating the means. %\blue{Thus, double-shrinkage is employed in NEST to address this issue.}

\subsection{The proposed approach and main contributions}
In the homogeneous setting, Tweedie's formula for Gaussian data estimates $\mu_i$ using the score function at $Y_i$, which is the gradient of the log marginal density of the sufficient statistic $Y_i$. However, this approach does not immediately extend to heterogeneous data. A significant challenge in the heterogeneous setting is how to pool information from different study units effectively while accounting for the heterogeneity captured by the possibly unknown nuisance parameter $\tau_i$. In this article, we address this issue by proposing a double shrinkage method that simultaneously incorporates the structural information encoded in both the primary (e.g. $Y_i$) and auxiliary (e.g. $S_i$) data. We develop a two--step approach, ``Nonparametric Empirical Bayes Structural Tweedie'' (NEST), which first estimates the bivariate score function at the primary and auxiliary data, %for data from a two--parameter exponential family, 
and second predicts $\mu_i$ under a weighted squared error loss function using a generalized version of Tweedie's formula that effectively incorporates the structural information encoded in the {nuisance parameter. Unlike the squared error loss, which has been widely used for large-scale shrinkage estimation, the weighted squared error loss provides a more suitable performance metric for analyzing heterogeneous data by utilizing a weighting scheme that captures the underlying heterogeneity across the study units. %the weighted squared error loss utilizes a weighting scheme that is sensitive to the underlying heterogeneity and employs a relatively larger weight on the estimation error for study units that have a smaller variance. 
	The corresponding shrinkage estimator that minimizes the expected weighted squared error loss naturally encompasses the ability to differentiate between the study units both with respect to $\mu_i$ and the heterogeneity captured by the unknown nuisance parameter $\tau_i$.} 

%A significant challenge in the heterogeneous setting is how to pool information from different study units effectively while accounting for the heterogeneity captured by the possibly unknown nuisance parameters. NEST addresses the issue by proposing a double shrinkage method that simultaneously incorporates the structural information encoded in both the primary (e.g. $Y_i$) and auxiliary (e.g. $S_i$) data. 
%Unlike existing $f$-modeling approaches, NEST employs a kernelized Stein's discrepancy measure \citep{liu2016kernelized,banerjee2019general} to recast the compound estimation problem as a convex optimization problem that directly estimates the optimal shrinkage directions {\bf I have no idea what this means---need some intuition here}. 
%NEST departs from linear shrinkage and other techniques that focus on specific forms of the prior distribution.  The computational algorithm is quick and scalable. The resulting estimator enjoys strong numerical and theoretical properties. 
%that the bivariate distribution varies with $\theta$, so we must estimate a two--dimensional function. NEST addresses this issue using a kernel which weights observations by their distance in both the $x$ and $\theta$ dimensions. The intuition here is that the density function should change smoothly as a function of $\theta$, so observations with variability close to $\theta$ can be used to estimate $f_{\theta}(x)$.  Once we obtain this two--dimensional density function, we simply apply Tweedie's formula to estimate the parameter corresponding to any particular combination of $x$ and $\theta$. 

NEST has several clear advantages. 
First, it simultaneously handles the two main selection biases introduced by heterogeneity: one, selection bias in the primary data (e.g. sample means), which cannot be effectively corrected without also correcting for, two, selection bias in the auxiliary data (e.g. sample variances).
By producing more accurate estimates for the nuisance parameters, NEST in general renders improved shrinkage factors for estimating the primary parameters.
Second, NEST makes no parametric assumptions about the prior since it uses a nonparametric method to directly estimate the bivariate score function. Third, NEST exploits the structure of the entire sample and avoids the information loss that occurs in the discretization step used in grouping methods \citep{Weinsteinetal18}. {Fourth, we develop a rigorous theoretical framework that establishes that within the class of Lipschitz continuous functions with bounded Lipschitz constants, NEST is asymptotically as good as the optimal Bayes rule that uniquely minimizes the weighted squared error loss.} Finally, NEST provides a general estimation framework for members of the two-parameter exponential family %. It is fast to implement, produces stable estimates, 
and exhibits robust performance against model mis-specification. We demonstrate numerically that NEST can provide high levels of estimation accuracy relative to a host of benchmark  methods.%, \blue{including the recent NPMLE method of \citet{gu2017empirical, gu2017unobserved}}. 
\subsection{Connection to existing works}
{For empirical Bayes (EB) estimation of the means, recent works such as \cite{Xieetal12,Tan15,jing2016sure,KouYang15} assume a parametric prior distribution for the means. This is in contrast to the setting where such a prior is unknown and in those settings there are two main modeling strategies for EB estimation. They are known as the $g$-modeling and $f$-modeling strategies in the terminology of \cite{Efr14-two}.} 
The idea of $g$-modeling is to first obtain a deconvolution estimate of the unknown prior distribution of $\mu_i$, and then predict $\mu_i$ by plugging this estimate into Bayes rule for various loss functions. The deconvolution estimate can be constructed via the nonparametric maximum likelihood estimate (NPMLE; \citealp{kiefer1956consistency, laird1978nonparametric}), or by modeling the unknown prior as a low-dimensional exponential family distribution \citep{efron2016empirical}. Some notable works along this line include \cite{JiangZhang09}, \cite{KoeMiz14}, \cite{gu2017empirical}, \cite{saha2020nonparametric}, and \cite{soloff2021multivariate}. In contrast, the $f$-modeling strategy usually assumes that the nuisance parameter, such as the variance, is known and sidesteps the need of deconvolution estimation by directly predicting $\mu_i$ based on Tweedie’s formula (or its generalized version), which only depends on the score function at the sufficient statistic $Y_i$ and the known nuisance parameter. Notable works along this line include \cite{BroGre09} and \cite{Efr11}, {both of which rely on fixed and known variances. In particular, \cite{BroGre09} use kernel density estimation techniques for separately estimating the marginal density and its gradient, and then taking their ratio to construct an estimate of the score function. \cite{Efr11}, on the other hand, represent the marginal density as a $K-$ parameter exponential family and use Lindsey's method \citep{lindsey1974comparison} for estimating the $K$ parameters.} NEST adopts the $f$-modeling strategy. However, in contrast to existing $f$-modeling methods, we allow the variances to be unknown and develop a convex optimization approach that directly provides consistent estimate of the score function and is capable of incorporating various structural constraints in the data-driven NEST estimator.  %Our numerical results show that the NEST offers substantial improvement in the estimation risk over other $f$-modeling methods under the squared error loss. 
%Second, we provide in Proposition \ref{prop:oraclerisks} precise conditions under which the oracle NEST estimator dominates Tweedie’s estimator with sample variances (cf. Definition \ref{def:twfs2}).

The $g$-modeling approach via the NPMLE provides an excellent alternative for EB estimation of heteroskedastic means. %NPMLE is competitive to NEST in most of our numerical studies.
However, to the best of our knowledge, the asymptotic properties of the NPMLE are highly nontrivial to establish and often require strong assumptions. For instance, the analysis in \cite{saha2020nonparametric} only works for a limited class of  covariance structures, and the theory on the rate of convergence is applicable only when the degree of heteroskedasticity is ``mild’’; alternatively the analysis in  \cite{soloff2021multivariate} assumes that $\mu_i$ are independent of  $\sigma_i$, which is often violated in practice (\citealp{Weinsteinetal18}). In contrast, we establish the asymptotic properties of NEST without assumptions on the degree of heteroskedasticity or independence between $\mu_i$ and $\sigma_i$.

A key advantage of \(g\)-modeling is its capability to address a wider range of problems, particularly those in which the direct use of the marginal density of the sufficient statistic cannot yield a solution. Meanwhile, the \(f\)-modeling approach, which often has a simple and intuitive form (e.g., Tweedie's formula), is attractive when only information on the marginal distribution is needed to solve the problem of interest. We develop a novel theory for \(f\)-modeling in Section \ref{Sec:theory} that employs techniques very different from those used in existing \(g\)-modeling approaches. 

\subsection{Organization}

The rest of the paper is structured as follows. In Section~\ref{Sec:method} we present our hierarchical Gaussian model where both the mean and variance parameters are unknown. %, and then discuss a version of Tweedie's formula that uses sample variances in Section~\ref{sec:twfs2}. 
{In Section \ref{sec:weighted_loss} we discuss the weighted squared error loss function and in Section \ref{GT:subsec} we rely on the natural-parameter Tweedie's formula for our hierarchical model to introduce the oracle NEST estimator under the weighted squared error loss} in Definition \ref{def:oraclenest}.
%first derives a generalized Tweedie's formula for a hierarchical Gaussian model where both the mean and variance parameters are unknown. 
We then develop a convex optimization approach in Section \ref{sec:opt_criteria} for estimating the unknown shrinkage factors in the oracle NEST formula. In Section \ref{Sec:theory}, we describe the theoretical setup, justify the optimization criterion, and finally establish asymptotic theories for the proposed NEST estimator. %Extensions to other members of the two-parameter exponential family are discussed in Appendix \ref{sec:extensions}. 
Simulation studies are presented in Section \ref{sec:simulations} to compare NEST with competing methods. The article concludes with a discussion in Section \ref{sec:discuss}. Two real-data applications are included in Section \ref{sec:realdata} of the Appendix. The methodological details for the regular squared loss, proofs of the main theorems and auxiliary results, as well as additional numerical results, extensions, and technical discussions are provided in the Appendix. 

%%%%%%%%%%%%%%%%%%%%%%%%%%%%%%%%
\section{Double shrinkage estimation on heteroskedastic Normal data}\label{Sec:method}
%%%%%%%%%%%%%%%%%%%%%%%%%%%%%%%%
In the main text of this article, we focus on the normal means problem. Extensions of the methodology to other members of the two-parameter exponential family are discussed in Section \ref{sec:extensions}. 

Suppose we collect $m_i$ observations for the $i^{th}$ study unit, $i=1, \ldots, n$. Conditional on $m_i$, the data are normally distributed obeying the following hierarchical model: %is special case of the hierarchical model \eqref{eq:expmodel}, where  $\bm \psi_i=(\mu_i,\tau_i)$, $\bm \eta_i=(\tau_i\mu_i,\tau_i)$ and, 
%where
% \begin{align}\label{eq:normmodel}
% \begin{split}
% Y_{ij}~|~\mu_i,\tau_i \stackrel{i.i.d}{\sim} N(\mu_i, 1/\tau_i),\ j = 1, \ldots, m_i, \\
% \mu_i~|\tau_i \stackrel{ind}{\sim} G_\mu(\cdot|\tau_i), \quad \tau_i\stackrel{i.i.d}{\sim}H_\tau(\cdot).
% \end{split}
%\end{align}
%\begin{align}\label{eq:normmodel}
\begin{eqnarray}
\label{eq:normmodel}
Y_{ij}~|~\mu_i,\tau_i &\stackrel{i.i.d.}{\sim}& N(\mu_i, 1/\tau_i),\quad j = 1, \ldots, m_i, \\
\nonumber \mu_i~|\tau_i &\stackrel{ind.}{\sim}& G_\mu(\cdot|\tau_i), \quad \tau_i\stackrel{i.i.d.}{\sim}H_\tau(\cdot), \quad i =1,\ldots, n,
\end{eqnarray}
\red{where $m_i$'s are a random sample from a distribution with mass function $q(\cdot)$ independent of $(\mu_i,\tau_i)$}. Let $Y_i=m^{-1}_i\sum_{j=1}^{m_i}Y_{ij}$ and $S_i^2=(m_i-1)^{-1}\sum_{j=1}^{m_i}(Y_{ij}-Y_i)^2$ respectively denote the sample mean and sample variance under Model \eqref{eq:normmodel}. Further, let $\bm Y=(Y_{1}, \ldots, Y_{n})$ and $\bm S=(S_1^2, \ldots, S_{n}^2)$ be the vectors of summary statistics, and $\bm y=(y_{1}, \ldots, y_{n})$ and $\bm s=(s_1^2,\ldots,s_n^2)$ the observed values. We view $\mu_i$ and $\tau_i$, both of which are unknown, as the primary and nuisance parameters, respectively. The prior distributions $G_\mu(\cdot|\tau_i)$ and $H_\tau(\cdot)$ are unspecified. When the precisions $\tau_i$ are known in Model \eqref{eq:normmodel}, compound estimation of the means $\bm \mu=(\mu_1,\ldots,\mu_n)^T$ under the squared error loss has received significant attention in recent years (see for example \cite{Weinsteinetal18,Xieetal12,cai2021nonparametric} and the references therein) and the estimator minimizing the expected squared error loss is,
\begin{equation}
\label{eq:twftau}
\mathbb{E}(\mu_i~|~Y_i={y}_i, \tau_i,m_i)=y_i+\dfrac{1}{m_i\tau_i}w_1(y_i;m_i,\tau_i),%\dfrac{1}{m_i\tau_i}\dfrac{\partial}{\partial y}\log f_{m_i,\tau_i}(y_i)=y_i+\dfrac{1}{m_i\tau_i}\ell^{(1)}_{f,\tau_i}(y_i;m_i),
\end{equation}
where $w_1(y;m,\tau)\coloneqq \dfrac{\partial}{\partial y}\log f_{m,\tau}(y)$ and $f_{m,\tau}(\cdot)$ is the pdf of the marginal distribution of $Y$. Equation \eqref{eq:twftau} is the celebrated Tweedie's formula with known variances \citep{Efr11} which forms the basis for $f$-modeling strategies for estimating $\mu_i$, and only requires estimation of the score functions $w_1(y_i;m_i,\tau_i)$ in order to compute the
estimator. This is particularly appealing in large-scale studies where one observes thousands of $(Y_i,\tau_i)$, making it possible to obtain an accurate estimate of $w_1(y_i;m_i,\tau_i)$ (see Section~\ref{sec:extensions} of the appendix for more details). Here, we develop a nonparametric empirical Bayes method for estimating $\bm \mu$ under the {weighted} squared error loss (Section \ref{sec:weighted_loss}) when $\tau_i$ are unknown. %jointly estimating the mean and variance under the squared error loss. 
%The special setting of two-parameter exponential families {with a known nuisance parameter} is considered in Appendix~\ref{Sec:Extensions}. 
\subsection{The weighted squared error loss}
\label{sec:weighted_loss}
Let $\bm{\delta}=(\delta_1, \ldots, \delta_n)^T$ be an estimator for $\bm{\mu}$ based on $(\bm Y,\bm S)$. Consider a class of loss functions
\begin{equation}
\label{eq:loss_funcs}
\ell^{(p)}(\mu_i,\delta_i;\tau_i)=\tau_i^p(\mu_i-\delta_i)^2,
\end{equation}
for $p\in\{0,1\}$, where $\ell^{(0)}(\mu_i,\delta_i;\tau_i)$ represents the usual squared error loss and $\ell^{(1)}(\mu_i,\delta_i;\tau_i)=\tau_i(\mu_i-\delta_i)^2$ is the weighted squared error loss with weight given by the precision $\tau_i$. 
\begin{figure}[!t]
	\centering
	\includegraphics[width=0.9\linewidth]{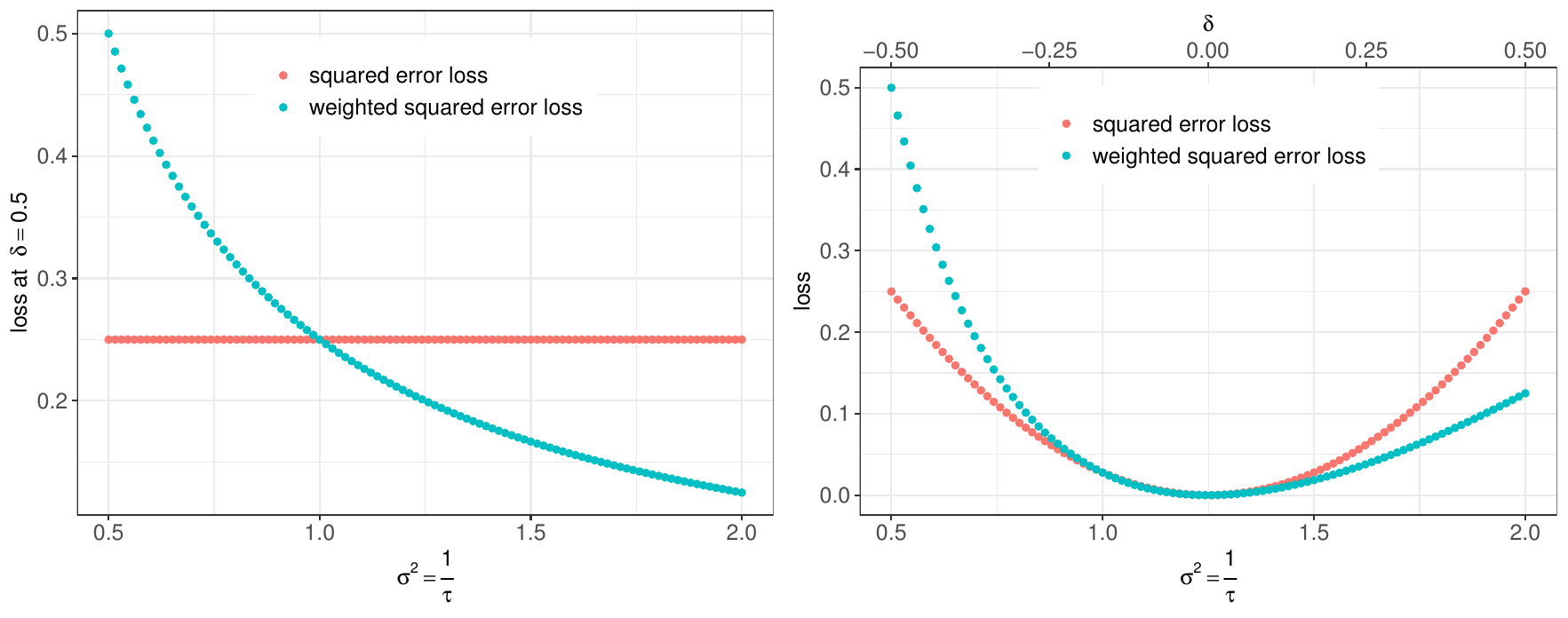}
	\caption{A toy illustration of the squared error and weighted squared error losses. Here $n=100$ and $\mu_i=0$ for all $i=1,\ldots,n$. The variances $\sigma_i^2=1/\tau_i$ are $n$ equispaced numbers between $0.5$ and $2$. In the right panel $\delta_i$ are $n$ equispaced numbers between $-0.5$ and $0.5$. The red dots represent the squared error losses while the green dots represent the corresponding weighted squared error losses for estimating $\mu_i$ using $\delta_i$ for $i=1,\ldots,n$.}
	\label{fig:newloss}
\end{figure}	

While the squared error loss treats any two study units equally with respect to their precisions $\tau_i$, the weighted squared error loss $\ell^{(1)}(\mu_i,\delta_i;\tau_i)$ is a particularly natural loss function which effectively captures the underlying heterogeneity across the study units and appropriately adjusts their contributions to the total estimation error. Consider Model \eqref{eq:normmodel}. 
Unlike the squared error loss, \(\ell^{(1)}(\mu_i,\delta_i;\tau_i)\) is invariant under a linear transformation of the data from \(Y_{ij}\) to \(\tilde{Y}_{ij}\); that is, 
\[
\ell^{(1)}(\mu_i,\delta_i;\tau_i)=\ell^{(1)}(\tilde{\mu}_i,\tilde{\delta}_i;\tilde{\tau}_i), 
\]
where 
$\tilde{Y}_{ij}=a_i+b_iY_{ij},$ $\tilde{\mu}_{i}=a_i+b_i\mu_{i}, \tilde{\tau}_i=\frac{\tau_i}{b_i^2}$, and $\tilde{\delta}_i=a_i+b_i\delta_i, $
for \((a_i,b_i)\in \mathbb{R}^2\) with \(b_i \ne 0\).

Moreover, the weighting scheme alleviates potential concerns regarding the estimation accuracy achieved for each study unit, since accurately estimating \(\mu_i\) when \(\tau_i\) is small is more challenging than when \(\tau_i\) is large. Intuitively, the weighted loss down-weights the impact of observations with extremely large variances, allowing for a relatively larger estimation error when the precision parameters are small while penalizing the estimation error more when the precision parameters are large. In Figure \ref{fig:newloss} we present a toy illustration of this phenomenon. We fix $n=100$ and $\mu_i=0$ for all $i=1,\ldots,n$. The variances $\sigma_i^2=1/\tau_i$ are $n$ equispaced numbers between $0.5$ and $2$. The red dots in Figure \ref{fig:newloss} represent the squared error losses while the green dots represent the corresponding weighted squared error losses for estimating $\mu_i$ using $\delta_i$ for $i=1,\ldots,n$. In the left panel of Figure \ref{fig:newloss} we see that when the precisions are large, the same $\delta_i$, say $\delta_i=0.5$, incurs a relatively larger loss under $\ell^{(1)}(\mu_i,\delta_i;\tau_i)$ than when the precisions are small. In the right panel, we pair each $\sigma_i^2$ with a $\delta_i$, where $\delta_i$ are $n$ equispaced numbers between $-0.5$ and $0.5$, and calculate the two losses. Both the left and the right panels reveal that $\ell^{(0)}(\mu_i,\delta_i;\tau_i)$ does not adjust the estimation error in response to the underlying heteroskedasticity represented by the precisions. 

In the main text we focus on the weighted squared error loss. The next subsection presents the oracle NEST estimator of $\bm \mu$ under this loss; the analysis under the usual squared error loss is presented in Section \ref{sec:squared_loss_app}.

%%%%%%%%%%%%%%
\subsection{The oracle NEST estimator under the weighted squared error loss}
\label{GT:subsec}
Denote the average loss $l_n^{(p)}(\bm{\mu},\bm \delta;\bm \tau)=n^{-1}\sum_{i=1}^{n}\ell^{(p)}(\mu_i,\delta_i;\tau_i)$. The compound Bayes risk is
\begin{equation}\label{eq:risk_se}
r_p(\bm{\delta}, \mathcal{G}) =  \mathbb{E}\{l_n^{(p)}(\bm\mu,\bm\delta;\bm \tau)\}=\dfrac{1}{n}\sum_{i=1}^{n}\int \int\int \ell^{(p)}(\mu_i,\delta_i;\tau_i)f_{m_i}(y, s^2 \mid\bm\psi_i) \mathrm{d}{y}\mathrm{d}{s^2}\mathrm{d}\mathcal{G}(\bm \psi_i),
\end{equation}
where $\bm \psi_i=(\mu_i,\tau_i)$, $\mathcal{G}(\bm\psi_i)=G_\mu(\mu_i|\tau_i)H_\tau(\tau_i)$, and $f_{m_i}(y, s^2 \mid\bm\psi_i)$ is the likelihood function of $(Y_i, S_i^2)$. % conditioned on $\bm\psi_i$}.
The Bayes estimator that minimizes Equation \eqref{eq:risk_se} is given by $\pmb\delta^\pi_{(p)}=(\delta^\pi_{1,(p)}, \ldots, \delta^\pi_{n,(p)})$, where 
\begin{equation}
\label{eq:bayes}
\delta^\pi_{i,(p)}\coloneqq \delta^\pi_{(p)}({y}_i,s_i^2,m_i)=\dfrac{\mathbb{E}(\tau_i^p\mu_i~|~{y}_i, s_i^2,m_i)}{\mathbb{E}(\tau_i^p~|~{y}_i, s_i^2,m_i)}.   
\end{equation}

Denote $\zeta_i\coloneqq\tau_i\mu_i$. In Definition \ref{def:oraclenest} below, we present the oracle NEST estimator of $\bm \mu$ under the weighted squared error loss which is the Bayes estimator $\bm \delta^{\pi}_{(1)}$. The key idea is to exploit the exponential family representation of the posterior distribution of $(\mu_i,\tau_i)$ to construct a Tweedie-type \red{formula} for $\delta^{\pi}_{i,(1)}$.
\begin{definition}
	\label{def:oraclenest}	
	Consider hierarchical Model~\eqref{eq:normmodel} with $m_i>3$. Then the oracle NEST estimator of $\mu_i$ under the weighted squared error loss is $\delta_{i,(1)}^{\pi}$ where
	\begin{eqnarray}
	\label{eq:oraclenest1}
	\delta_{i,(1)}^{\pi}\coloneqq \delta^{\pi}_{(1)}(y_i,s_i^2,m_i)=\dfrac{\mathbb{E}(\zeta_i|y_i,s_i^2,m_i)}{\mathbb{E}(\tau_i|y_i,s_i^2,m_i)}
	%&=& y_i+\dfrac{1}{m_i\hat{\tau}^{\pi}(y_i,s_i^2,m_i)}w_1(y_i,s_i^2,m_i)\nonumber\\
	&=&y_i+\dfrac{s_i^2}{m_i}\gamma(y_i,s_i^2,m_i)w_1(y_i,s_i^2,m_i), 
	\end{eqnarray}
	$$ \mbox{and} \quad \gamma(y_i,s_i^2,m_i) = \dfrac{m_i-1}{m_i-3-2s_i^2w_2(y_i,s_i^2;m_i)}.$$
\end{definition}
We first discuss the oracle NEST estimator in Equation \eqref{eq:oraclenest1} and then explain its genesis.

In Equation \eqref{eq:oraclenest1} $\delta_{i,(1)}^{\pi}$ is a ratio of $\hat{\zeta}^{\pi}_i\coloneqq \mathbb{E}(\zeta_i|y_i,s_i^2,m_i)$ and $\hat{\tau}_i^{\pi}\coloneqq \mathbb{E}(\tau_i|y_i,s_i^2,m_i)$, and it involves the shrinkage factor $\gamma_i\coloneqq \gamma(y_i,s_i^2,m_i)$ that should be applied to $s_i^2$ when $\tau_i$ is unknown. This  shrinkage factor is, by construction, positive and depends on $w_{2,i}\coloneqq w_2(y_i,s_i^2;m_i)$ that controls the magnitude of shrinkage that is applied to the sample variance $s_i^2$. In practical applications, however, the score functions $w_{1,i}$ and $w_{2,i}$ are unknown. In Section \ref{sec:opt_criteria}, we develop a data-driven NEST estimator with estimated scores.

{The particularly useful representation of $\delta_{i,(1)}^{\pi}$ in Equation \eqref{eq:oraclenest1} relies on the construction of Tweedie's formulae for the natural parameters $(\zeta_i,\tau_i)$ under Model \eqref{eq:normmodel}. Specifically, we represent the posterior distribution of $(\mu_i,\tau_i)$ as a two-parameter exponential family with natural parameters $(\zeta_i,\tau_i)$ which allows one to explicitly compute the posterior means of $\zeta_i$ and $\tau_i$ as follows:}
\begin{eqnarray}
\hat{\zeta}^{\pi}_i\coloneqq \hat{\zeta}^{\pi}(y_i,s_i^2,m_i)=\mathbb{E}(\zeta_i|{y_i},s^2_i,m_i)&=&{y_i}\mathbb{E}(\tau_i|{y_i},s^2_i,m_i)+m^{-1}_iw_{1}({y_i},s^2_i;m_i),\label{eq:generalized-Tweedie} \\
\hat{\tau}_i^{\pi}\coloneqq \hat{\tau}^{\pi}(y_i,s^2_i,m_i)=	\mathbb{E}(\tau_i|{y_i},s^2_i,m_i)&=&\dfrac{m_i-3}{(m_i-1)s^2_i}-\dfrac{2}{m_i-1}w_{2}({y_i},s^2_i;m_i).\label{eq:generalized-Tweedie-var}
\end{eqnarray}

Equations \eqref{eq:generalized-Tweedie} and \eqref{eq:generalized-Tweedie-var}, which are proven in Section \ref{sec:prf_eqs}, represent the natural parameter Tweedie's formulae for the parameters $(\zeta_i,\tau_i)$ under Model \eqref{eq:normmodel} and are significant for several reasons. First, they generalize Tweedie's formula with known variances [Equation \eqref{eq:twftau}], which can be recovered from Equation \eqref{eq:generalized-Tweedie} by treating $\tau_i$ as a known constant and dividing both sides by $\tau_i$. Second, $\hat{\tau}_i^{\pi}$ in Equation \eqref{eq:generalized-Tweedie-var} has an interesting interpretation. Apart from being the Bayes estimate of $\tau_i$ under the squared error loss, $1/\hat{\tau}_i^{\pi}$ is the Bayes estimate of $1/\tau_i$ under Stein's loss \citep{JamSte61}.
Third, Equations \eqref{eq:generalized-Tweedie} and \eqref{eq:generalized-Tweedie-var} can be used for compound estimation of the parameters $\theta_i=T(\zeta_i,\tau_i,m_i)$, where $T(\cdot)$ is a known function of $(\zeta_i,\tau_i,m_i)$. For instance, in finance applications the Sharpe ratio for a portfolio $i$ is given by $T(\zeta_i,\tau_i,m_i)=\zeta_i/\sqrt{\tau_i}$ and one can construct an estimator $\hat{\theta}_i\coloneqq T(\hat{\xi}_i^{\pi},\hat{\tau}_i^{\pi},m_i)$ using Equations \eqref{eq:generalized-Tweedie} and \eqref{eq:generalized-Tweedie-var} to estimate $\theta_i$. In Section \ref{sec:mfdata} of the appendix we present a real data application that is dedicated to the compound estimation of Sharpe ratios for a large number of mutual fund portfolios. However we note that $\hat{\theta}_i$ contructed in the aforementioned fashion may not, in general, be the optimal estimator of $\theta_i$ under the squared error or the weighted squared error loss considered here.

%%%%%%%%%%%%%%%%%%%%%%%%%%%%%

\section{The data-driven NEST estimator}
\label{sec:opt_criteria}
{In this section we discuss the estimation of the shrinkage factors and introduce the data-driven NEST estimator in Definition \ref{def:neb} below.} We begin by introducing some notation. Let $\bm x=({y},s^2)$ be a generic pair of observations from the distribution with marginal density $f_m(\bm x)$, which we assume is continuously differentiable with support on $\mathcal{X}\subseteq\mathbb{R}\times \mathbb{R}^{+}$. Denote the score function 
\begin{equation}
\label{eq:score_func}
\bm w(\bm x;m)=\nabla_{\bm x} \log f_m(\bm x)\coloneqq \left\{w_1(\bm x;m),w_2(\bm x;m)\right\}.   
\end{equation}
Next, for $i=1, \ldots, n,$ let $\bm z^{i} =(\bm x^{i},m_i)$, where $\bm x^{i}$ denotes the observed value of random variable $\bm X^{i}$ with density $f_{m_i}(\bm x)$.

%----------------------------------------------------------
%We  denote $z_k^i$ to be the $k^{th}$ element of $\bm z^i$ where $k=1,2$. 

\subsection{Convex optimization}
\label{sec:convex_opt}
We first describe the methodology and then provide explanations. Let
\begin{equation}
\label{eq:rbf_kernel}
\mathcal K_\lambda(\bm z^i,\bm z^j)=\exp\Bigr\{-\dfrac{1}{2\lambda^2}(\bm z^i-\bm z^j)^T\hat \Omega(\bm z^i-\bm z^j)\Bigl\}
\end{equation} 
be the Radial Basis Function (RBF) kernel with bandwidth parameter $\lambda>0$, with $\hat{\Omega}$ being the inverse of the sample covariance matrix of $(\bm z^{1},\ldots,\bm z^{n})$. Denote
\begin{equation}\label{eq:W0}
\mathcal W^{n\times 2}_0=\left\{\bm w(\bm x^1;m_1),\ldots, \bm w(\bm x^n;m_n)\right\}^T,
\end{equation}
where $\bm w(\bm x^i;m_i)=\nabla_{\bm x} \log f_{m_i}(\bm x)\Big|_{\bm x=\bm x_i}\coloneqq\left\{w_1(\bm x^i;m_i),w_2(\bm x^i;m_i)\right\}.$ 
We denote $w_k(\bm x^i;m_i)$ by $w_{k,i}$, $k=1,2$, for the remainder of this article.

Let $\nabla_{z_k^j}\mathcal K_\lambda(\bm z^i,\bm z^j)$ be the partial derivative of $\mathcal K_\lambda(\bm z^i,\bm z^j)$ with respect to the $k^{th}$ component of $\bm z^j$. The following matrices are needed in our proposed estimator:
$$
\bm K_\lambda^{n\times n} = [K_{ij,\lambda}]_{1\leq i\leq n, 1\leq j\leq n}, \quad \bm \nabla\bm K_\lambda^{n\times 2}=\left[\nabla K_{ik,\lambda}\right]_{1\leq i\leq n, 1\leq k\leq 2},
$$ 
where $K_{ij,\lambda}=\mathcal{K}_\lambda(\bm z_i,\bm z_j)$ and $\nabla K_{ik,\lambda}=\sum_{j=1}^{n}\nabla_{z_k^j}\mathcal K_\lambda(\bm z^i,\bm z^j)$. Next we formally define our proposed Nonparametric Empirical-Bayes Structural Tweedie (NEST) estimator.
\begin{definition}
	\label{def:neb}	
	Consider  Model~\eqref{eq:normmodel} with $m_i>3$. For a fixed bandwidth parameter $\lambda>0$, let $\hat{\mathcal W}_n(\lambda)=\{\hat{\bm w}_{\lambda,n}(1),\ldots,\hat{\bm w}_{\lambda,n}(n)\}^T$, where $\hat{\bm w}_{\lambda,n}(i) = \{\hat{w}^{(1)}_{\lambda,n}(i), \hat{w}^{(2)}_{\lambda,n}(i)\}^T$, be the solution to the following quadratic optimization problem:
	\begin{eqnarray}
	\label{eq:quad_opt}
	&&\min_{\mathcal W\in \mathbb{R}^{n\times 2}}\dfrac{1}{n^2}{\sf trace}\Big(\mathcal W^{T}\bm K_\lambda\mathcal W+2\mathcal W^{T}\bm \nabla\bm K_\lambda\Big).%+\rho_n(\mathcal W;\lambda),%\dfrac{\lambda}{n^2}\|\mathcal W\|_F^2,
	\end{eqnarray}	
	%where $\rho_n(\mathcal W;\lambda)$ is a penalty on the elements of $\mathcal W$. %$\|\mathcal W\|_F$ is the Frobenius norm of $\mathcal W$. 
	Then the NEST estimator for $\mu_i$ is
	{\begin{eqnarray}
		\label{eq:proposedest_theta}
		\delta^{\sf ds}_{i,n}(\lambda)&=&{y_i}+\dfrac{s_i^2}{m_i}\hat{\gamma}_{i,n}(\lambda)\hat{w}^{(1)}_{\lambda,n}(i),%{y_i}+\dfrac{1}{m_i{\tau}^{\sf ds}_i(\lambda)}\hat{w}_{1,\lambda}^i,
		\\
		\label{eq:proposedest_tau}
		\text{where }\hat{\gamma}_{i,n}(\lambda)&=&\dfrac{m_i-1}{m_i-3-2s_i^2\hat{w}^{(2)}_{\lambda,n}(i)},%{\tau^{\sf ds}_i(\lambda)}&=&\dfrac{m_i-3}{(m_i-1)s_i^2}-\dfrac{2}{m_i-1}\hat{w}_{2,\lambda}^i,
		\end{eqnarray}} 
	with the superscript {\small \sf ds} denoting ``double shrinkage''. 
\end{definition}	

\begin{remark} 
	A key characteristic of $\delta^{\sf ds}_{i,n}(\lambda)$ in Equation \eqref{eq:proposedest_theta} is that it exploits the joint structural information available in both ${Y}_i$ and $S_i^2$ through $\hat{\mathcal{W}}_n(\lambda)$. Since the weighted loss function involves both the mean and the variance, we perform shrinkage on both these dimensions. Inspecting Equations \eqref{eq:proposedest_theta} and \eqref{eq:proposedest_tau}, we expect that the improved accuracy achieved by {$\hat{\gamma}_{i,n}(\lambda)$} %$\tau^{\sf ds}_i(\lambda)$
	will lead to better shrinkage factors for $\delta^{\sf ds}_{i,n}(\lambda)$ and hence additional reduction in the estimation risk. Our numerical results in Sections \ref{sec:simulations} and \ref{sec:realdata} confirm this intuition, demonstrating that the proposed NEST estimator dominates other shrinkage estimators across many settings.  In Section~\ref{sec:extensions} of the appendix we adopt a similar strategy to extend the estimation framework presented in Definition~\ref{def:neb} to other distributions in the two-parameter exponential family where the nuisance parameter is known.
\end{remark} 

Although not immediately obvious, we show in Section~\ref{Sec:theory} that, under the compound estimation setting, minimizing the objective function \eqref{eq:quad_opt} is asymptotically equivalent to minimizing the kernelized Stein’s discrepancy (KSD;  \citealp{liu2016kernelized,chwialkowski2016kernel}). Roughly speaking, the KSD measures how far a given $n\times 2$ matrix $\mathcal W$ is from the true score matrix $\mathcal W_0$. A key property of the KSD is that it is always non-negative and is equal to 0 if and only if $\mathcal W$ and $\mathcal W_0$ are equal. Hence, solving the convex program \eqref{eq:quad_opt} is equivalent to finding a $\hat {\mathcal W}$ that is as close as possible to $\mathcal W_0$. Since the {oracle NEST estimator in Definition \ref{def:oraclenest}} 
is constructed based on $\mathcal W_0$, we can expect that the {data-driven} NEST estimator based on $\hat{\mathcal W}_n(\lambda)$ would be asymptotically close to its oracle counterpart. 
Theory underpinning this intuition, as well as the asymptotic optimality of $\delta^{\sf ds}_i(\lambda)$, 
are established in Section~\ref{Sec:theory}.

%The second term in Equation \eqref{eq:quad_opt}, $\rho_n(\mathcal W;\lambda)$, is a penalty function that increases as the elements of $\mathcal W$ move further away from $0$. In this article, we take 
%$\rho_n(\mathcal W;\lambda)=(\lambda/n^2)\sum_{i=1}^{n}\sum_{k=1}^{2}\mathcal W_{ik}^2=(\lambda/n^2)\|\mathcal{W}\|_F^2,$ where $\mathcal W_{ik}$ are the elements of matrix  $\mathcal W$ and $\|\mathcal W\|_F$ is the Frobenius norm of $\mathcal W$. An alternative approach, as pursued in  \cite{james2020irrational}, is to penalize the lack of smoothness in $\bm w$ as a function of $\bm x$. \red{While the first term in Equation \eqref{eq:quad_opt} is an asymptotically unbiased estimator of the KSD (Theorem \ref{thm:spq}), the regularization parameter $\lambda$ in the second term controls the bias-variance trade-off in the score function estimate. For instance, a small value of $\lambda$ allows unbiasedness but the resulting $n$ dimensional score function estimator has more variance compared to when $\lambda$ is large, which forces the estimated shrinkage factors towards $0$, and in the limit the NEST estimate is simply the unbiased estimate $y_i$. In Section \ref{Subsec:bandwidth} we discuss a data-driven approach to choose $\lambda$ for practical implementation of the NEST estimator.}

The bandwidth parameter \(\lambda\) in the RBF kernel \(\mathcal K_\lambda\) [Equation \eqref{eq:rbf_kernel}] controls the bias–variance trade-off in the score function estimate. For instance, a small value of \(\lambda\) reduces the bias, but the resulting \(n\)-dimensional score function estimator has greater variance compared to when \(\lambda\) is large. This drives the estimated shrinkage factors toward \(0\); in the limit, the NEST estimate becomes simply the unbiased estimate \(y_i\). In Section \ref{Subsec:bandwidth}, we discuss a data-driven approach to choosing \(\lambda\) for the practical implementation of the NEST estimator.

We end this section with a discussion of the simpler case of equal sample sizes, i.e. $m_i=m$ for all $i$. For instance, the leukemia dataset analyzed in \cite{jing2016sure} consists of the expression levels of $n=5{,}000$ genes for $m=27$ acute lymphoblastic
leukemia patients. The heterogeneity across the $n$ units in this case is due to the intrinsic variability instead of the varied number of replicates. When the $m_i$'s are equal, the RBF kernel $\mathcal{K}_\lambda(\cdot,\cdot)$ needs to be modified to avoid a singular sample covariance matrix. Denote $\mathcal K_\lambda(\bm x^i,\bm x^j)=\exp\{-0.5\lambda^{-2}(\bm x^i-\bm x^j)^T\hat{\Omega}(\bm x^i-\bm x^j)\}$ the modified RBF kernel with $\hat{\Omega}^{2\times 2}$ being the inverse of the sample covariance matrix of $(\bm x^{1},\ldots,\bm x^{n})$.
Correspondingly in Definition~\ref{def:neb}, $\hat{\mathcal W}_n(\lambda)$ are the estimates of the shrinkage factors $\mathcal W^{n\times 2}_0=\left\{\bm w(\bm x^1;m),\ldots, \bm w(\bm x^n;m)\right\}^T$, where $\bm w(\bm x^i;m)=\left\{w_1(\bm x^i;m),w_2(\bm x^i;m)\right\}^T\coloneqq(w_{1,i}, w_{2,i})^T$. %and equations \eqref{eq:proposedest_theta} and \eqref{eq:proposedest_tau} continue to provide the double shrinkage estimate of $\mu_i$ with $m_i$ replaced by $m$. 

\subsection{Details around implementation}
\label{Subsec:bandwidth}

In this section we discuss details around the implementation of NEST. First note that Equation \eqref{eq:quad_opt} can be solved separately for the two columns of $\mathcal{\bm W}$, which respectively yield the estimates for $w_{1,i}$ and $w_{2,i}$. Next, the solution to Equation \eqref{eq:quad_opt} is available in the closed form of $\hat{\mathcal W}_n(\lambda)=-\bm K_\lambda^{-1}\bm \nabla\bm K_\lambda$. However, in our implementation the closed form solution is replaced by a convex program that directly solves \eqref{eq:quad_opt} with the following two constraints: 
\begin{description}
	\item (1) $\mathcal{\bm W}\bm a\preceq \bm b$, where  $\bm a = (0,1)^T$ and $\bm b =(b_1,\ldots,b_n)$ with $b_i= \frac 1 2(m_i-3)/s_i^2-\kappa$ for some $\kappa>0$;
	\item (2) $\bm 1^T\mathcal W\bm e_k=0,~k=1,2$, where $\bm e_k$ is the canonical basis vector with $1$ at coordinate $k$. 
\end{description}
Inspecting Equation~\eqref{eq:proposedest_tau} shows that adding the first constraint guarantees that {$\hat{\gamma}_{i,n}(\lambda)<\infty$}. Hence adding this constraint is desirable in numerical implementations; similar ideas have been used in the seminal work of \cite{KoeMiz14}. The second constraint ensures that the sample mean of the $k^{th}$ estimated shrinkage factor is zero since the expectation of the gradient of the log marginal density is indeed zero.

The practical implementation requires a data-driven scheme for choosing $\lambda$. 
We propose to use a variation of the modified cross validation scheme of \cite{brown2013poisson}, which involves splitting $Y_{ij}$ into two parts:  
{$
	U_{ij}=Y_{ij}-(1/\alpha)\epsilon_{ij}$ and $V_{ij}=Y_{ij}+\alpha\epsilon_{ij}, 
	$}
where $\epsilon_{ij} \sim N(0, \tau_i^{-1})$, and $U_{ij}$ and $V_{ij}$ are used to construct the estimator and to choose the tuning parameter, respectively. However in our setup $\tau_i$ is unknown; hence we {define $U_{ij}=Y_{ij}-(1/\alpha)\sqrt{S_i^2}\epsilon_{ij}$, $V_{ij}=Y_{ij}+\alpha\sqrt{S_i^2}\epsilon_{ij}$ and sample $\epsilon_{ij}$ independently from $N(0,1)$.} Let $\bar{V}_i=m_i^{-1}\sum_{j=1}^{m_i}V_{ij}$, $\bar{U}_i=m_i^{-1}\sum_{j=1}^{m_i}U_{ij}$, $\mathcal U=\{U_{ij}: 1\leq i \leq n, 1\leq j\leq m_i\}$ and $\mathcal V=\{V_{ij}: 1\leq i\leq n, 1\leq j\leq m_i\}$. {Then conditional on $(\mu_i,\tau_i)$, $\bar{U}_i$ and $\bar{V}_i$ are uncorrelated with mean $\mu_i$ and variances $(1+\alpha^{-2})/(m_i\tau_i)$ and $(1+\alpha^2)/(m_i\tau_i)$, respectively. Define
	$$
	{\vartheta}_n(\lambda;\mathcal U,\mathcal V)=\dfrac{1}{n}\sum_{i=1}^{n}\dfrac{\left\{\bar{V}_i-\delta_{i,n}^{\sf ds}(\bar{U}_i;\mathcal{U},\lambda)\right\}^2}{S_{V,i}^2},
	$$ where $\delta_{i,n}^{\sf ds}(\bar{U}_i;\mathcal{U},\lambda)$ is the NEST estimator of $\mu_i$ based on $\mathcal{U}$ and $S^2_{V,i}$ is the sample variance of $\{V_{ij}:1\le j\le m_i\}$.}
The tuning parameter will be chosen as  $\hat{\lambda}\coloneqq\argmin_{\lambda\in\Lambda}{\vartheta}_n(\lambda)$. 
In our numerical studies of Section \ref{sec:simulations}, we set $\alpha=1/2$, {$\Lambda=[0.5,10^{2}]$}. The tuning parameter $\hat{\lambda}$ is obtained from this scheme and then used to estimate $\mu_i$ via Equation \eqref{eq:proposedest_theta}.

\begin{remark}
	\label{rem:1}
	{Note that for large $m$, $\bar{V}_i$ and $S_{V,i}^2$ are independent and $\mathbb{E}[{\vartheta}_n(\lambda;\mathcal U,\mathcal V)]$ can be well approximated by the following quantity: 
		$$n^{-1}\sum_{i=1}^{n}\mathbb E[\ell^{(1)}\{\mu_i,\delta_{i,n}^{\sf ds}(\bar{U}_i;\mathcal{U},\lambda);\tau_i/(1+\alpha^2)\}]+n^{-1}\sum_{i=1}^{n}\mathbb E(\bar{V}_i-\mu_i)^2/S^2_{V,i}$$ for any fixed $\lambda>0$ and $n$. The first term depends on $\lambda$ while the second term is independent of it. We choose the value of $\lambda$ that minimizes ${\vartheta}_n(\lambda;\mathcal U,\mathcal V)$ to construct the data-driven NEST estimator of $\bm \mu$ based on the original sample $(\bm Y,\bm S)$. For $\alpha\to 0$, this scheme is equivalent to choosing a value of $\lambda$ that minimizes the true weighted squared error loss of $\delta_{i,n}^{\sf ds}(\bar{U}_i;\mathcal{U},\lambda), i=1,\ldots,n$. We note in Section~\ref{sec:simulations} that when $m$ is small, the risk of the data-driven NEST estimator is generally higher than the risk of the NEST estimator that chooses $\lambda$ by minimizing the true weighted squared error loss of $\bm \delta^{\sf ds}_{n}(\lambda)$ based on $(\bm Y,\bm S)$ (see for example the left panels in Figures \ref{fig:simind_wse} and \ref{fig:simdep_wse_part1}). This is not unexpected since $\vartheta_n(\lambda;\mathcal U,\mathcal V)$ is not an unbiased estimator of the true risk. However, ${\vartheta}_n(\lambda;\mathcal U,\mathcal V)$ provides a practical criterion for choosing $\lambda$ and works well empirically in our numerical studies for larger $m$.} The development of an {unbiased risk estimate, such as a SURE-type criterion,} for this setting is a challenging  topic requiring further research. See also \cite{ignatiadis2019covariate} for related discussions.
\end{remark}

The \texttt{R} code that implements the NEST estimator and reproduces the numerical results in this paper can be downloaded from the link:
{\url{https://www.dropbox.com/sh/5yptcj4epxdgbqs/AADAyHZNDCv4Hqagsi97vC2Da?dl=0}}.

{\begin{remark}
		In the theory section below, we assume that (a) both the true density and the true score function are Lipschitz continuous, and that (b) the objective function is minimized over a class of score functions confined within a Lipschitz ball. While unconstrained optimization is performed in practice, it is necessary for the theoretical proofs to restrict the optimization to a tractable function class. See Remark \ref{rem:lip} in Section \ref{theory:subsec} for further discussions. 
	\end{remark}
}

%%%%%%%%%%%%%%%%%%%%%%%%%%%%%%%%
\section{Theory}
\label{Sec:theory}
%%%%%%%%%%%%%%%%%%%%%%%%%%%%%%%%
In this section we introduce the Kernelized Stein's Discrepancy (KSD) measure \citep{liu2016kernelized,chwialkowski2016kernel} and discuss its connection to the quadratic program \eqref{eq:quad_opt}. While the KSD has been used in various contexts including goodness of fit tests \citep{liu2016kernelized,yang2018goodness}, variational inference \citep{liu2016stein} and Monte Carlo integration \citep{oates2017control}, its connections to compound estimation and empirical Bayes methodology was established only recently \citep{banerjee2019general,luotransfer}. The analysis in this and following sections is geared towards the case $m_i=m$ for $i=1,\ldots,n$. Under this setting, $(\bm X^{1},\ldots,\bm X^{n})$ constitute an i.i.d random sample from $f_m(\bm x)$. The case of unequal $m_i$'s can be analyzed in a similar fashion %by assuming that $m_i$'s are a random sample from a distribution with mass function $q(\cdot)$. Then 
\red{where} $\bm Z = (\bm X,m)$ has distribution with density $p(\bm z)\coloneqq q(m)f_m(\bm x)$ \red{and with} %where 
$\bm Z^{i}=(\bm X^{i},m_i)$, $(\bm Z^{1},\ldots,\bm Z^n)$ are an i.i.d random sample from $p(\bm z)$.

\subsection{Kernelized Stein's Discrepancy}
\label{sec:ksd}
Suppose $\bm X$ and $\bm X'$ are i.i.d. copies from the marginal distribution of $(Y,S^2)$ that has density $f$ wherein the dependence on $m$ is implicit. Denote \(\bm w(\bm X)\) and \(\bm w(\bm X')\), as defined in Equation \eqref{eq:score_func}, to be the score functions at \(\bm X\) and \(\bm X'\), respectively. Suppose $\tilde{f}$ is an arbitrary density function on the support of $(Y,S^2)$, for which we similarly define $\tilde{\bm w}(\bm X)$. The KSD, formally defined as 
\begin{equation}
\label{eq:ksd}
\mathcal{S}_\lambda(f,\tilde{f})=\mathbb{E}_{\bm X,{\bm X'}\sim f}\Big[\Bigl\{\tilde{\bm w}(\bm X)-\bm w(\bm X)\Bigr\}^T\mathcal{K}_\lambda(\bm X,\bm X')\Bigl\{\tilde{\bm w}(\bm X')-\bm w(\bm X')\Bigr\}\Big],
\end{equation}
provides a discrepancy measure between $f$ and $\tilde{f}$ in the sense that $\mathcal{S}_\lambda(f,\tilde{f})$ tends to increase when there is a bigger disparity between $\bm w$ and $\tilde{\bm w}$ (or equivalently, between $f$ and $\tilde f$), and
$$
\mbox{$\mathcal{S}_\lambda(f,\tilde{f})\ge0$ and $\mathcal{S}_\lambda(f,\tilde{f})=0$ if and only if $f=\tilde{f}$}.
$$
The direct evaluation of $\mathcal{S}_\lambda(f,\tilde{f})$ is difficult because $\bm w$ is unknown. \cite{liu2016kernelized} introduced an alternative representation of the KSD that does not directly involve $\bm w$: 
\begin{eqnarray}
\label{eq:sqp_kappa}
\mathcal{S}_\lambda(f,\tilde{f}) &=&\mathbb{E}_{\bm X,{\bm X'}\sim f}\kappa_\lambda[\tilde{\bm w}(\bm X),\tilde{\bm w}(\bm X')](\bm X,\bm X')\\
& = & \mathbb{E}\left\{\dfrac{1}{n(n-1)}\sum_{i=1}^{n}\sum_{j=1}^{n}\kappa_\lambda[\tilde{\bm w}(\bm X^i),\tilde{\bm w}(\bm X^j)](\bm X^i,\bm X^j){I}(i\ne j)\right\}\nonumber\\
\label{eq:sqp}
&=&\mathbb{E}\big[\bar{\mathbb{M}}_{\lambda,n}(\tilde{\mathcal W})\big]\coloneqq \mathbb{M}_\lambda(\tilde{\mathcal W}),
\end{eqnarray}
where $\mathbb X\coloneqq\{\bm X^1,\ldots,\bm X^n\}$ is an i.i.d. random sample from $f$, $\mathbb{E}$ denotes expectation under the joint distribution of $\mathbb X$ and 
\begin{eqnarray*}
	%	\label{eq:kappa}
	\kappa_\lambda[\tilde{\bm w}(\bm x),\tilde{\bm w}(\bm x')](\bm x,\bm x')&=&\tilde{\bm w}(\bm x)^T\tilde{\bm w}(\bm x') \mathcal{K}_\lambda(\bm x,\bm x')   
	+    \tilde{\bm w}(\bm x)^T \nabla_{\bm x'} \mathcal{K}_\lambda(\bm x,\bm x')
	+\nabla_{\bm x} \mathcal{K}_\lambda(\bm x,\bm x')^T \tilde{\bm w}(\bm x)\nonumber\\ 
	&+&  \text{trace}(\nabla_{\bm x}\nabla_{\bm x'} \mathcal{K}_\lambda(\bm x,\bm x')),
\end{eqnarray*}
is a smooth and symmetric positive definite kernel function associated with the U-statistic $\bar{\mathbb{M}}_{\lambda,n}(\tilde{\mathcal W})$. 
\begin{remark}
	\label{rem:2}
	The representation of the KSD [Equation \eqref{eq:ksd}] in terms of $\kappa_{\lambda}$ in Equation \eqref{eq:sqp_kappa} is stated as Theorem 3.6 in \cite{liu2016kernelized} and its proof hinges upon an important property of the RBF kernel $\mathcal K_\lambda$, namely, both $\mathcal K_\lambda(\bm x,\cdot)$ and $\mathcal K_\lambda(\cdot,\bm x)$ are in the Stein class of $f$ for any fixed $\bm x\in\mathcal X$ [cf. Definition 3.4 of \cite{liu2016kernelized}]. Formally, a function $\varphi:\mathcal X\to\mathbb R$ is in the Stein class of $f$ if $\varphi$ is smooth and satisfies $\int_{\mathcal X}\nabla_{\bm x}\{\varphi(\bm x)f(\bm x)\}\mathrm d\bm x=\bm 0$. 
\end{remark}
The implementation of the NEST estimator in Definition \ref{def:neb} boils down to the estimation of $\mathcal W_0$ via the convex program \eqref{eq:quad_opt}, which corresponds to minimizing   
\begin{equation}
\label{eq:samplecrit}
\hat{\mathbb{M}}_{\lambda,n}(\tilde{\mathcal W}) = \dfrac{1}{n^2}\sum_{i=1}^{n}\sum_{j=1}^{n}\kappa_\lambda[\tilde{\bm w}(\bm X^i),\tilde{\bm w}(\bm X^j)](\bm X^i,\bm X^j)%+\dfrac{\lambda(n)}{n^2}\|\tilde{\mathcal W}\|_F^2
\end{equation}
w.r.t. $\tilde{\mathcal W}$. A key observation is that if the empirical criterion $\hat{\mathbb{M}}_{\lambda,n}(\tilde{\mathcal W})$ is asymptotically equal to the population KSD criterion $\mathbb{M}_\lambda(\tilde{\mathcal W})$, then minimizing $\hat{\mathbb{M}}_{\lambda,n}(\tilde{\mathcal W})$ with respect to $\tilde{\mathcal W}$ is effectively the process of finding a $\tilde{\mathcal W}$ that is as close as possible to $\mathcal{W}_0$ in Equation \eqref{eq:W0}. This intuitively justifies the NEST estimator in Definition \ref{def:neb}. 

In Proposition \ref{thm:spq} below, we formally establish the asymptotic consistency of the sample criterion $\hat{\mathbb{M}}_{\lambda,n}(\tilde{\mathcal W})$ around the population criterion ${\mathbb{M}}_\lambda(\tilde{\mathcal W})$ for any fixed $\lambda>0$ and under the following regularity condition.% where $f$ denotes the density function of the joint marginal distribution of $(Y,S^2)$ and $\mathbb{E}_f$ denotes expectation under $f$.
\begin{assumption}
	\label{assump:1}
	%$\mathbb{E}_f|\kappa[\tilde{\bm w}(\bm X),\tilde{\bm w}(\bm X')](\bm X,\bm %X')|^2<\infty$.
	$\mathbb{E}_{\bm X^i,\bm X^j\sim f}|\kappa_\lambda[\tilde{\bm w}(\bm X^i),\tilde{\bm w}(\bm X^j)](\bm X^i,\bm X^j)|^2<\infty$ for any $(i,j)\in\{1,\ldots,n\}$.
\end{assumption}
{Assumption \ref{assump:1} is a standard moment condition on $\kappa_\lambda[\tilde{\bm w}(\bm X^i),\tilde{\bm w}(\bm X^j)](\bm X^i,\bm X^j)$  [see, for example, Section 5.5 in \cite{serfling2009approximation}] , which is needed for establishing the Central Limit Theorem for the U-statistic $\bar{\mathbb{M}}_{\lambda,n}(\tilde{\mathcal W})$ in Equation \eqref{eq:sqp}.}

\begin{proposition}
	\label{thm:spq}
	Suppose Assumption \ref{assump:1} holds. For any fixed $\lambda>0$, we have
	$$\big|\hat{\mathbb{M}}_{\lambda,n}(\tilde{\mathcal W})-{\mathbb{M}}_\lambda(\tilde{\mathcal W})\big|=O_p\big({n}^{-1/2}\big).$$
\end{proposition}

Along with the fact that ${\mathbb{M}}_\lambda({\mathcal W_0})=0$, Proposition \ref{thm:spq} justifies $\hat{\mathbb{M}}_{\lambda,n}(\tilde{\mathcal W})$ as an appropriate optimization criterion.
Next we show that the NEST estimator in Definition \ref{def:neb} is asymptotically close to its oracle counterpart. 

\subsection{Asymptotic Properties of NEST}\label{theory:subsec}

This section studies the asymptotic properties of the NEST estimator under the setting where $\lambda\coloneqq \lambda(n)$ varies with $n$. We begin by recalling the oracle NEST estimator %defining an oracle estimator 
$\bm \delta^{\pi}_{(1)}=(\delta^{\pi}_{1,(1)},\ldots,\delta^{\pi}_{n,(1)})$ for $\bm \mu$ in Equation \eqref{eq:oraclenest1}, where
\begin{equation*}
%	\label{eq:oracle_estimator}
\delta_{i,(1)}^{\pi}\coloneqq\delta^{\pi}_{(1)}(y_i,s_i^2,m) = \dfrac{\hat{\zeta}^{\pi}(y_i,s_i^2,m)}{\hat{\tau}^{\pi}(y_i,s_i^2,m)}={y_i+\dfrac{s_i^2}{m}\gamma(y_i,s_i^2;m_i)w_1(y_i,s_i^2,m)},%y_i+\dfrac{1}{m\hat{\tau}^{\pi}(y_i,s_i^2,m)}w_1(y_i,s_i^2,m),
\end{equation*}
and $\hat{\zeta}^{\pi}$ and $\hat{\tau}^{\pi}$ are respectively the Bayes estimators of $\tau\mu$ and $\tau$ as defined in Equations \eqref{eq:generalized-Tweedie} and \eqref{eq:generalized-Tweedie-var}. Viewing the proposed NEST estimator $\delta_{i,n}^{\sf ds}(\lambda)$ as a data-driven approximation to $\delta^{\pi}_{i,(1)}$, we study the quality of this approximation for large $n$ and fixed $m$. 
%In Proposition \ref{thm:w}, we have established the consistency of the estimated score functions $\hat{\mathcal W}_n(\lambda)$, which are obtained as the solution to the quadratic optimization problem in Equation \eqref{eq:quad_opt}. This theory will be strengthened in this section, where we establish a uniform version of the above result over a class of score functions contained within a Lipschitz ball. 

We briefly summarize the notations before proceeding. In what follows, \(\bm w_0^{(k)}=(w_{k,1},\ldots,w_{k,n})\) and \(\hat{\bm w}^{(k)}_{\lambda,n}=\{\hat{w}^{(k)}_{\lambda,n}(1),\ldots,\hat{w}^{(k)}_{\lambda,n}(n)\}\) are used to denote, respectively, the \(k^{th}\) (with \(k\in\{1,2\}\)) columns of \(\mathcal{W}_0\) and \(\hat{\mathcal W}_n(\lambda)\). The Frobenius norm of a matrix \(\bm A\) is denoted by \(\|\bm A\|_F\). For two sequences \(a_n\) and \(b_n\), the notation \(a_n\asymp b_n\) means that there exist constants \(d_1\) and \(d_2\) with \(d_2\ge d_1 >0\) such that \(d_1a_n\le b_n\le d_2a_n\) for sufficiently large \(n\).

The following regularity conditions are needed in our technical analysis. 

%\begin{remark}
%The oracle estimator $\bm \delta^{\pi}_{(1)}$, which serves as the benchmark in our theoretical analysis, is different from the Bayes estimator $\bm \delta^{\pi}$ defined in Equation \eqref{eq:bayes}: $\bm \delta^{\pi}$  has full knowledge of the prior distributions $G_\mu(\cdot|\tau)$ and $H_\tau(\cdot)$, whereas $\pmb \delta^{*}$ only has the information on the true score functions $(w_{1,i},w_{2,i})$. Under the special setting of conjugate priors, $\pmb \delta^{*}$ and  $\bm \delta^{\pi}$ coincide (Corollary \ref{cor:normal_conj}).  
%\end{remark}
%Thereafter, in corollary \ref{cor:normal_conj}, we evaluate $\delta_i^{\sf ds}(\lambda)$ under a simple setting of a Normal-Gamma conjugate prior for which $\delta^{*}$ coincides with the Bayes estimator $\delta^{\pi}$.
\begin{assumption}
	\label{assump:2}
	The joint density $f$ and its score functions are Lipschitz continuous.
\end{assumption}
\begin{assumption}
	\label{assump:eigen_bound}
	The eigenvalues of $\Omega$ are finite and bounded away from $0$.
\end{assumption}

\begin{comment}
\begin{assumption}
\label{assump:steinclass}
For any fixed $\lambda>0$ and $k\in\{1,2\}$, the estimated scores $\hat{\mathcal W}_n(\lambda)$ in Definition \ref{def:neb} satisfy the following:
\begin{enumerate}[label={(\alph*.)},
ref={\theassumption(\alph*)}]
% \dfrac{1}{n(n-1)}\mathbb E\Big[\sum_{i=1}^{n}\sum_{j\ne i=1}^{n}w_{k,i}\mathcal K_\lambda(\bm X^{i},\bm X^j)\hat{w}^{(k)}_{\lambda,n}(j)+\nabla_{X^{i}_k}\{K_\lambda(\bm X^{i},\bm X^j)\hat{w}^{(k)}_{\lambda,n}(j)\}\Big]=0\nonumber
\item \label{assump:steinclass_b}$\dfrac{1}{n(n-1)}\sum_{i=1}^{n}\sum_{j\ne i=1}^{n}\mathbb E_{\mathbb X\setminus \bm X^i}\Big[\int_{\mathcal X}\nabla_{x^{i}_k}\Bigl\{\mathcal K_\lambda(\bm x^{i},\bm X^j)\hat{w}^{(k)}_{\lambda,n}(j)f(\bm x^i)\Bigr\}\mathrm d\bm x^i\Big]=0$.
\item \label{assump:steinclass_a} Conditional on $\bm X^i$, $\mathbb E_{\mathbb X\setminus \bm X^i}\Big[\dfrac{1}{n}\sum_{j=1}^{n}\Bigl\{\mathcal K_{\lambda}(\bm X^i,\bm X^j)\hat{w}_{\lambda,n}^{(k)}(j) +\nabla_{X^{j}_k}\mathcal K_{\lambda}(\bm X^i,\bm X^j)\Bigr\}\Big]=0,~i=1,\ldots,n.$
\item \label{assump:steinclass_c}$ \dfrac{1}{n(n-1)}\sum_{i=1}^{n}\sum_{j\ne i=1}^{n}\mathbb E\Bigl\{\hat{w}_{\lambda,n}^{(k)}(j)\nabla_{X^{i}_k}\mathcal K_{\lambda}(\bm X^i,\bm X^j)
+\nabla_{X^{i}_k}\nabla_{X^{j}_k}\mathcal K_{\lambda}(\bm X^i,\bm X^j)\Bigr\}=0$. 
\item \label{assump:constrained} For all large $\|\bm X^i\|_2$, $|\hat{w}_{\lambda,n}^{(k)}(i)|=O(\|\bm X^i\|_2)$, $i=1,\ldots,n$.
\end{enumerate}
\end{assumption}
\end{comment}

{To facilitate a rigorous analysis, we consider a class of score functions \(\mathbb{W}\) consisting of all differentiable functions \(W\) that are Lipschitz continuous with Lipschitz constants bounded by \(L\) and that satisfy \(\Vert W(0)\Vert_{\infty}\leq U\), where \(L, U > 0\) are fixed constants chosen sufficiently large so that the true score \(\mathcal{W}_0\) belongs to \(\mathbb{W}\). We employ the optimization criterion \eqref{eq:quad_opt} with \(\lambda \coloneqq \lambda_n=O_p( (\log n)^{-3/4}n^{-1/8})\).

	\begin{remark}\label{rem:lip}
		Our theoretical analysis assumes that, instead of performing unconstrained optimization, we minimize over the function class \(\mathbb{W}\) to obtain the estimated score function \(\hat{\mathcal{W}}\). Although this departure from practical implementation may seem restrictive, under the regularity condition that the true score function lies within \(\mathbb{W}\), such a constraint is both reasonable and necessary. Importantly, this restriction only reflects the complexity inherent in the theoretical analysis and may not be viewed as a limitation of the scope of our proposed methodology. Specifically, constraining both the true and estimated scores to a Lipschitz ball allows us to leverage established results from empirical process theory, thereby ensuring the rigor of our analysis; without this restriction, the analysis would become intractable. 
	\end{remark}
}

{
	\begin{theorem}
		\label{thm:w}
		Under Assumptions \ref{assump:2} -- \ref{assump:eigen_bound}, for any sup-exponential $f$, there exists $\lambda_n$ such that 
		$$\lim_{n\to\infty}\mathbb{P}\Bigl\{\dfrac{1}{n}\big\|\hat{\mathcal W}_n(\lambda_n)-{\mathcal W}_0\big\|_F^2\ge \epsilon \, (\log n)^2 n^{-1/8}\Bigr\}=0,~\text{for any }\epsilon>0.$$
	\end{theorem}
	
	The proof of Theorem \ref{thm:w} consists of three key steps:  
	\begin{description}
		\item  (a) We establish the uniform convergence of the empirical measures for \(L_1\), \(L_2\), and KSD metrics to their population counterparts over the function class \(\mathbb{W}\). This is done in Lemma~\ref{lemma.g1}.
		
		\item (b) We derive inequalities relating the KSD and \(L_p\) norms, which hold uniformly over all score functions in the class. This derivation critically depends on the Lipschitz continuity of the function class (i.e., a Lipschitz ball) and the sub-exponential properties of the true density. Notably, this represents the most intricate and innovative aspect of our analysis, as it establishes a rigorous connection between the KSD for all Gaussian kernels and the \(L_p\) norm, which may be of independent interest.
		This is done in Lemma~\ref{lemma.g2}.
		
		\item (c) Leveraging the optimization criterion, we observe that the sample KSD (actually a very close approximation of it) of the selected score function achieves the minimal value over the class. Combining the results from steps (a) and (b), we demonstrate convergence in probability of the \(L_2\) error of the selected score function. %At this stage, we also optimize the asymptotic choice of the kernel bandwidth \(\lambda\) to achieve the best possible rate of convergence. 
		
	\end{description}
	
	\begin{remark}
		We would like to mention that in Theorem~\ref{thm:w}, the rate of convergence is polynomial in \(n\) (the sample size), but it is relatively slow compared to the parametric rate. This slower rate arises primarily due to transitioning from the KSD distance to the $L_2$ distance; see discussion in Sec.~\ref{sec:bounded.f}. We recognize that, in the recent literature on score function estimation (cf. \citealp{zhang2024minimax,dou2024optimal}), it is well established that sharper rates are attainable if the true density \(f\) is bounded away from zero. In our setting,  when \(f\) is bounded below, it follows directly from our proof of Theorem~\ref{thm:w} that the $L_2$ error of the proposed estimator has a parametric rate (barring some logarithmic factors). In this context, our proof technique based on Lemmas~\ref{lemma.g1} and \ref{lemma.g2} is sharp. See  Lemma~\ref{lem:bounded.f} for more details. 
	\end{remark}
	
}

Next, we provide decision-theoretic guarantees on $\bm \delta_n^{\sf ds}(\lambda)$ in relation to $\bm \delta^{\pi}_{(1)}$. We impose the following moment condition on the prior distributions of $\mu$ and $\tau$.
\begin{assumption}
	\label{assump:4}
	%$\mathbb{E}_H|1/\tau|<\infty$ where the expectation is with respect to prior distribution $H$ of $\tau$ in equation \eqref{eq:normmodel}.
	For some $\epsilon_i\in(0,1), i=1,2,3$, $\mathbb{E}_G\{\exp(\epsilon_1|\mu|)\}<\infty$, $\mathbb{E}_H\{\exp(\epsilon_2/\tau)\}<\infty$ and $\mathbb{E}_H\{\exp(\epsilon_3\tau)\}<\infty$.
\end{assumption}
The moment conditions in Assumption \ref{assump:4} ensure that with high probability $|\mu|\le \log n$ and $1/\log n\le \tau\le \log n$ as $n\to \infty$. It is likely that Assumption \ref{assump:4} can be further relaxed but we do not seek the full generality here.
{
	\begin{theorem}
		\label{thm:bayesrisk}
		Suppose Assumptions \ref{assump:2} -- \ref{assump:4} hold. For any sub-exponential $f$ and $\gamma < 1/8$ there exists $\lambda_n $ such that 
		$n^{-1}\big\|{\bm\delta}^{\sf ds}_n(\lambda_n)-\bm\delta^{\pi}_{(1)}\big\|_2^2=o_p(n^{-\gamma})$. Furthermore, under the same conditions, 
		$$\big|l_n^{(1)}\{\bm \mu,\bm \delta^{\sf ds}_n(\lambda_n);\bm \tau\}-l_n^{(1)}(\bm \mu,\bm \delta^{\pi}_{(1)};\bm \tau)\big|=o_p(n^{-\gamma/2})~\text{as}~n\to\infty. $$
	\end{theorem}
}

Theorem \ref{thm:bayesrisk} establishes the optimality theory of $\bm \delta^{\sf ds}_n(\lambda_n)$ by showing that (a) the {average weighted squared error between  ${\bm \delta}^{\sf ds}_n(\lambda_n)$ and $\bm \delta^{\pi}_{(1)}$ is vanishingly small (in an asymptotic sense)}, and (b) the estimation loss of NEST converges in probability to that of its oracle counterpart as $n\to \infty$. 

%%%%%%%%%%%%%%%%%%%%%%%%%%%%%%%%
\section{Numerical experiments}
\label{sec:simulations}
In this section we assess the performance of the NEST estimation framework for compound estimation of Normal means under the weighted squared error loss.
We focus on the hierarchical Model of Equation \eqref{eq:normmodel} and compare the following four approaches for estimating $\bm \mu$ when the precisions $\bm\tau$ are unknown: (1) the proposed NEST method where the tuning parameter $\lambda$ is chosen using a modified cross-validation approach described in Section \ref{Subsec:bandwidth}, (2) the NEST method which estimates $\lambda$ by minimizing the true loss (NEST Orc. $\lambda$), (3) the g-modelling approach of \citet{gu2017empirical, gu2017unobserved} which first estimates the joint prior distribution of $(\mu_i,\tau_i)$ using nonparametric
maximum likelihood estimation (NPMLE) techniques \citep{kiefer1956consistency,laird1978nonparametric} and then plugs them into the Bayes rule for estimating $\mu_i$ under the weighted squared error loss, and (4) the oracle Bayes estimator $\bm \delta^{\pi}_{(1)}$ in Equation \eqref{eq:oraclenest1} that has full knowledge of the prior distributions of $(\mu_i,\tau_i)$. For NPMLE we use the R-function \texttt{WGLVmix}, available within the R-package \texttt{REBayes} \citep{koenker2017rebayes}, to estimate the joint prior distribution of $(\mu_i,\sigma_i^2)$. To compute $\bm \delta^{\pi}_{(1)}$ we rely on the R-package \texttt{RStan} \citep{rstan,carpenter2017stan}.

\begin{figure}[!t]
	\centering
	\begin{subfigure}[t]{1\textwidth}
		\includegraphics[width=1\linewidth]{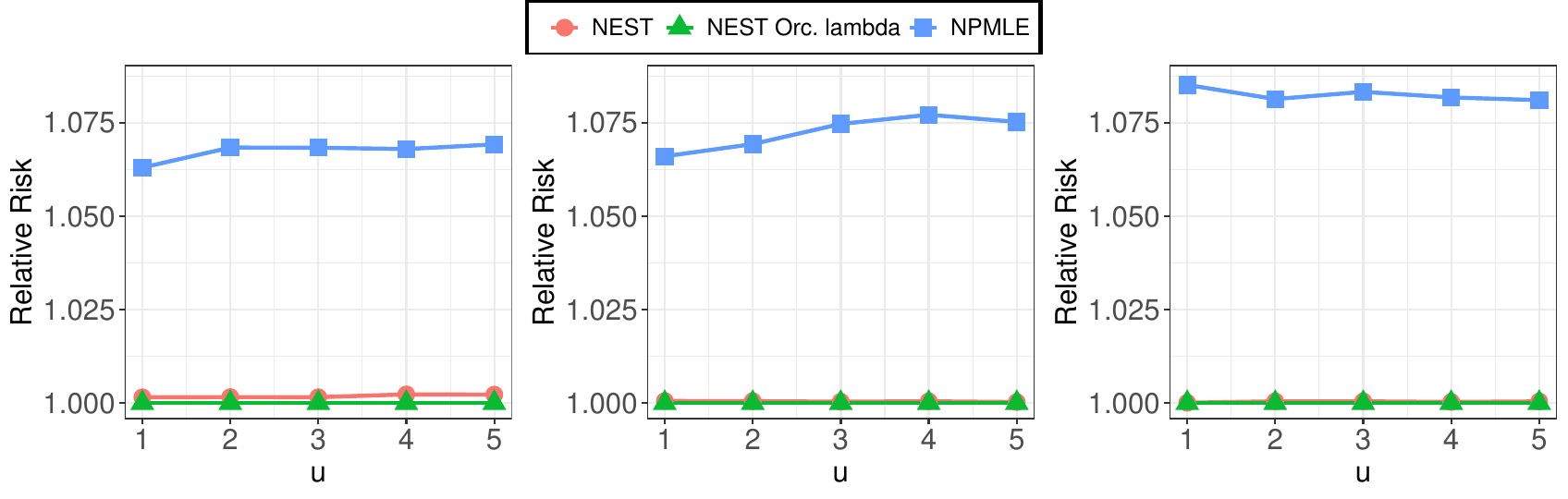}
		\caption{Setting 1: $\mu_i\stackrel{i.i.d}{\sim} N(0,3)$ and $\sigma_i^2\stackrel{i.i.d}{\sim}U(0.25,u)$.}
		\label{fig:simInd_1_wse} 
	\end{subfigure}
	\begin{subfigure}[t]{1\textwidth}
		\includegraphics[width=1\linewidth]{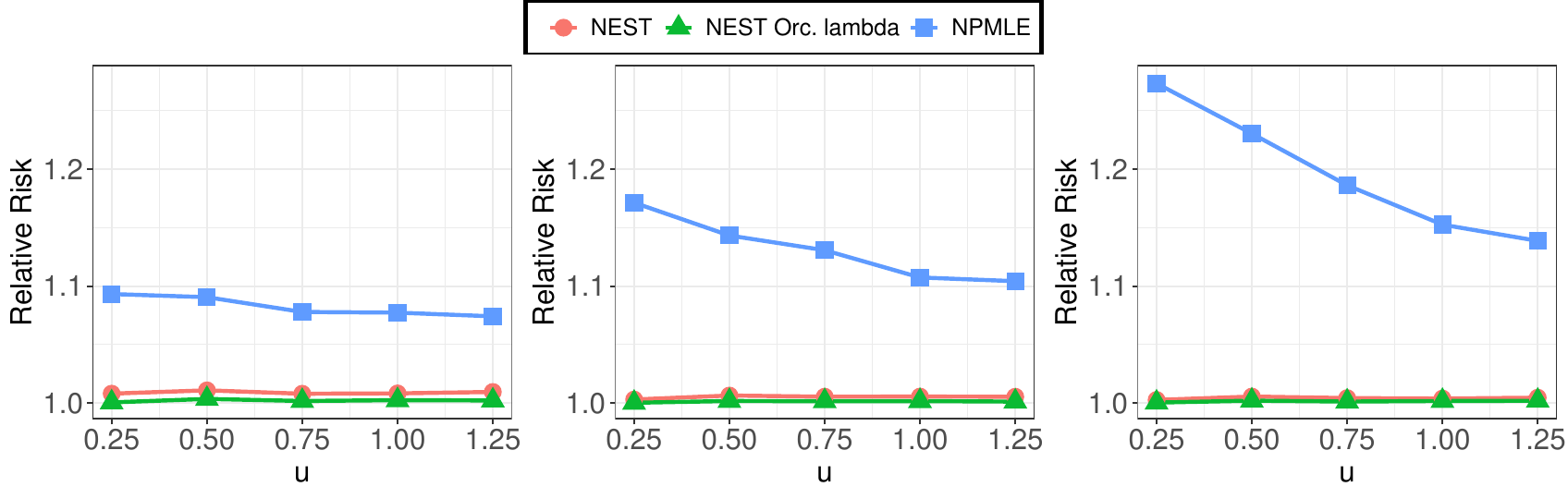}
		\caption{Setting 2: Here $\sigma_i^2\stackrel{i.i.d}{\sim}N(u,1)$ truncated below $0.1$ and $\mu_i\stackrel{i.i.d}{\sim}~{\sf Laplace}({\sf location}=0,{\sf scale}=1)$}
		\label{fig:simInd_2_wse}
	\end{subfigure}
	\caption{Comparison of relative risks when $(\mu_i, \sigma_i^2)$ are independent. Plots show $m=10,15,20$ left to right.}
	\label{fig:simind_wse}
\end{figure}

\begin{figure}[!t]
	\centering
	\begin{subfigure}[b]{1\textwidth}
		\includegraphics[width=1\linewidth]{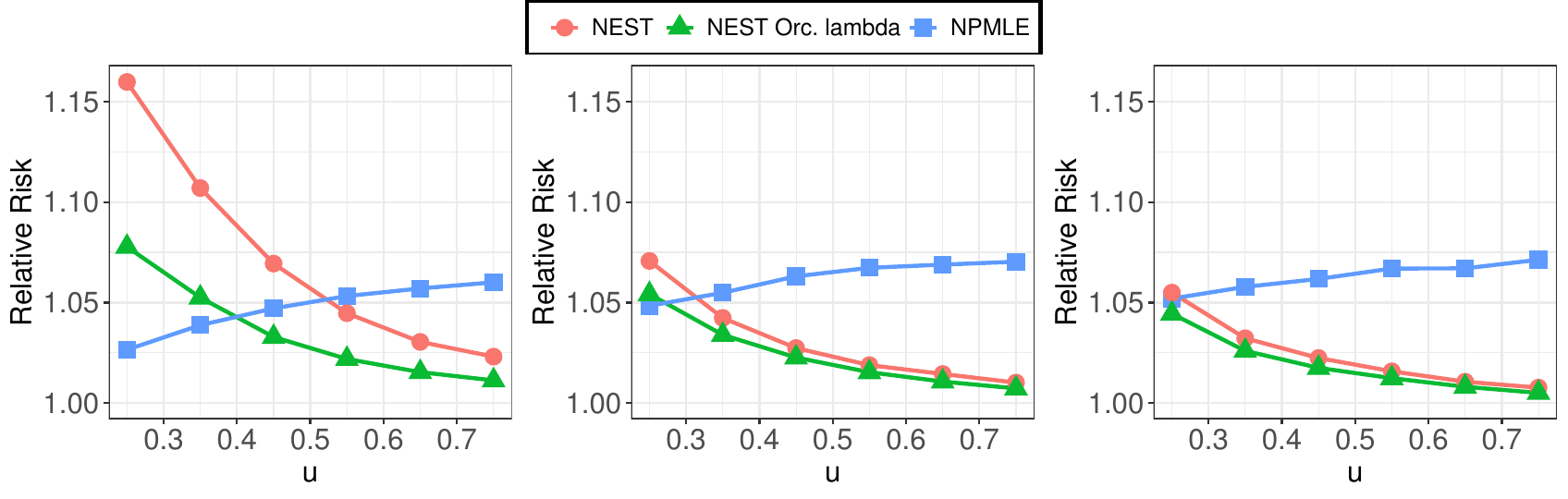}
		\caption{ Setting 3: $\sigma_i^2\stackrel{i.i.d}{\sim}U(0.25,3)$ and $\mu_i~|~\sigma_i^2\stackrel{ind.}{\sim}~0.5~N(-\sigma_i^2,u)+0.5~N(\sigma_i^2,u)$.}
		\label{fig:simdep_1_wse} 
	\end{subfigure}
	\begin{subfigure}[b]{1\textwidth}
		\includegraphics[width=1\linewidth]{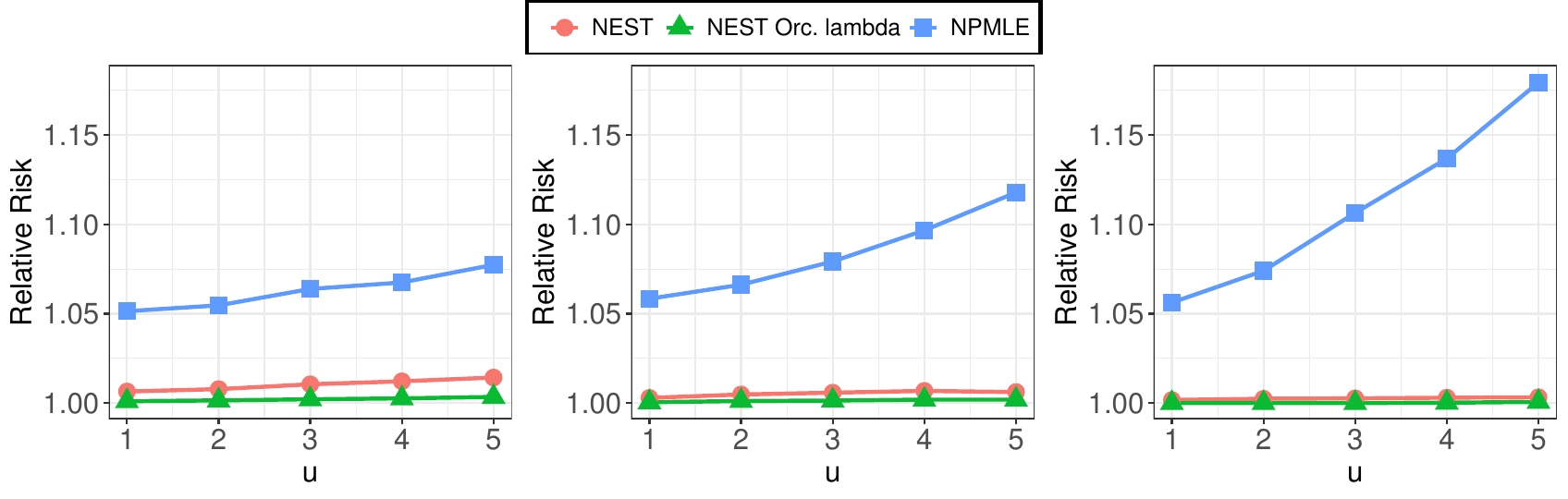}
		\caption{Setting 4: $\sigma_i^2\stackrel{i.i.d}{\sim}U(0.25,u)$ and $\mu_i~|~\sigma_i^2\stackrel{ind.}{\sim}~0.5~N(\sigma_i^2,\sigma_i^2)+0.5~N(-\sigma_i^2,\sigma_i^2)$.}
		\label{fig:simdep_2_wse}
	\end{subfigure}
	\centering
	\begin{subfigure}[b]{1\textwidth}
		\includegraphics[width=1\linewidth]{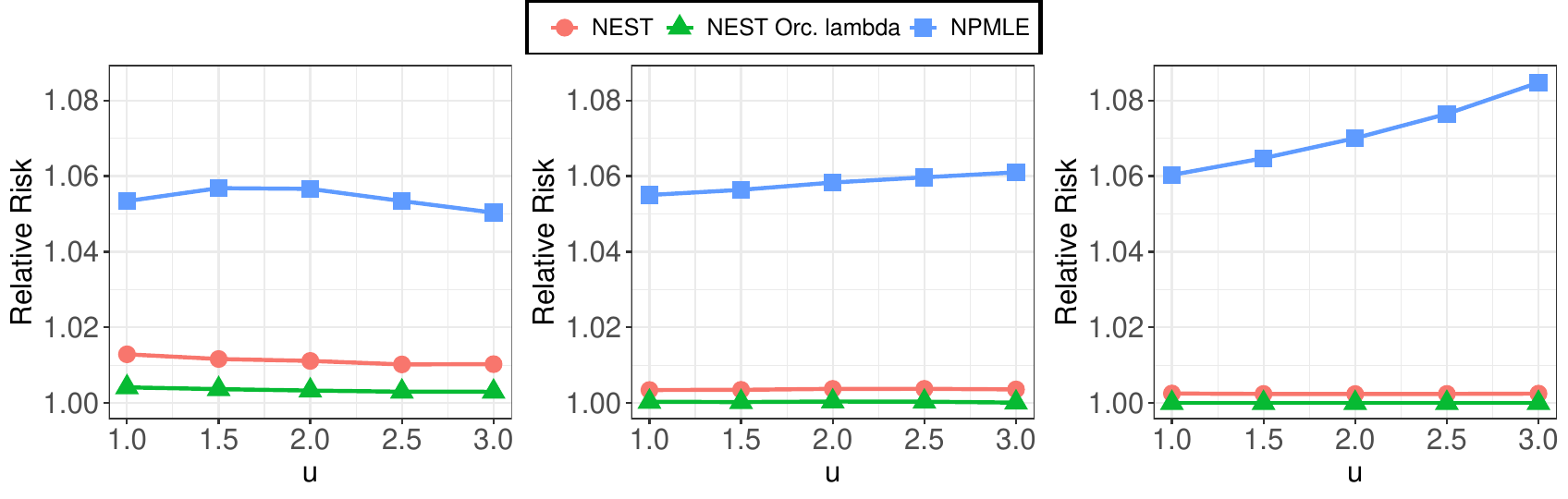}
		\caption{Setting 5: $\sigma_i^2\stackrel{i.i.d}{\sim}{\sf IG}({\sf shape}=5,~{\sf rate}=u)$ and $\mu_i~|~\sigma_i^2\stackrel{ind.}{\sim}~0.5~N(\sigma_i^2,\sigma_i^2)+0.5~N(-\sigma_i^2,\sigma_i^2)$.}
		\label{fig:simdep_3_wse} 
	\end{subfigure}
	\caption{Comparison of relative risks when $(\mu_i, \sigma_i^2)$ are dependent - Settings 3, 4 and 5. Plots show $m=10,15,20$ left to right.}
	\label{fig:simdep_wse_part1}
\end{figure}

\begin{figure}[!t]
	\centering
	\begin{subfigure}[b]{1\textwidth}
		\includegraphics[width=1\linewidth]{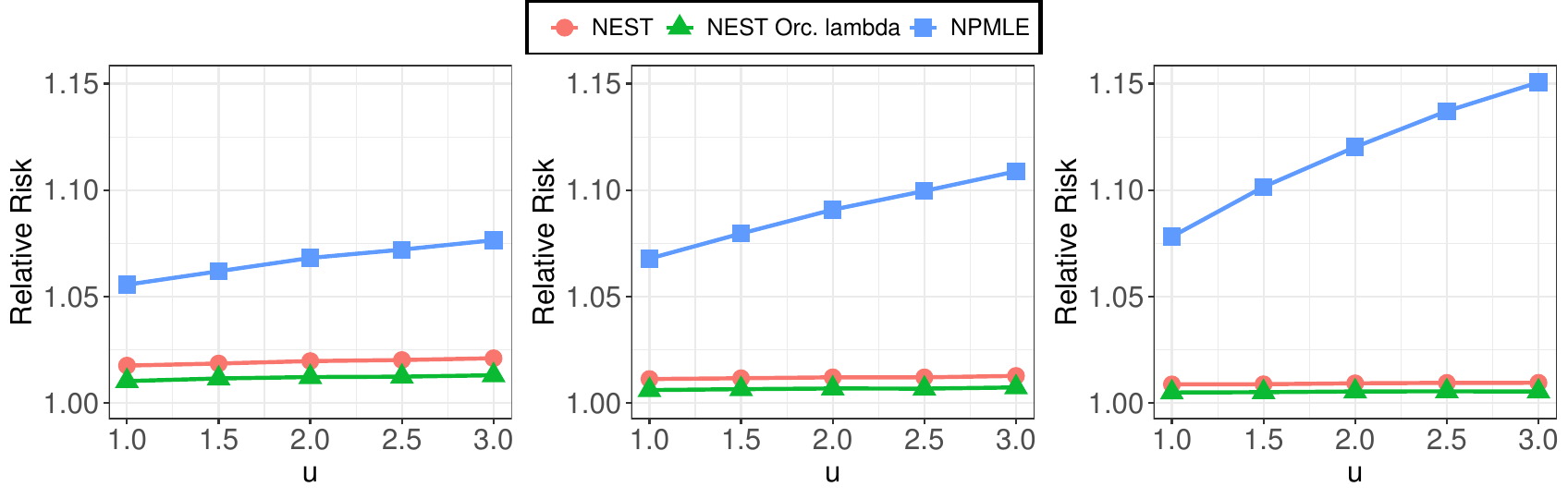}
		\caption{Setting 6: $\sigma_i^2\stackrel{i.i.d}{\sim}U(0.1,u)$ and $\mu_i~|~\sigma_i^2\stackrel{ind.}{\sim}~{\sf Laplace}({\sf location}=0,{\sf scale}=\sqrt{\sigma_i^2/2})$}
		\label{fig:simdep_4_wse}
	\end{subfigure}
	\begin{subfigure}[b]{1\textwidth}
		\includegraphics[width=1\linewidth]{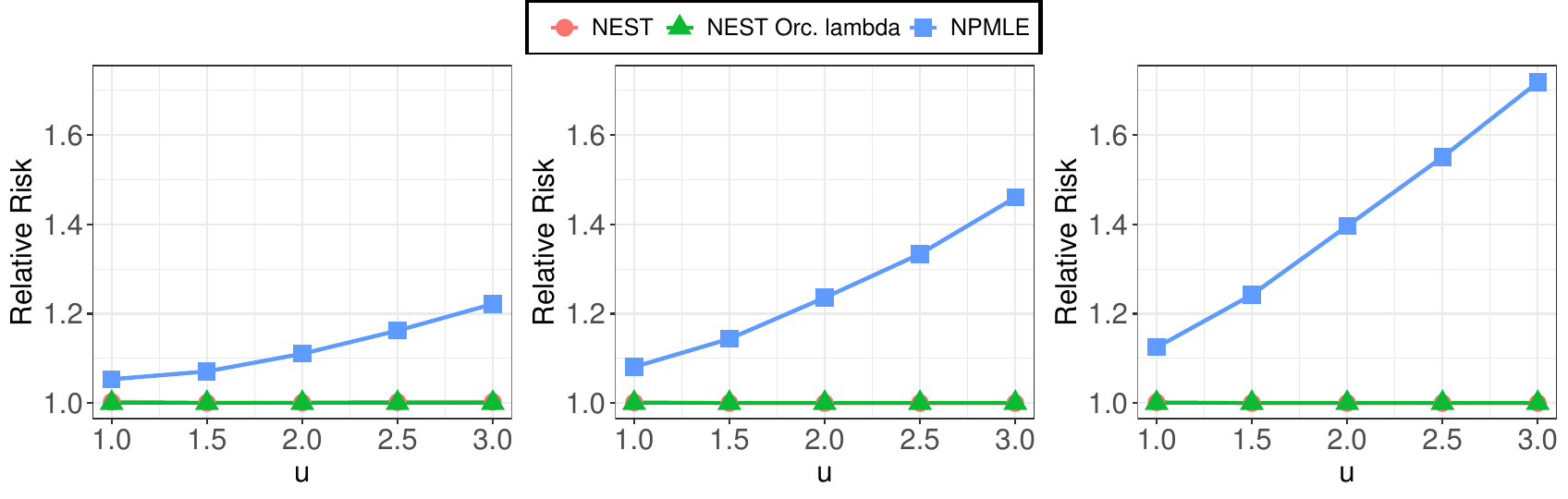}
		\caption{Setting 7: $\sigma_i^2\stackrel{i.i.d}{\sim} 0.5~{\sf Trunc.} N(0.5,0.1^2;~0.1)+0.5~{\sf Trunc.}N(u,0.1^2;~0.1)$ where ${\sf Trunc.} N(a,b;~c)$ represents a truncated Normal distribution with mean $a$, variance $b$ and truncated below $c$. Here $\mu_i|\sigma_i^2\stackrel{ind.}{\sim}{\sf Logistic}({\sf location} = \sigma_i^2,~{\sf scale}=\sqrt{\sigma_i^2})$.}
		\label{fig:simdep_5_wse}
	\end{subfigure}
	\caption{Comparison of relative risks when $(\mu_i, \sigma_i^2)$ are dependent - Settings 6 and 7. Plots show $m=10,15,20$ left to right.}
	\label{fig:simdep_wse_part2}
\end{figure}

\begin{figure}[!t]
	\centering
	\begin{subfigure}[!h]{1\textwidth}
		\includegraphics[width=1\linewidth]{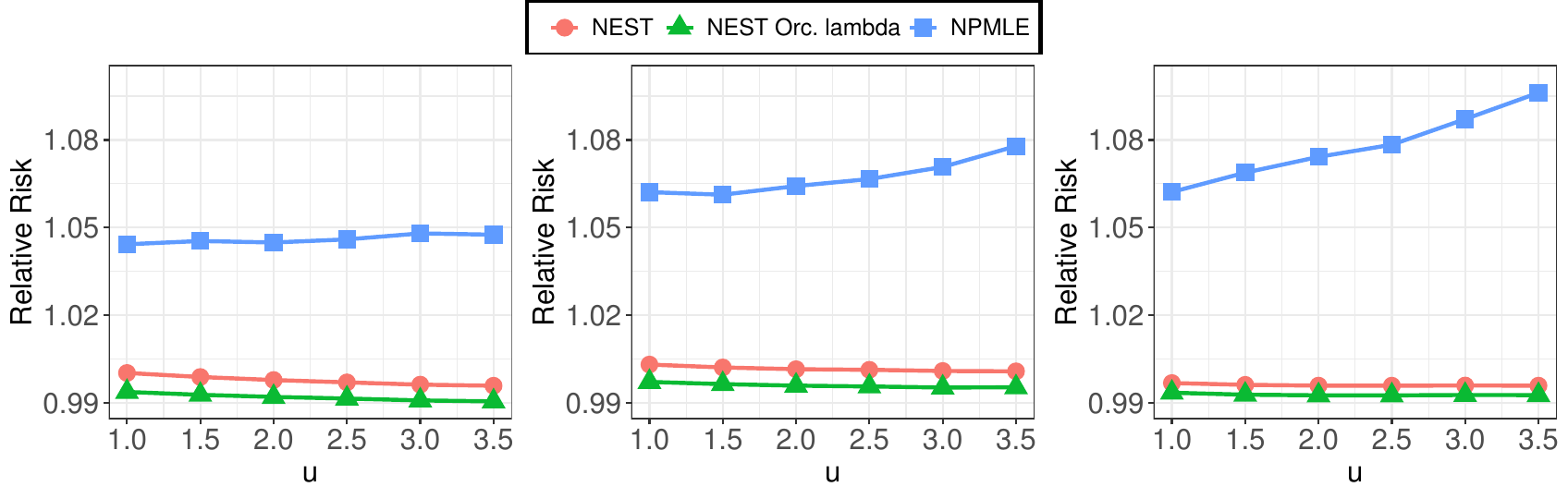}
		\caption{Setting 8: $(\mu_i,~\sigma_i^2)$ generated according to Setting 5 and $Y_{ij}|(\mu_i, \sigma^2_i)\stackrel{i.i.d.}{\sim}{\sf Laplace}({\sf location}=\mu_i,{\sf scale}=\sqrt{\sigma_i^2/2})$. }
		\label{fig:nonnormal_1_wse} 
	\end{subfigure}
	\begin{subfigure}[!h]{1\textwidth}
		\includegraphics[width=1\linewidth]{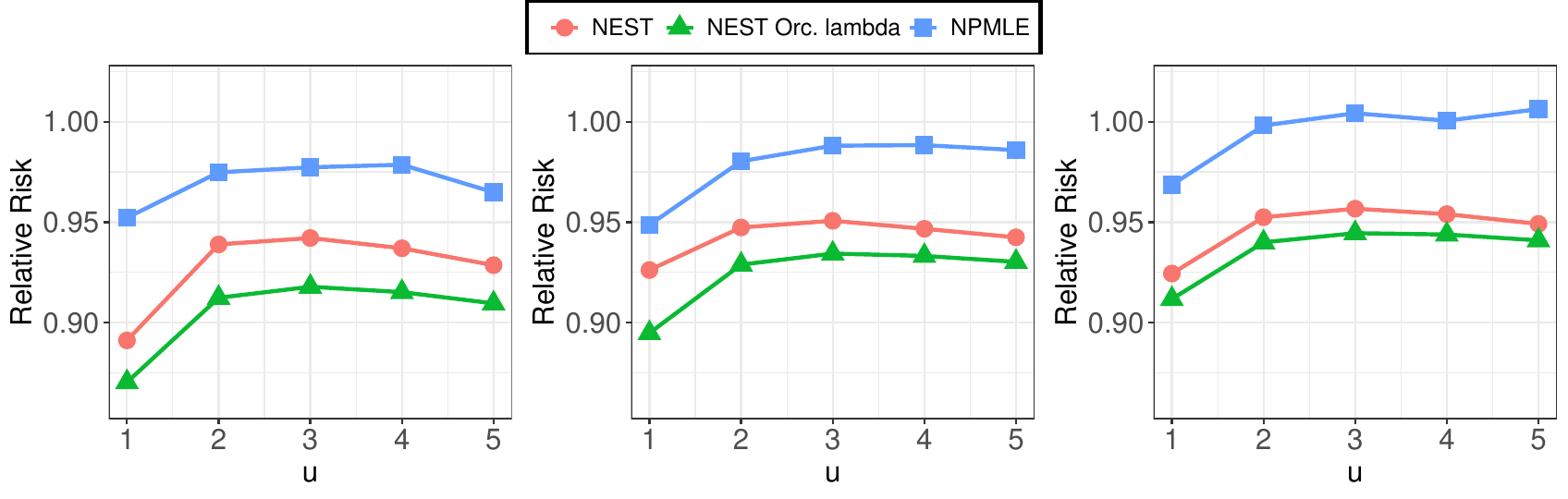}
		\caption{Setting 9: $(\mu_i,~\sigma_i^2)$ generated according to Setting 4 and $Y_{ij}|(\mu_i, \sigma^2_i)=\mu_i+\xi_i$ where $\xi_i$ are independently sampled from a central $t-$distribution with 4 degrees of freedom.}
		\label{fig:nonnormal_2_wse}
	\end{subfigure}
	\caption{Comparison of relative risks when the data $Y_{ij}|(\mu_i, \sigma^2_i)$ are not normally distributed. Plots show $m=10,15,20$ left to right.}
	\label{fig:nonnormal_wse}
\end{figure}

The aforementioned four approaches are evaluated on nine different simulation settings, with the goal of assessing the relative performance of the competing estimators as the heterogeneity in the variances $\sigma_i^2$ is varied while keeping the sample sizes $m_i$ fixed at $m$. The nine simulation settings can be categorized into three types: two settings where mean and variances are independent, five settings where mean and variance are correlated and two settings that represent departures from the Normal data-generating model. For each setting we set $n=1{,}000$ and compute the average Bayes risk for each competing estimator of $\bm\mu$ across $50$ Monte Carlo repetitions. Figures \ref{fig:simind_wse} to \ref{fig:nonnormal_wse} plot the relative risk of each competing estimator, which is the ratio of the average Bayes risk of the competing estimator to that of the oracle Bayes estimator $\bm \delta^{\pi}_{(1)}$.

The first two settings, whose results are displayed in Figures \ref{fig:simInd_1_wse} and \ref{fig:simInd_2_wse}, correspond to the independent case. We detail these settings below.

\begin{itemize}
	\item Setting 1: $\mu_i\stackrel{i.i.d}{\sim} N(0,3)$ and $\sigma_i^2\stackrel{i.i.d}{\sim}U(0.25,u)$ where we let $u$ to vary across five levels, $\{1,2,3,4,5\}$. 
	\item Setting 2: $\mu_i\stackrel{i.i.d}{\sim}~{\sf Laplace}({\sf location}=0,{\sf scale}=1)$ and $\sigma_i^2\stackrel{i.i.d}{\sim}N(u,1)$ truncated below $0.1$ where $u$ varies across $\{0.25,0.5,0.75,1,1.25\}$. 
\end{itemize}
The three plots in Figures \ref{fig:simInd_1_wse} and \ref{fig:simInd_2_wse} show the relative risks as $u$ varies for $m = 10, 15$ and $20$ (left to right). We see that under Setting 1, NEST and NEST Orc. $\lambda$ have relative risks substantially closer to $1$ than NPMLE. As $m$ increases, the performance of NEST and NEST Orc. $\lambda$ improves as their relative risks almost coincide with $1$. The NPMLE, on the other hand, exhibits a slightly higher relative risk at $m=20$ than at $m=10$ and $15$. 

The next five settings, Figures \ref{fig:simdep_wse_part1} and \ref{fig:simdep_wse_part2}, correspond to the correlated case. We describe these five settings below.
\begin{itemize}
	\item Setting 3: $\sigma_i^2\stackrel{i.i.d}{\sim}U(0.25,3)$ and $\mu_i~|~\sigma_i^2\stackrel{ind.}{\sim}~0.5~N(-\sigma_i^2,u^2)+0.5~N(\sigma_i^2,u^2)$ where $u$ varies across $\{0.25,0.35,0.45,0.55,0.65,0.75\}$.
	\item Setting 4: $\sigma_i^2\stackrel{i.i.d}{\sim}U(0.25,u)$ and $\mu_i~|~\sigma_i^2\stackrel{ind.}{\sim}~0.5~N(\sigma_i^2,\sigma_i^2)+0.5~N(-\sigma_i^2,\sigma_i^2)$ where we let $u\in\{1,2,3,4,5\}$. 
	\item Setting 5: we sample $\sigma_i^2$ from an Inverse Gamma (IG) distribution with shape parameter fixed at $5$ and rate at $u$ where $u\in\{1,1.5,2,2.5,3\}$. Conditional on the variances, the means $\mu_i$ are sampled from the design described in Setting 4.
	\item Setting 6: $\sigma_i^2\stackrel{i.i.d}{\sim}U(0.1,u)$ and $\mu_i~|~\sigma_i^2\stackrel{ind.}{\sim}~{\sf Laplace}({\sf location}=0,{\sf scale}=\sqrt{\sigma_i^2/2})$ where $u\in\{1,1.5,2,2.5,3\}$. 
	\item Setting 7: $\sigma_i^2$ are independently sampled from a two component mixture of truncated Normal distributions that are truncated below $0.1$ with means $0.5$ and $u$, standard deviation $0.1$ and mixing weights $0.5$. Conditional on the variances, the means $\mu_i$ are sampled from a Logistic distribution with location $\sigma_i^2$ and scale $\sqrt{\sigma_i^2}$. We let $u$ to vary across five levels, $\{1,1.5,2,2.5,3\}$. 
\end{itemize}
Setting 3 is presented in Figure \ref{fig:simdep_1_wse} where $u$ controls the spread of the prior distribution of $\mu_i$ given $\sigma_i^2$ and reveals that when $u$ is small, NPMLE has a relatively better risk performance than NEST. This is expected because when $u$ is small, the prior distribution of $\mu_i$ can be well approximated by a discrete distribution with just two mass points and the NEST estimator finds it particularly difficult to estimate the means well in these scenarios because such a discrete prior on $\mu_i$ introduces potential multimodality of the underlying marginal distribution of $(Y_i,S_i^2)$ in which case the true score functions may no longer be Lipschitz continuous, an assumption that is needed for the KSD approach to produce consistent estimates of the true scores in Section \ref{sec:opt_criteria}. The NPMLE, on the other hand, has a near parametric rate of convergence in Gaussian denoising problems when the underlying prior on the means is discrete \citep{saha2020nonparametric}, which explains its improved performance in Setting 3, especially when $u\to0$.

Settings 4 to 7, Figures \ref{fig:simdep_2_wse}, \ref{fig:simdep_3_wse} and \ref{fig:simdep_wse_part2}, represent scenarios where NEST dominates NPMLE uniformly for all values of $u$ and $m$. The relative risk of NPMLE in these scenarios is sometimes as high as 1.6 while NEST and NEST Oracle $\lambda$ exhibit relative risk profiles substantially closer to $1$.

The last two settings, displayed in Figures \ref{fig:nonnormal_1_wse} and \ref{fig:nonnormal_2_wse}, correspond to the scenarios where the data $Y_{ij}|(\mu_i, \sigma^2_i)$ are not normally distributed.

In Figure \ref{fig:nonnormal_1_wse}, $(\mu_i,\sigma_i^2)$ are sampled according to Setting 5 but $Y_{ij}|(\mu_i, \sigma^2_i)$ are generated independently from a Laplace distribution with location $\mu_i$ and scale $\sqrt{\sigma_i^2/2}$. Similarly, in Figure \ref{fig:nonnormal_2_wse}, $(\mu_i,\sigma_i^2)$ are sampled according to Setting 4 but $Y_{ij}|(\mu_i, \sigma^2_i)=\mu_i+\xi_i$ where $\xi_i$ are independently sampled from a central $t-$distribution with 4 degrees of freedom.
Across both these settings the proposed NEST method demonstrates robustness to departures from the Normal model. Proposition 7 in \cite{barp2019minimum} guarantees that, in general, the influence function of minimum KSD estimators, such as the NEST estimator, is bounded under data corruption and the behavior of the NEST estimator in Settings 8 and 9 is potentially an example of such a robustness property. The scenario in Setting 9 is particularly interesting because we note from Figure \ref{fig:nonnormal_2_wse} that the NEST estimator exhibits a smaller estimation risk than the Bayes oracle across all values of $u$. This is not surprising because in Settings 8 and 9, the Bayes oracle estimator $\bm \delta^{\pi}_{(1)}$ in Equation \eqref{eq:oraclenest1}, while having full knowledge of the underlying prior distributions, continues to rely on the incorrect Gaussian data generation process for $Y_{ij}$ to estimate the means. The NEST estimator, on the other hand, exhibits potential robustness to such mis-specification for estimating the shrinkage factors from the data.% of minimum KSD estimators.

In summary, the results of the preceding nine simulation settings reveal that when the variances are unknown, the NEST estimation framework, in general, enjoys a relatively better risk performance for estimating $\bm \mu$ under the weighted squared error loss than the NPMLE based $g$-modeling approach. However, as Setting 3 reveals, the NPMLE may exhibit a substantially better risk performance when the underlying prior distribution on the means can be well approximated by a discrete mixture with a finite number of mass points. In this scenario, the NEST estimator is unable to effectively account for the potential multimodality of the marginal distribution while estimating the shrinkage factors and therefore may return a poorer estimate of the means.

In Section \ref{sec:add_numexp} of the appendix, we present additional numerical experiments for the following cases: compound estimation of Normal means under the squared error loss (Section \ref{sec:nomrsimuMeans_se}), under unequal and small sample sizes $m_i$ (Section \ref{sec:mi}) and compound estimation of ratios under the squared error loss (Section \ref{num:simuRatios}). 

Moreover, two real data examples are discussed in Section \ref{sec:realdata}.

\section{Discussion}
\label{sec:discuss}
In this article we propose a nonparametric empirical Bayes framework, NEST, for large scale estimation of normal means when the corresponding precisions are unknown. We show that the data-driven NEST estimator is asymptotically as good as the optimal Bayes rule for estimating the means under the weighted squared error loss. Under the usual squared error loss our analysis in Section \ref{sec:squared_loss_app} reveals that the oracle NEST estimator dominates Tweedie's formula that relies on sample variances for shrinkage estimation of the means. Furthermore, in our numerical experiments with both simulated and real data, we show that NEST delivers a better risk performance, often substantially, than competing shrinkage estimators under both the weighted squared error loss and the squared error loss.  

A key component of the NEST framework is the consistent estimation of the bivariate shrinkage factors that appear in the oracle NEST estimator. Our approach for estimating these shrinkage factors relies on the KSD measure and we solve a convex optimization problem to derive the shrinkage factors. However, the advantage of the KSD approach for empirical Bayes methods extends far beyond the hierarchical Model \eqref{eq:normmodel} considered in this paper. For instance, a natural extension of Model \eqref{eq:normmodel} is in large-scale regression problems where the goal is to estimate a large number of low-dimensional regression coefficients when some potentially useful side information on these unknown regression coefficients are available. In this setting the KSD approach is a natural candidate for estimating the multivariate shrinkage factors that arise in this context. Our future research will be directed towards studying this problem and developing empirical Bayes methods for the same.
% Appendix here
% Options are (1) APPENDIX (with or without general title) or
%             (2) APPENDICES (if it has more than one unrelated sections)
% Outcomment the appropriate case if necessary
%
% \begin{APPENDIX}{<Title of the Appendix>}
% \end{APPENDIX}
%
%   or
%
% \begin{APPENDICES}
% \section{<Title of Section A>}
% \section{<Title of Section B>}
% etc
% \end{APPENDICES}

%%
%\theendnotes

% Acknowledgments here
%\ACKNOWLEDGMENT{The authors thank the Editor, the Associate Editor and the anonymous referees for their valuable comments and feedback that have significantly improved the quality and presentation of this paper. T. Banerjee was partially supported by the University of Kansas General Research Fund allocation \#2302216. W. Sun was supported in part by NSF grant DMS-2015339 during the initial part of the project period. G.M. and W.S. thank Dr. Zeyu Yao at Zhejiang University for helpful discussions on the theory. }

% CASE 1: BiBTeX used to constantly update the references
%   (while the paper is being written).
\bibliographystyle{chicago} % outcomment this and next line in Case 1
\bibliography{refs} % if more than one, comma separated
\newpage
\appendix
\begin{center}
	\Large{Appendix for ``Nonparametric Empirical Bayes Estimation On Heterogeneous Data''}
\end{center}

This Appendix is organized as follows. 

\begin{description}
	
	\item (a) Additional methodological details and proofs: In Section \ref{sec:squared_loss_app}, we present an analysis of the oracle NEST estimator (cf. Definition \ref{def:oraclenest}) under the (usual) squared error loss. Our main theories, namely Theorem \ref{thm:w} and Theorem \ref{thm:bayesrisk}, are proven in Sections \ref{proof-thm1} and \ref{sec:proof-thm2}, respectively. Sections \ref{proof-other-main} and \ref{proof-pro-ec} contain the technical proofs for the auxiliary theories in the main text and the appendix, respectively. 
	
	\item (b) Additional numerical results: In Section \ref{sec:add_numexp}, we provide results for additional numerical experiments for the following scenarios: compound estimation of Normal means under the squared error loss (Section \ref{sec:nomrsimuMeans_se}), under unequal and small sample sizes $m_i$ (Section \ref{sec:mi}) and compound estimation of ratios under the squared error loss (Section \ref{num:simuRatios}). Two real data examples are presented in Section \ref{sec:realdata}.

	\item (c) Additional technical details: In Section \ref{sec:extensions} we discuss extensions of our methodology to several well known members in the two-parameter exponential family when the nuisance parameter is known. A discussion on the computational complexity of NEST is provided in Section \ref{sec:complexity}. We conclude by reporting the impact of $\alpha$, the modified cross-validation parameter from Section \ref{Subsec:bandwidth}, on the estimation of $\lambda$ in Section \ref{sec:alpha}.
	
\end{description} 

\section{Analysis under the squared error loss}
\label{sec:squared_loss_app}
%%%%%%%%%%%%%%
In this section we present an analysis of the oracle NEST estimator $\bm \delta^{\pi}_{(1)}$ with respect to the squared error loss. We first introduce a counterpart to Tweedie's formula for $\mu_i$ (Equation \eqref{eq:twftau}) that relies on the sample variances $S_i^2$ instead of the unknown variances. Thereafter, we compare these two shrinkage estimators under the squared error loss.

When the precisions are unknown %Model \eqref{eq:normmodel} in its full generality does not allow a closed-form expression for $\delta_{i,(0)}^\pi$ given in Equation \eqref{eq:bayes}. In this setting a Tweedie-type formula, such as the one in Equation \eqref{eq:twftau}, is not readily available. In such a scenario 
existing shrinkage methods, such as \citet{Xieetal12,Weinsteinetal18}, usually rely on the sample variance $S_i^2$, a consistent estimator of the unknown population variance $1/\tau_i$, for practical implementation. For instance, in Definition \ref{def:twfs2} we present the oracle {Pseudo-}Tweedie's formula with sample variance which is a natural counterpart to Equation \eqref{eq:twftau} when the variance is unknown.
\begin{definition}(Oracle {Pseudo-}Tweedie's formula with sample variances)
	\label{def:twfs2}	
	Consider the hierarchical Model~\eqref{eq:normmodel}. Then an estimator for $\mu_i$ is $\delta_i^{\sf TF}$ where,
	\begin{eqnarray*}
		\label{eq:oraclenest}
		\delta_i^{\sf TF}\coloneqq\delta^{\sf TF}(y_i,s_i^2,m_i) =y_i+\dfrac{s_i^2}{m_i} w_1(y_i,s_i^2;m_i),
	\end{eqnarray*} 
	and $w_1({y},s^2;m)\coloneqq \dfrac{\partial}{\partial{y}}\log f_m({y},s^2)$.
\end{definition}
The oracle Pseudo-Tweedie's formula in Definition \ref{def:twfs2} has a striking similarity to the oracle NEST estimator $\delta_{i,(1)}^{\pi}$ (Definition \ref{def:oraclenest}) in the sense that both $\delta_i^{\sf TF}$ and $\delta_{i,(1)}^{\pi}$ involve an unbiased estimate $y_i$ of $\mu_i$ plus a shrinkage factor. The key difference, however, is that (1) while the shrinkage factor in $\delta_{i,(1)}^{\pi}$ relies on $\gamma_i$, $\delta_i^{\sf TF}$ directly uses sample variances $s_i^2$, and (2) while it is not immediately clear what is the underlying loss function that $\delta_i^{\sf TF}$ is minimizing for estimating $\mu_i$, $\delta_{i,(1)}^{\pi}$ uniquely minimizes the expected weighted squared error loss.

Notwithstanding, Proposition \ref{prop:twf} establishes that the oracle {Pseudo-}Tweedie's estimator $\bm \delta^{\sf TF}=(\delta_1^{\sf TF},\ldots,\delta_n^{\sf TF})$ dominates the sample mean estimator under a squared error loss function.
\begin{proposition}
	\label{prop:twf}
	Denote $w_1^{'}({y},s^2;m)$ as the partial derivative of $w_1({y},s^2;m)$ with respect to $y$. Suppose $f_m(y,s^2)$ is a log-concave density and {$w_1{'}(y,s^2;m)$ is a non-decreasing function of $s^2$.} Then under Model \eqref{eq:normmodel} and for $m_i>3$, we have,
	$$r_0(\bm \delta_{(0)}^{\pi},\mathcal{G})\le r_0(\bm \delta^{\sf TF},\mathcal{G})< r_0(\bm Y,\mathcal{G}).$$ 
\end{proposition}
However, when a large number of units are investigated simultaneously, traditional sample variance estimators may suffer from selection bias \citep{jing2016sure,kwon2018f}, and their direct use may lead to severe deterioration in the MSE for estimating the means. We present Proposition \ref{prop:oraclerisks} which shows that under the squared error loss function the risk of the oracle NEST estimator is uniformly smaller than that of the Pseudo-Tweedie's estimator $\bm \delta^{\sf TF}$ (Definition \ref{def:twfs2}).%The oracle NEST estimator, in contrast, relies on carefully constructed shrinkage factors for both the sample mean and sample variance. %,  which ultimately provide a much improved approximation to $\delta_i^{\pi}$.
%%%%%%%%%%%%%%%%%%%%%%%%%
%\red{Denote $\bm{\delta}=(\delta_1, \ldots, \delta_n)^T$  an estimator for $\bm{\mu}$ based on $(\bm Y,\bm S)$. Consider the squared error loss $l_n(\bm{\mu},\bm \delta)=n^{-1}\sum_{i=1}^{n}\ell(\mu_i,\delta_i)$, where  $\ell(\mu_i,\delta_i)=(\delta_i-\mu_i)^2$. The compound Bayes risk is given by Equation \eqref{eq:risk_se}. }

We impose the following regularity conditions for comparing the estimation risks of $\bm \delta_{(1)}^{\pi}$ and $\bm\delta^{\sf TF}$ in Proposition \ref{prop:oraclerisks}.
%\begin{assumption}
%	\label{assump:5}
%	The shrinkage factor $\gamma(y,s^2,m)$ satisfies	(a) $\mathbb{E}\Big[S^2\gamma(Y,S^2,m)\Big|\mu,\tau\Bigr]\le 1/\tau$, (b) $s^2\gamma(y,s^2,m)$ is non-decreasing in $s^2$, and (c) $\gamma(y,s^2,m)$ is non-increasing in $s^2$.
%\end{assumption}
\begin{assumption}
	\label{assump_5}
	%	The shrinkage factor $\gamma(y,s^2,m)$ satisfies (a) $s^2\gamma(y,s^2,m)$ is non-decreasing in $s^2$, and (b) $\gamma(y,s^2,m)$ is non-increasing in $s^2$.
	The shrinkage factor $\gamma(y,s^2,m)$ is non-increasing in $s^2$.
\end{assumption}
\begin{assumption}
	\label{assump_6}
	%	Let $\omega(y,s^2;m)\coloneqq w_1(y,s^2;m)\dfrac{\partial}{\partial y}w_2(y,s^2;m)$. Then (a) $\omega(y,s^2;m)\le 0$ and, (b) $\omega(y,s^2;m)$ is non-decreasing in $s^2$.
	Let $\omega(y,s^2;m)\coloneqq w_1(y,s^2;m)\dfrac{\partial}{\partial y}w_2(y,s^2;m)$. Then, $\omega(y,s^2;m)\le 0$.
\end{assumption}
Assumptions \ref{assump_5} and \ref{assump_6} are regularity conditions on the behavior of the shrinkage factor $\gamma$ and the score functions $w_1,w_2$. For instance, Assumptions \ref{assump_5} enforces monotonicity on shrinkage factor $\gamma(y,s^2,m)$, which is satisfied, for example, when $(\mu,\tau)$ have a conjugate prior under Model \eqref{eq:normmodel}. Similarly, Assumption \ref{assump_6} holds under conjugate priors and is also true when the prior on $\tau$ is discrete with just one mass point. In Remarks \ref{rem:sqloss_1} and \ref{rem:sqloss_2} we discuss these assumptions in more details. Proposition \ref{prop:twf} can be extended as follows.
%In Appendix \ref{app:more_oraclenest} we discuss an additional result that provides an upper bound on the compound risk of $\bm \delta^{\pi}_{(1)}$ when Assumptions \ref{assump:5} and \ref{assump:6}(b) are relaxed.
\begin{proposition}\label{prop:oraclerisks}
	Suppose $f_m(y,s^2)$ is a log-concave density and {$w_1{'}(y,s^2;m)$ is a non-decreasing function of $s^2$.} If Assumptions \ref{assump_5} -- \ref{assump_6} hold then, under Model \eqref{eq:normmodel} and for $m_i>5$, we have,
	$$r_0(\bm \delta_{(0)}^{\pi},\mathcal{G})\le r_0(\bm \delta_{(1)}^{\pi},\mathcal{G})\le r_0(\bm \delta^{\sf TF},\mathcal{G})<r_0(\pmb Y, \mathcal G).$$
\end{proposition}
Note that the oracle NEST estimator $\delta^{\pi}_{i,(1)}$ is, in general, different from $\delta^{\pi}_{i,(0)}$: %$\delta^{\pi}_i$ has full knowledge of the prior distributions $G_\mu(\cdot|\tau)$ and $H_\tau(\cdot)$, whereas $\delta^{*}_i$  has only the information on the true score functions $(w_{1,i},w_{2,i})$. 
$\delta^{\pi}_{i,(1)}$ is the Bayes estimator under the weighted squared error loss, while $\delta^{\pi}_{i,(0)}$ is the Bayes estimator under the usual squared error loss. This follows since from Equation \eqref{eq:loss_funcs}, and dropping subscript $i$, 
$$ \delta^{\pi}_{(1)} = \dfrac{\mathbb{E}(\tau\mu|y,s^2)}{\mathbb{E}(\tau|y,s^2)}=\dfrac{\mathbb{E}[\tau\mathbb{E}(\mu|y,s^2,\tau)|y,s^2]}{\mathbb{E}(\tau|y,s^2)}\ne \mathbb{E}(\mu|y,s^2)=\delta^{\pi}_{(0)}.
$$ 
However, $r_0(\bm \delta_{(0)}^{\pi},\mathcal{G})= r_0(\bm \delta_{(1)}^{\pi},\mathcal{G})$ when
$\mu$ and $\tau$ are conditionally independent given $y$ and $s^2$, in which case $\delta^{\pi}_{(1)}=\delta^{\pi}_{(0)}$.
In particular, Corollary \ref{cor:normal_conj} shows that when $\mathbb{E}(\mu_i|y_i,s_i^2,\tau_i)$ is independent of $\tau_i$ %$G_\mu(\cdot|\tau)$ is $N(\mu_0,1/\tau)$ and $H_\tau(\cdot)$ is $\Gamma(\alpha,\beta)$ 
then the data-driven NEST estimator is asymptotically close to $\bm\delta^{\pi}_{(0)}$.
\begin{corollary}
	\label{cor:normal_conj}
	Consider hierarchical Model \eqref{eq:normmodel} and suppose $\mathbb{E}(\mu_i|y_i,s_i^2,\tau_i)$ is independent of $\tau_i$. %where $G_\mu(\cdot|\tau)$ is $N(\mu_0,1/\tau)$ and $H_\tau(\cdot)$ is $\Gamma(\alpha,\beta)$. %With $m>3$, let $\bm \delta^{\pi}$ denote the Bayes estimator of $\bm \mu$. 
	Then, under the conditions of Theorem \ref{thm:bayesrisk}, we have
	%$$ \dfrac{1}{n}\big\|{\bm\delta}^{\sf ds}(\lambda)-\bm\delta^{\pi}\big\|_2^2=O_p(n^{-1/2})\text{ and }\big|l_n(\bm \delta^{\sf ds}(\lambda),\bm \mu)-l_n(\bm \delta^{\pi},\bm \mu)\big|=O_p(n^{-1/4}).$$
	$ \dfrac{c_n}{n}\big\|{\bm\delta}^{\sf ds}_n(\lambda)-\bm\delta^{\pi}_{(0)}\big\|_2^2=o_p(1)~\text{as}~n\to\infty.$ Furthermore, under the same conditions, 
	$c_n^{1/2}\big|l_n^{(0)}\{\bm \delta^{\sf ds}_n(\lambda),\bm \mu;\bm\tau\}-l_n^{(0)}\{\bm \delta^{\pi}_{(0)},\bm \mu;\bm \tau\}\big|=o_p(1)~\text{as}~n\to\infty.$
\end{corollary}
Corollary \ref{cor:normal_conj} is a straightforward consequence of Theorem \ref{thm:bayesrisk} and the fact that when $\mathbb{E}(\mu_i|y_i,s_i^2,\tau_i)$ is independent of $\tau_i$, $\delta^{\pi}_{i,(1)}=\delta^{\pi}_{i,(0)}$. %$G_\mu(\cdot|\tau),~H_\tau(\cdot)$ are as specified in Corollary \ref{cor:normal_conj} then under Model \eqref{eq:normmodel}, $\bm \delta^{\pi}_{(1)}=\bm \delta^{\pi}_{(0)}$.}
A popular setting where $\mathbb{E}(\mu_i|y_i,s_i^2,\tau_i)$ is indeed independent of $\tau_i$ is the scenario where $G_\mu(\cdot|\tau)$ and $H_\tau(\cdot)$ belong to the family of conjugate priors under Model \eqref{eq:normmodel}. Under this setting the posterior expectation of $\mu$ is a linear combination of the prior expectation of $\mu$ and the maximum likelihood estimate $y$ \citep{diaconis1979conjugate}, and the weights in this linear combination are proportional to $m$ and the prior sample size. For instance, if $G_\mu(\cdot|\tau)$ is $N(\mu_0,1/\tau)$ and $H_\tau(\cdot)$ is $\Gamma({\sf shape}=\alpha,{\sf rate}=\beta)$ in Model \eqref{eq:normmodel} then standard calculations give $\delta_{(0)}^{\pi}=(\mu_0+my)/(m+1)$ where $\mu_0$ is the prior expectation of $\mu$ and $m_0=1$ is the prior sample size.
Moreover, in this setting we also have
$$\log f_m(y_i,s_i^2) = c_0+\dfrac{m-3}{2}\log s_i^2-(\alpha+m/2)\log\Bigl\{\beta+0.5(m-1)s_i^2+0.5\dfrac{m}{m+1}(y_i-\mu_0)^2\Bigr\},
$$ where $c_0$ is a constant independent of $(y_i,s_i^2)$. From the above display, 
\begin{eqnarray}
w_1(y_i,s_i^2) &=& -\dfrac{(\alpha+m/2)\{{m}/(m+1)\}(y_i-\mu_0)}{\beta+0.5(m-1)s_i^2+0.5\{{m}/(m+1)\}(y_i-\mu_0)^2}\nonumber\\
w_2(y_i,s_i^2) &=& \dfrac{(m-3)}{2s_i^2}-\dfrac{0.5(\alpha+m/2)(m-1)}{\beta+0.5(m-1)s_i^2+0.5\{{m}/(m+1)\}(y_i-\mu_0)^2}.\nonumber
\end{eqnarray}
Substituting these expressions for $w_1(y_i,s_i^2), w_2(y_i,s_i^2)$ in Equation \eqref{eq:oraclenest1} gives $\delta^{\pi}_{i,(1)}=\delta^{\pi}_{i,(0)}$. Another example where $\delta^{\pi}_{(1)}=\delta^{\pi}_{(0)}$ is when $G_\mu(\cdot|\tau)$ and $H_\tau(\cdot)$ are independent discrete distributions with just one mass point, respectively. These two examples are also the settings, among others, under which the first inequality in the left hand side of Proposition \ref{prop:twf} is strict, that is, $r_0(\bm \delta_{(0)}^{\pi},\mathcal{G})< r_0(\bm \delta^{\sf TF},\mathcal{G})$.

In Section \ref{sec:nomrsimuMeans_se} we present several numerical experiments to compare the data-driven NEST estimator against other shrinkage estimators for compound estimation of the Normal means from Model \eqref{eq:normmodel} under the squared error loss. Our empirical results corroborate Proposition \ref{prop:oraclerisks} and suggest that the efficiency gain of the data-driven NEST estimator over competing linear shrinkage methods \citep{jing2016sure,Weinsteinetal18} and Pseudo-Tweedie's formula (Definition \ref{def:twfs2}) is substantial across many settings. We end this section with the following remarks:
\begin{remark}
	\label{rem:sqloss_1}
	Here we provide three examples of $\mathcal{G}$ that satisfy Assumptions \ref{assump_5} and \ref{assump_6}.
	\begin{enumerate}
		\item $G_\mu(\cdot|\tau)$ and $H_\tau(\cdot)$ belong to the family of conjugate priors under Model \eqref{eq:normmodel}. 
		\item $H_\tau(\cdot)$ is a distribution with just one mass point, say $\tau_0$, and $G_\mu(\cdot|\tau_0)$ has a log-concave density.
		\item $H_\tau(\cdot)$ has a log-concave density and $G_\mu(\cdot|\tau)$ is a discrete distribution with just one mass point, say $\mu_0$.
	\end{enumerate} 
	In Figure \ref{fig:contourplot_satisfy} we consider specific choices of $\mathcal G$ for each of these three examples. In particular, the left panel of Figure \ref{fig:contourplot_satisfy} provides a contour plot of $\mathcal G$ for example 1 where $G_\mu(\cdot|\tau)$ is $N(0,1/\tau)$ and  $H_\tau(\cdot)$ is $\Gamma({\sf shape}=10,{\sf rate}=5)$. The center panel presents the contour plot of the marginal density $f_m$ for example 2 where $G_\mu(\cdot|\tau)$ is $N(0,1/\tau)$ and  $H_\tau(\cdot)$ has all its mass concentrated at $\tau_0=1$. Finally, the right panel presents the contour plot of $f_m$ for example 3 where $\mu_0=0$ and  $H_\tau(\cdot)$ is $\Gamma({\sf shape}=10,{\sf rate}=5)$.
	\begin{figure}[!t]
		\centering
		\includegraphics[width=1\linewidth]{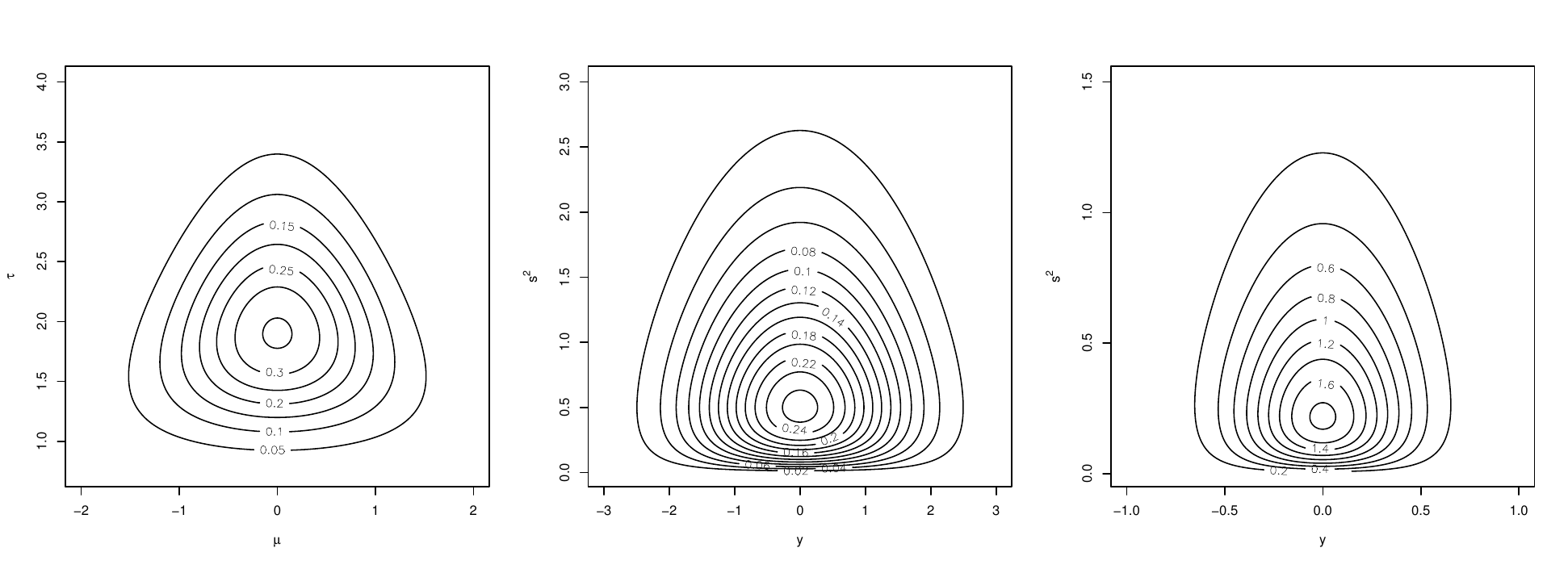}
		\caption{Remark \ref{rem:sqloss_1} contour plots with $m=5$. Left to Right: contour plot of $\mathcal G$ for example 1 where $G_\mu(\cdot|\tau)$ is $N(0,1/\tau)$ and  $H_\tau(\cdot)$ is $\Gamma({\sf shape}=10,{\sf rate}=5)$. Contour plot of the marginal density $f_m$ for example 2 where $G_\mu(\cdot|\tau)$ is $N(0,1/\tau)$ and  $H_\tau(\cdot)$ has all its mass concentrated at $\tau_0=1$. Contour plot of $f_m$ for example 3 where $\mu_0=0$ and  $H_\tau(\cdot)$ is $\Gamma({\sf shape}=10,{\sf rate}=5)$.}
		\label{fig:contourplot_satisfy}
	\end{figure}
\end{remark}
\begin{remark}
	\label{rem:sqloss_2}
	Here we provide two examples of $\mathcal G$ that induce marginal densities $f_m$ which are no longer log-concave and, hence, do no satisfy Assumption \ref{assump_5}. In these examples, log-concavity of $f_m$ is lost because $w_1^{'}(y,s^2;m)$ may be positive on its domain.
	\begin{enumerate}
		\item $H_\tau(\cdot)$ is a distribution with just one mass point, say $\tau_0$, and $G_\mu(\cdot|\tau_0)$ has mass points at $2\tau_0$ and $-2\tau_0$ with equal probability.
		\item $H_\tau(\cdot)$ has mass points at $1$ and $2$ with equal probability, and $\mu=4/\tau$.
	\end{enumerate}
	Figure \ref{fig:contourplot_nosatisfy} presents the contour plots of $f_m$ for these two examples.
	\begin{figure}[!h]
		\centering
		\includegraphics[width=1\linewidth]{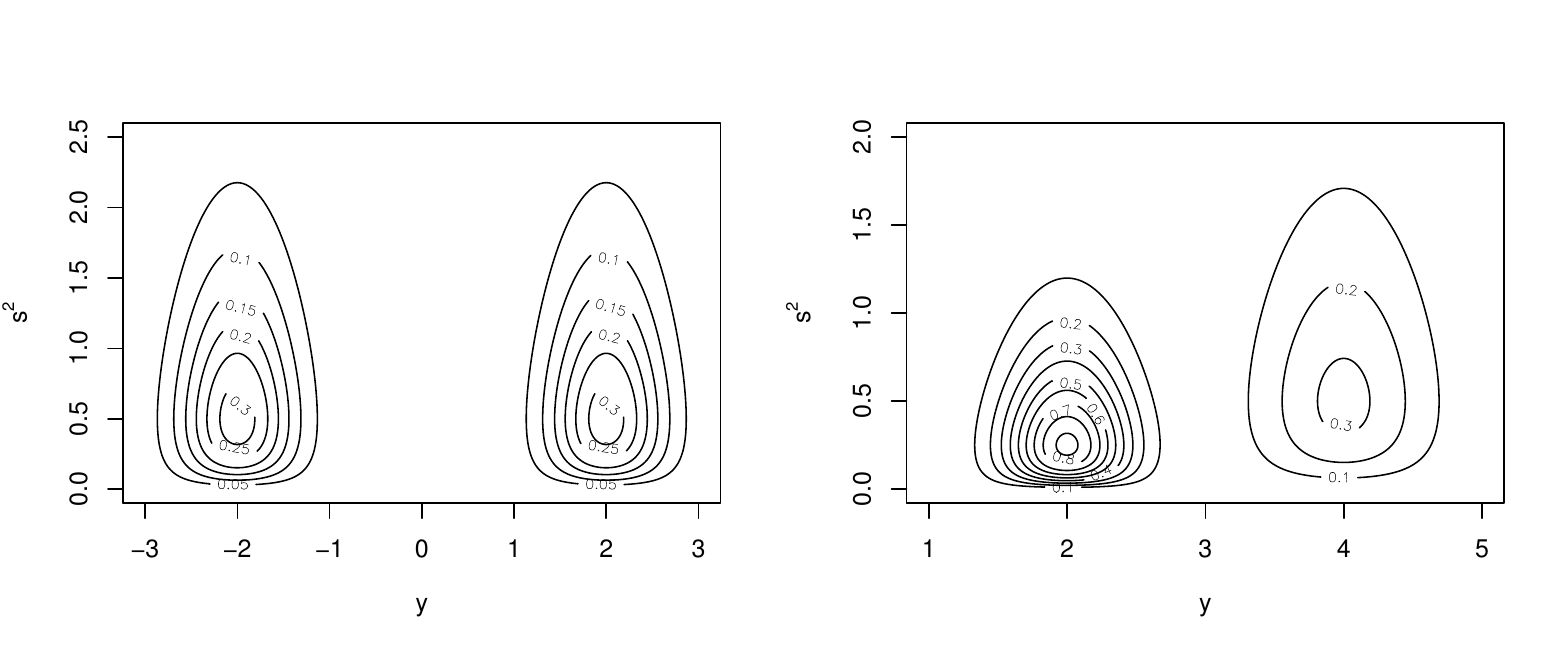}
		\caption{Remark \ref{rem:sqloss_2} contour plots. Left: contour plot of the marginal density $f_m$ for example 1 where $\tau_0=1$ and $m=5$. Right: contour plot of $f_m$ for example 2 with $m=5$.}
		\label{fig:contourplot_nosatisfy}
	\end{figure}
\end{remark}

{
	%%%%%%%%%%%%%%%%%%%%%%%%%%%%%%%%%%%%%%%%%%%%%%%%%%%%%%%%%%
	\section{Proof of Theorem  \ref{thm:w}}\label{proof-thm1}
	%%%%%%%%%%%%%%%%%%%%%%%%%%%%%%%%%%%%%%%%%%%%%%%%%%%%%%%%%%

	\subsection{Overview of the main ideas and summary of notations}
	
	We begin by outlining the main ideas and key steps, and define important notations along the way.

	Consider a set of score functions $\fclass$ that contains all  differentiable functions $h$ that are Lipschitz continuous with Lipschitz constants less than $L>0$ and $||h(0)||_\infty\leq U$, where $L>0$ and $U > 0$ are pre-fixed constants. Let $f$ denote the true marginal density, which is sub-exponential. Denote by $\mathcal W_0$ its score function; to explicitly related $f$ and its score, we use the notation $h_f$ and $\mathcal W_0$ interchangeably for presentation ease. By Assumption \ref{assump:2}, we have $\mathcal W_0 \in \fclass$. Suppose we observe $\mathbb X=(\bm X^1,\ldots,\bm X^n)$, where $\bm X^i$'s are i.i.d. random variables with density $f$. Denote $\widehat{\mathcal{W}}$ the selected score function obtained by minimizing \eqref{eq:samplecrit} over $\mathbb{W}$. For any $h \in \fclass$, and for $p=1,2$, define:
	$$
	\Delta^{(p)}(h)={\mathbb{E}_{X\sim f} \vert \vert h(X) - \mathcal W_0(X) \vert \vert_p^p}.
	$$ 
	This is the $L_p$ distance between $h$ from the true score function $\mathcal W_0$. Define its sample analogue as 
	$$
	{\hat{\Delta}^{(p)}_n(h)} =n^{-1} \sum_{i=1}^{n} \vert \vert h(\bm X^i) - \mathcal W_0(\bm X^i) \vert \vert^p_p.
	$$ 
	\begin{remark}
		For presentation convenience, we henceforth omit boldface notation for vectors. Note that in our applications, we have \(X^i \in \mathbb{R}^2\), but our proof holds for any fixed dimension. Therefore, we present the results for a general \(K\)-dimensional density $f$ and specialize to \(K=2\) at the end to derive the rates relevant to our application. 
	\end{remark} 
	
	In our proof, we first consider the situation where $\Omega=I$ is known, in which case we consider the kernel with $\hat \Omega = I$. The situation with a general $\hat \Omega$ can be handled without essential difficulty; see the end of Section \ref{proof-thm1-main} for a detailed discussion. 
	
	For any $\lambda > 0$, we denote the Kernelized Stein's discrepancy of $h$ from $\mathcal W_0$ as 
	$$
	D_{\lambda}(h)=\mathbb E_{(U,V)\sim f^2} [ \mathcal K_{\lambda}(U,V)(h(U)-\mathcal W_0(U))'(h(V)-\mathcal W_0(V))],
	$$ 
	where $\log \mathcal K_{\lambda}(u,v)=-||u-v||^2/(2\lambda)$. Moreover, define its sample version as 
	$$\hat D_{\lambda,n}(h)=n^{-1}(n-1)^{-1} \sum_{i \neq j} \kappa_\lambda[ {h}](X^i, X^j), 
	$$ where the term $\kappa_\lambda[h](u, v)$ is given by
	\[
	\kappa_\lambda[h](u, v)=\mathcal K_\lambda(u, v) {h^T}(u){h}(v) + {h}^T(v) \nabla_{u} \mathcal K_\lambda(u, v) + {h}^T(u) \nabla_{v} \mathcal K_\lambda(u, v)  + \text{trace}(\nabla_{u} \nabla_{v} \mathcal K_\lambda(u,v))~.
	\]
	It is important to note that, as an empirical version of \(D_{\lambda}(h)\), the evaluation of \(\hat D_{\lambda,n}(h)\) does not depend on knowledge of \(\mathcal{W}_0\). This property allows us to use a slightly modified version of \(\hat D_{\lambda,n}(h)\) as the objective criterion for constructing the data-driven NEST estimator. 
	
	The proof proceeds in three steps:
	\begin{description}
		\item[(a)] We establish uniform convergence, over \(\fclass\), of $L_1$ norm, $L_2$ norm and KSD from the empirical measures to the corresponding population measures. 
		\item[(b)] We establish relevant theories that characterize the relationships between KSD norm and the \( L_p \) norm. 
		\item[(c)] We determine the rate at which the upper bound from step (b) converges to \(0\) as \(n \to \infty\).
	\end{description}
	The details of the above steps are spelled out below.

	\subsection{Key lemmas and the main proof}\label{proof-thm1-main}

	For any $h \in \fclass$, Stein's lemma implies that $\mathbb E_{(U,V)\sim f^2 }\{\kappa_\lambda[h](U, V)\}=D_{\lambda}(h)$; hence, we would expect $\hat D_{\lambda,n}(h) \to D_{\lambda}(h)$ \text{ as } $n \to \infty$. 
	
	Next, we establish the uniform convergence over $\fclass$ of three distances ($L_1$, $L_2$ and KSD) from the empirical measures to the population measures.
	
	\begin{lemma}\label{lemma.g1}
		Suppose Assumption \ref{assump:2} holds. Let
		$$\mathcal{G}_n=\max\bigg\{\sup_{h \in \fclass} \big| \hat \Delta_n^{(1)}(h)- \Delta^{(1)}(h) \big|, \,\sup_{h \in \fclass} \big| \hat \Delta_n^{(2)}(h)- \Delta^{(2)}(h) \big|, \,
		\sup_{h \in \fclass} \left|\hat D_{\lambda,n}(h)- D_{\lambda}(h)\right|
		\bigg\}.$$
		Then for any $f$ that is sub-exponential, we have $\mathcal{G}_n=O_p(n^{-1/2}\log n)$. 
		%$\exp\, \mathcal{G}_n \leq c_0\, n^{-1/2}(\log n)^{1/2}$. 
		%Also, for any $\delta \in (0,1)$, $$\mathbb{P}\big(\sqrt{n}\, \mathcal{G}_n \leq c_0\, (\log n)^{1/2}+c_1 (\log \delta^{-1})^{1/2}\big)\geq 1-\delta,$$ where, $c_0$ and $c_1$ are constants.  
	\end{lemma}
	
	The above lemma establishes that \(\mathcal{G}_n\) converges in probability at a parametric rate (up to poly-logarithmic factors). The proof of this lemma is provided in Section \ref{sec:prf_lemma.g1}. It primarily relies on two key properties: (a) that $f$ is sub-exponential; and (b) that \(\fclass\) is a Lipschitz ball.
	
	\begin{remark}
		The almost sure convergence of the \(L_1\) distance in \(\mathcal{G}_n\) (without rates) follows directly from Theorem 2.4.3 of \citet{van1996weak}, since the sub-exponentiality of \(f\) ensures that \(\fclass\) possesses a measurable envelope, and its Lipschitz continuity implies polynomial entropy bounds. For the other two distances, however, an additional truncation step is required. For the proof of Theorem \ref{thm:w}, we need explicit convergence rates as stated in Lemma~\ref{lemma.g1}, which necessitates the more detailed argument presented in Section \ref{sec:prf_lemma.g1}.
	\end{remark}

	To derive an identity between various distances, we define an intermediate norm 
	$$
	\bar{\Delta}(h)=\mathbb E_{X\sim f} \{||h(X)-\mathcal W_0(X)||_2^2f(X)\}.
	$$ 
	Moreover, consider euclidean based $L_p$ norms: for $p > 0$
	$$
	\dot \Delta^{(p)}(h)=\{\mathbb{E}_{X\sim f} ||h(X)-\mathcal W_0(X)||_2^p\}^{1/p}.
	$$
	Note that $\Delta^{(2)}(h)=[\dot \Delta^{(2)}(h)]^2$, but the regular-$L_1$ based norm satisfies $\Delta^{(1)}(h)\leq\dot \Delta^{(1)}(h)$.

	The next lemma establishes a result that connects the KSD norm with the \( L_p \) norm. This important result is formulated in terms of norms defined over the population distribution and holds uniformly for all \( h \in \fclass \). Denote by $\Phi_K(b)=\PP(\chi(K)<b)$, the probability that a $\chi$-variable with $K$ degrees of freedom is less than $b$. 
	
	\begin{lemma}\label{lemma.g2}
		Under Assumption \ref{assump:2}, there exists a constant $C>0$ such that, for all $\lambda, b > 0$ and $d>1$ we have: 
		\begin{align}\label{eq:lem2.bound}
		\sup_{h \in \fclass} \quad \Phi_K(b) \,\bar{\Delta}(h)
		- C \bigg\{ \lambda^{-K} \big[D_\lambda(h) +  e^{-b^2/2}\, \Delta^{(2)}(h)\,\big] +  \lambda\,b\, \Delta^{(2)}(h) + \lambda \, b^{K+1} \dot \Delta^{(1)}(h)\bigg\}<0~.
		\end{align}
	\end{lemma}
	
	The proof of the above lemma, which relies critically on the sub-exponentiality of \( f \),
	is quite intricate and is presented in Section \ref{sec:prf_lemma.g2}. We will apply this lemma with \( b = O(\log n) \) and \( \lambda = n^{-\alpha} \) with $\alpha\in [0,1/4)$ as \( n \to \infty \) and for large \( d \). Thus, we will be considering $\lambda\coloneqq \lambda_n$ approaching $0$ but at a controlled rate such that $\liminf_{n \to \infty} \lambda n^{1/4}=\infty$. 
	This helps establish a relationship between the KSD distance \( D_{\lambda}(h) \), the \( L_p \) distance, and the intermediate metric \( \bar{\Delta}(h) \). However, it is important to note that these distances are defined with respect to the population distribution. In practice, the sample criterion produces a bound on \( D_{\lambda}(h)\) for the selected score function, subject to sampling variation, the order of which is controlled by Lemma~\ref{lemma.g1}. Taking these considerations into account, we select the kernel bandwidth that minimizes the bound on \(\Delta^{(1)}(h)\) for the chosen score function, based on our analysis.
	
	Next, recall that the criterion we used for optimization problem defined in \eqref{eq:samplecrit} is
	\[ \widehat {\mathbb M}_{\lambda,n} (h)=n^{-2}
	\sum_{i, j} \kappa_\lambda[h](X^i, X^j)=(1-n^{-1})\hat D_{\lambda,n}(h)+n^{-2} \sum_{i=1}^n  \kappa_\lambda[h](X^i, X^i).
	\]
	Define the following quantity
	\begin{equation}\label{eqn:Jn}
	\hat{\mathcal{J}}_n=\sup_{h \in \fclass} \big|\hat {\mathbb M}_{\lambda,n}(h) - (1-n^{-1}) {D}_\lambda(h)-n^{-1}\mathbb E_f ||h(X)||_2^2\big|.
	\end{equation}
	By construction, the selected score function $\widehat{\mathcal W}$ satisfies: 
	$$
	\widehat{\mathbb M}_{\lambda, n} (\widehat{\mathcal W}) \leq \inf_{h \in \fclass} \widehat {\mathbb M}_{\lambda,n} (h).
	$$
	Next, by definition of $\hat{\mathcal{J}}_n$ in \eqref{eqn:Jn}, we have 
	\begin{align*}
	\widehat{\mathbb M}_{\lambda, n} (\widehat{\mathcal W}) & \geq (1-n^{-1}) {D}_\lambda(\widehat{\mathcal W})-n^{-1}\mathbb E_f ||\widehat{\mathcal W}(X)||^2_2 - \hat{\mathcal{J}}_n, \text{ and } \\
	\widehat {\mathbb M}_{\lambda,n} (\mathcal W_0) & \leq (1-n^{-1}) {D}_\lambda(\mathcal W_0)-n^{-1}\mathbb E_f ||\mathcal W_0(X)||_2^2 + \hat{\mathcal{J}}_n.
	\end{align*}
	Combining the above three inequalities and noting that ${D}_\lambda(\mathcal W_0)=0$, we arrive at:
	\begin{equation}\label{equ:D-lam}
	{D}_\lambda(\widehat{\mathcal W})\leq (n-1)^{-1} \{\mathbb E_f ||{\widehat{\mathcal W}(X)||_2^2} - \mathbb E_f ||\mathcal W_0(X)||_2^2\} + 2 \hat{\mathcal{J}}_n.
	\end{equation}

	To control the first term on the right side of \eqref{equ:D-lam}, we can show that there exists constants $c_3$ and $c_4$ such that
	$$
	\mathbb E_f ||\widehat{\mathcal W}(X)||_2^2 - \mathbb E_f ||\mathcal W_0(X)||_2^2 \leq c_3\, \Delta^{(1)}(\widehat{\mathcal W})+c_4.
	$$
	The proof of the above equation follows along similar lines to the proof of the subsequent lemma, which is established in Section \ref{sec:prf_lem.a2}. Hence, we do not repeat the arguments here. \begin{lemma}\label{lem.a2}
		For sub-exponential $f$ with rate $\lambda_f$ for any $p >1$ and $b >1$, there exists a constant $C$ such that,
		$$\sup_{h \in \fclass} \big[\dot \Delta^{(p)}(h) - C \{ b \,[\,\{\dot \Delta^{(1)}(h)\}^{1/p} + \exp(-\lambda_f \, p^{-1} b)\,]\,\}\big]<0.$$
		In particular, there exist constants $C_1$ and $C_2$ such that for all large $n$,
		\begin{align*}
		& \sup_{h \in \fclass} \dot \Delta^{(p)}(h) - C_1  (\log n) [\{\dot \Delta^{(1)}(h)\}^{1/p} + n^{-1}]<0 \text{ for all } p >0, \text{ and, } \\ 
		& \sup_{h \in \fclass} \big[\Delta^{(2)}( h) - C_2 (\log n)\{ \dot \Delta^{(1)}( h) + n^{-1}\}\big] < 0 .
		\end{align*}
	\end{lemma}
	
	To control the second term on the right side of \eqref{equ:D-lam}, we will need the following lemma. 
	\begin{lemma}\label{lemma.Jn}
		$\hat{\mathcal{J}}_n = O_p(n^{-1/2} \log n)$ when $\liminf_{n \to \infty} \lambda_n n^{1/4}=\infty$. 
	\end{lemma}
	
	The proof of this lemma, which is presented in Section \ref{sec:prf_Jn}, is fundamentally based on Lemma~\ref{lemma.g1}, but it involves additional analysis. 
	
	Thus, combining the above results, we arrive at: 
	$$
	{D}_\lambda(\widehat{\mathcal W}) \leq c_3 n^{-1} \Delta^{(1)}(\widehat{\mathcal W})+ O_p(n^{-1/2} \log n).
	$$
	
	We use \eqref{eq:lem2.bound} with \( b = 2\log n \) and \( d = K\varepsilon \) for any fixed \(\varepsilon > 0\), and plug in the above upper bound on \({D}_\lambda(\widehat{\mathcal W})\). Note that \(1 - \Phi_k(2\log n) = o(n^{-2})\). We have
	\begin{align}\label{eq:temp.A3}
	\bar{\Delta}(\widehat{\mathcal W}) \leq C \left[ \lambda^{-K} \left \{\frac{\Delta^{(1)}(\widehat{\mathcal W})}{n} + \frac{a_n}{n^{1/2}} + \frac{\Delta^{(2)}(\widehat{\mathcal W})}{n^2}  \right \}+ a_n \lambda \Delta^{(2)}(\widehat{\mathcal W}) + a_n^{K+1} \lambda \dot \Delta^{(1)}(\widehat{\mathcal W})  \right](1+o_p(1)),
	\end{align}
	where $a_n=\log n$ and $C$ is a constant independent of $\lambda$, $b$, and $\widehat{\mathcal W}$. 
	
	As \(f\) is sub-exponential, by Lemma~\ref{lem.a2} we have
	\[
	\Delta^{(2)}(\widehat{\mathcal W}) \leq a_n (\dot \Delta^{(1)}(\widehat{\mathcal W}) + n^{-1}).
	\]
	Substituting the above bound in \eqref{eq:temp.A3} and collecting dominant terms on the right-hand side, we get:
	\[
	\bar{\Delta}(\widehat{\mathcal W}) \leq C \left[  \frac{a_n \lambda^{-K}}{n^{1/2}} + {a_n}^{K+1} \lambda \, \Delta^{(1)}(\widehat{\mathcal W}) \right] (1 + o_p(1)).
	\]

	Next, we state a lemma, which will also be used in Section \ref{sec:prf_lemma.g2}. 
	\begin{lemma}\label{lem.a3}
		For sub-exponential $f$ for any $p >1$ and $b >0$, there exists a constant $C$ such that,
		$$\sup_{h \in \fclass}\big[{\dot \Delta}^{(1)}(h)- C\{ (\log n)^{K/2}\bar{\Delta}^{1/2}(h)+n^{-1}\}\big]<0 \text{ for all large } n.$$
	\end{lemma}
	
	It follows from Lemma~\ref{lem.a3} that the following inequality holds: 
	\begin{align}\label{eq:temp.AA.1}
	\dot \Delta^{(1)}(\widehat{\mathcal W}) \leq \{{a_n}^K \bar{\Delta}(\widehat{\mathcal W})\}^{1/2} + n^{-1}.
	\end{align}
	
	Combining the above two inequalities, we get: 
	$$ \dot \Delta^{(1)}(\widehat{\mathcal W}) \lesssim \max\big\{ a_n^{(1-K)/2}\lambda^{-K/2}\,n^{-1/4},\lambda \,a_n\big\}.$$
	Now, setting $\lambda=a_n^{-1+1/(K+2)}\,n^{-2^{-1}(K+2)^{-1}}$ we have: 
	\[
	\dot \Delta^{(1)}(\widehat{\mathcal W}) \lesssim a_n^{(K+2)^{-1}}\,n^{-2^{-1}(K+2)^{-1}}.
	\]
	Noting $\Delta^{(1)}(\widehat{\mathcal W})\leq \dot \Delta^{(1)}(\widehat{\mathcal W})$ and using Lemma \ref{lemma.g1}, we get:
	$$
	\hat \Delta^{(1)}(\widehat{\mathcal W}) \lesssim a_n^{(K+2)^{-1}}\,n^{-2^{-1}(K+2)^{-1}}=(\log n)^{1/4}n^{-1/8},
	$$
	where, the last equality was obtained by substituting $a_n=\log n$ and $K=2$. 
	It follows from Lemmas~\ref{lem.a2} and \ref{lemma.g1} that the same rate holds for $\hat{\Delta}^{(2)}(\widehat{\mathcal W})$ with an additional $a_n$ factor, and so, we have 
	$$
	\hat \Delta^{(2)}(\widehat{\mathcal W})  \lesssim a_n^{(K+2)^{-1}+1}\,n^{-2^{-1}(K+2)^{-1}}=(\log n)^{5/4}n^{-1/8},
	$$
	when $\lambda=(\log n)^{-3/4}n^{-1/8}$. This completes the proof of the theorem for the case $\Omega=I$. 
	
	Note that if \(\Omega\) is known and not equal to \(I\), then by Assumption \ref{assump:2}, the above proof holds with \(f\) being the density of \(\Omega^{1/2}X\). If \(\Omega\) is unknown and \(\hat{\Omega}\) is used in the Gaussian kernel, first note that the kernel bandwidth $\lambda^2$ satisfies \(o(n^{-1/2})\), and we also have \(\|\hat{\Sigma} - \Sigma\| = O_p(n^{-1/2})\). Therefore, the objective criterion with the sample inverse covariance \(\widehat {\mathbb M}_{\lambda,n}(h;\hat{\Omega})\) and the (unknown) true inverse covariance \(\widehat {\mathbb M}_{\lambda,n}(h;\Omega)\) differs only by
	\[
	|\widehat {\mathbb M}_{\lambda,n}(h;\hat{\Omega}) - \widehat {\mathbb M}_{\lambda,n}(h;\Omega)| = O_p(n^{-1}).
	\]
	Thus, the proof of the theorem still holds. %\hfill$\blacksquare$
	
	\subsection{Improved convergence rate when $f$ is bounded below}\label{sec:bounded.f}
	
	If the true density \(f\) is bounded below, the rate of convergence of the \(L_2\) risk in Theorem~\ref{thm:w} can be significantly improved. Specifically, consider the case where \(f\) has bounded support and is bounded away from zero on its support. In this setting, we obtain the following result for the \(L_2\) risk of the proposed estimator. This result can be readily extended to the case where \(f\) is a mixture density with a flat core and sub-exponential tails.

	%away from zero—which would hold, for example, if \(f\) lies within a Lipschitz ball or is a mixture of sub-exponential and flat densities—then it follows
	
	\begin{lemma}\label{lem:bounded.f}
		Under Assumptions \ref{assump:2} -- \ref{assump:eigen_bound}, if $f$ has bounded support and is bounded below, then there exists $\lambda_n$ such that 
		$$\dfrac{1}{n}\big\|\hat{\mathcal W}_n(\lambda_n)-{\mathcal W}_0\big\|_F^2=O_p((\log n)^{10} n^{-1/2}).$$
	\end{lemma}
	
	\noindent \textbf{Proof.} If the true density $f$ is such that it is bounded below on its support, i.e.,  $\inf_{u\in \mathbb{R}^K}f(u) \times 1\{f(u)\neq 0\}>c$ for some constant $c >0$, then, instead on \eqref{eq:temp.AA.1} in the proof of Theorem~\ref{thm:w} we will have an much improved upper bound on $\dot \Delta^{(1)}(\widehat{\mathcal W})$ as
	\begin{align}\label{eq:temp.AA.2}
	\dot \Delta^{(1)}(\widehat{\mathcal W}) \lesssim \bar{\Delta}(\widehat{\mathcal W}).
	\end{align}
	This would imply by following the successive steps in the proof of Theorem~\ref{thm:w} that: 
	$$ \dot \Delta^{(1)}(\widehat{\mathcal W}) \lesssim  a_n\lambda^{-K}\,n^{-1/2}+\lambda \,a_n^{K+1} \dot \Delta^{(1)}(\widehat{\mathcal W}).$$
	Now, setting $\lambda=a_n^{-(K+2)}$ we get: 
	$$ \dot \Delta^{(1)}(\widehat{\mathcal W}) \lesssim a_n^{(K+1)^2}\,n^{-1/2}.$$
	Thus, we have, $ \dot \Delta^{(1)}(\widehat{\mathcal W}) \lesssim a_n^{9}\,n^{-1/2}$ for $\lambda=a_n^{-4}$. This implies:  
	$$ \hat \Delta^{(2)}(\widehat{\mathcal W}) \lesssim a_n^{10}\,n^{-1/2}=(\log n)^{10}n^{-1/2}.$$ 
	
	\textbf{Remark.} The poly-logarithmic terms in the rate of convergence stated in Lemma~\ref{lem:bounded.f} are not optimized but arise directly from the proof of Theorem~\ref{thm:w}. The difference between the convergence rates in Theorem~\ref{thm:w} and Lemma~\ref{lem:bounded.f} stems solely from the fact that \eqref{eq:temp.AA.2} provides a much sharper bound than \eqref{eq:temp.AA.1}. This sharper bound is achieved because \(f\) is bounded below. Consequently, in this case unlike Theorem~\ref{thm:w}, we do not incur a loss of precision (in the convergence rate, up to logarithmic terms) when transitioning from the KSD distance between score functions to the \(L_2\) distance.

	\subsection{Proof of Lemma~\ref{lemma.g1}}\label{sec:prf_lemma.g1} 
	For all $h\in \fclass$, denote $h-\mathcal{W}_0$ by $\dot{h}$. 
	Note that, $\dot h$ lies in a  Lipschitz ball with radius at most $2L$.  
	We next derive the rates of convergence for $\sup_{h \in \fclass}|G_n(h)|$, where
	\[
	G_n(h) = \frac{1}{n} \sum_{i=1}^n \|\dot{h}(X^i)\|_2^2 - \mathbb{E}\|\dot{h}(X)\|_2^2~.
	\]
	
	The convergence of the other two terms in \(\mathcal{G}_n\) follows along similar lines. Thus, obtaining the rate of convergence for \(\sup_{h \in \fclass}\left|\hat \Delta^{(1)}(h)-\Delta^{(1)}(h)\right|\) is even simpler than for \(\sup_{h \in \fclass}\left|G_n(h)\right|\), as \(||\dot{h}||_1\) is Lipschitz for all \(h \in \fclass\), whereas the set \(\{||\dot{h}||_2^2: h \in \fclass\}\) is not.
	
	For deriving the rate of convergence of $G_n$ we first consider truncating the distribution of \(X\) at \(M\). Let \(Y = X \, \mathbf{1}_{\{\|X\|_2 \leq M\}}\). Then, by the sub-exponentiality of \(f\), we have
	\[
	\begin{aligned}
	\mathbb{E}\|\dot{h}(X)\|_2^2 - \mathbb{E}\|\dot{h}(Y)\|_2^2
	&= \mathbb{E}\left[\|\dot{h}(X)\|_2^2 \, \mathbf{1}_{\{\|X\|_2 > M\}}\right] \\
	&\leq (2L)^2 \, \mathbb{E}\|X\|_2^2 \, \mathbf{1}_{\{\|X\|_2 > M\}} 
	\leq C \cdot M e^{-\lambda_f M},
	\end{aligned}
	\]
	where $\lambda_f$ is the rate parameter of $f$. 
	
	Next we state a lemma that is needed in our proof.
	\begin{lemma}\label{lem.a1}
		For sub-exponential $f$ and any $p >0$, we have  
		$$\sup_{h \in \fclass}\Delta^{(p)}(h)< \infty.$$
	\end{lemma}
	
	Using the proof of Lemma ~\ref{lem.a1}, \(\mathbb{E}\|X\|_2^4 < \infty\) due to sub-exponentiality of \(f\), and by the strong law of large numbers, we have
	\[
	\begin{aligned}
	&\frac{1}{n} \sum_{i=1}^n \|\dot{h}(X^i)\|_2^2 - \frac{1}{n} \sum_{i=1}^n \|\dot{h}(Y^i)\|_2^2
	= n^{-1} \sum_{i=1}^n \|\dot{h}(X^i)\|_2^2 \mathbf{1}_{\{\|X^i\|_2 > M\}} \\
	&\leq (2L)^2 \cdot n^{-1} \sum_{i=1}^n \|X^i\|_2^2 \, \mathbf{1}_{\{\|X^i\|_2 > M\}} \stackrel{a.s.}{\rightarrow} (2L)^2 \, \mathbb{E}\|X\|_2^2 \, \mathbf{1}_{\{\|X\|_2 > M\}} \quad \text{ as } n \to \infty,
	\end{aligned}
	\]
	
	Now, consider the following quantity
	\[
	\tilde{G}_n(h) = n^{-1} \sum_{i=1}^n \|\dot h(Y^i)\|_2^2 - \mathbb{E}\|\dot h(Y)\|_2^2
	\]
	and choose the truncation  \(M = \lambda_f^{-1} \log n\). Then, there exists a constant $C$ such that 
	\begin{align}\label{eq:l1.1}
	\sup_{h \in \fclass} |G_n(h) - \tilde{G}_n(h)| \leq C \, n^{-1}\log n \;\text{ a.s.}
	\end{align}
	We next concentrate on $\sup_{h \in \fclass}|\tilde G_n(h)|$. By the symmetrization lemma (\citealp{van1996weak}, Lemma 2.3.1), we have: 
	\begin{equation}\label{eqn:EG}
	\mathbb E \sup_{h \in \fclass}|\tilde G_n(h)|
	\leq 2\mathbb{E}\left[\,\sup_{h \in \fclass} \left| \frac{1}{n}\sum_{i=1}^n \epsilon_i ||\dot{h}(Y^i)||_2^2\right|\;\right],
	\end{equation}
	where \(\epsilon_i \overset{i.i.d.}{\sim} \mathrm{Rademacher}(\pm 1)\), and is independent of \(Y^i\). 
	
	Next, consider $||\dot{h}(Y)||^2$ as a composition of two functions $\phi_1$ and $\phi_2$,  where 
	$\phi_1: \mathbb{R} \to \mathbb{R}$ is $\phi(z)=z^2$ and $\phi_2: \mathbb{R}^K\to \mathbb{R}$
	is $\phi_2(z;\dot{h})=||\dot{h}(z)||_2$. 
	As $Y^i$ are bounded in $L_2$ norm by $M$, $\phi_1$ is Lipschitz with constant $2M.$
	
	Now, by applying the contraction inequality in Theorem 4.12 of \citet{ledoux1991probability}, the right side of \eqref{eqn:EG} is upper bounded by
	\begin{align}\label{eq:l1.2}
	4M \, \mathbb{E}\left[\;\sup_{h \in \fclass} \left| \frac{1}{n}\sum_{i=1}^n \epsilon_i ||\dot{h}(Y^i)||_2\right|\;\right]:=4M \mathcal{R}_n(\fclass).
	\end{align}
	Now, note that the class of all function $\{\phi_2(\,\cdot\,;\dot{h}): h \in \fclass\}$ lies in a Lipschitz ball $\fclass_{2L}$ of radius $2L$.  
	Using Dudley’s Entropy Integral [cf. Theorem 2.2.4 of \citep{van1996weak}], we obtain the following bound
	\[
	\mathcal{R}_n(\fclass)\leq \mathcal{R}_n(\fclass_{2L}) \lesssim \int_0^{\mathrm{diam}(\fclass_{2L})} \log^{1/2} \mathcal{N}(\epsilon, \fclass_{2L}, L_2(P_n))\,d\epsilon, 
	\]
	where \(\mathcal{N}(\epsilon, \fclass_{2L}, L_2(P_n))\) is the covering number and $P_n$ is the empirical measure. For \(\fclass_{2L}\), the covering number is upper bounded by:
	\[
	\log\mathcal{N}(\epsilon, \fclass_L, L_2(P_n)) \lesssim K\log(L/\epsilon).
	\]
	It follows that $\mathcal{R}_n(\fclass_{2L}) \lesssim LK^{1/2} n^{-1/2}$.
	This, along with \eqref{eq:l1.1}, implies that
	$$
	\sup_{h \in \fclass} |G_n(h)| = O_p(n^{-1/2} \log n),
	$$
	thus proving the desired result.
	
	\subsection{Proof of Lemma~\ref{lemma.g2}} \label{sec:prf_lemma.g2}
	
	We aim to establish an identity between KSD and $L_2$ norms that hold for all bandwidth $\lambda$ as well as for all $h \in \fclass$. For any $h \in \fclass$, define
	$$
	g(u, v; h) = \mathcal{K}_\lambda(u, v)\, \dot{h}(u)^T \dot{h}(v),
	$$
	where $\dot{h} = h - \mathcal{W}_0$. By definition, we have
	$D_\lambda(h) = \mathbb{E}_{(U,V) \sim f^2} \left\{ g(U, V; h) \right\}$. 
	
	Our strategy is to gradually approximate $D_\lambda(h)$ by intermediate quantities $I_\lambda^{(\cdot)}(h)$s towards $\bar{\Delta}(h)$ -- at every step, we will document an upper bound on the approximation error. 
	
	Now we start proving Lemma \ref{lemma.g2}. Let $\varepsilon = \lambda b$. Define $I_\lambda^{(1)}(h)$ as 
	$$
	\mathbb{E}_{(U,V) \sim f^2} \left\{ g(U, V; h)\, \mathbb{I}\{\|U - V\|_2 < \varepsilon\} \right\}.
	$$
	It follows that 
	\[
	D_\lambda(h) - I_\lambda^{(1)}(h) = \mathbb{E}\left[ \dot{h}(U)^T \dot{h}(v)\, \mathcal{K}_\lambda(U, V)\, \mathbb{I}\{\|U-V\|_2 > \varepsilon\} \right].
	\]
	
	Noting that,
	$\mathcal{K}_\lambda(U, V)\, \mathbb{I}\{\|U-V\|_2 > \varepsilon\} \leq e^{-b^2/2}$,
	and applying the Cauchy–Schwarz inequality, we arrive at
	\begin{align}\label{eq:2.1}
	|D_\lambda(h) - I_\lambda^{(1)}(h)| \leq \Delta^{(2)}(h)\, e^{-b^2/2}.
	\end{align}
	
	Next, consider the following quantity
	$$I_\lambda^{(2)}(h) = \mathbb{E}\left[ \mathcal{K}_\lambda(U, V)\, \|\dot{h}(U)\|_2^2\, \mathbb{I}\{\|U-V\|_2 < \varepsilon\} \right].$$
	By expanding
	$g(u, v; h)$ in  $I_\lambda^{(1)}(h)$ as  
	$$\mathcal{K}_\lambda(u, v)\, \|\dot{h}(u)\|_2^2 + \mathcal{K}_\lambda(u, v)\, \dot{h}(u)^T(\dot{h}(v)-\dot{h}(u)),$$
	we have $I_\lambda^{(1)}(h) = I_\lambda^{(2)}(h) + \mathbb{E}[R(U, V; h)]$, where 
	$$
	R(u, v; h) = \mathcal{K}_\lambda(u, v)\, \dot{h}(u)^T\big(\dot{h}(v)-\dot{h}(u)\big)\, \mathbb{I}\{\|u-v\|_2 < \varepsilon\}.
	$$
	Noting that, $\mathcal{K}_\lambda(u, v)<1$, we have:
	\begin{align}\label{eq:2.2}
	\big|\mathbb{E}\,R(U, V; h)\big| &\leq  \mathbb{E}_{(U,V)\sim f^2)}\, ||\dot{h}(U)||_2\,||\dot{h}(V)-\dot{h}(U)||_2\, \mathbb{I}\{\|U-V\|_2 < \varepsilon\}, \\
	&\leq 2L \cdot \mathbb{E}_{(U,V)\sim f^2)}\, ||\dot{h}(U)||_2\,||V-U||_2\, \mathbb{I}\{\|U-V\|_2 < \varepsilon\}.
	\end{align}
	By Fubini's theorem, the expectation above can be evaluated as:
	$$\int_{\mathbb{R}^K} ||\dot{h}(u)||_2 f(u) \bigg\{\int_{\mathbb{R}^K} ||v-u||_2 f(v) \mathbb{I}\{\|u-v\|_2 < \varepsilon\}\,dv\bigg\}du\, .$$
	The inner integral above is bounded by $\varepsilon \int_{\mathbb{R}^K} f(v)\, dv$ which is again upper bounded by $c_1 \varepsilon^{K+1}$ where $c_1$ is a constant that depends only on $K$ and is independent of $\lambda$ and $b$.
	Thus, we arrive at: 
	\begin{align}\label{eq:2.5}
	| I_\lambda^{(1)}(h) - I_\lambda^{(2)}(h)|
	\leq c_1\, \varepsilon^{K+1} \dot \Delta^{(1)}(h). 
	\end{align}

	Next, consider the following intermediate quantity:
	\[
	I_\lambda^{(3)}(h) = \int_{\mathbb{R}^K} \int_{\mathbb{R}^K} f^2(u)\,\mathcal{K}_\lambda(u,v)\,||\dot{h}(u)||_2^2\,\mathbb{I}\{\|u - v\|_2 < \varepsilon\}\,dv\,du.
	\]
	By Fubini's theorem, we rewrite it as
	\[
	I_\lambda^{(3)}(h) = \int_{\mathbb{R}^K} f^2(u)\,||\dot{h}(u)||_2^2 \left[\int_{\mathbb{R}^K} \mathcal{K}_\lambda(u,v)\,\mathbb{I}\{\|u - v\|_2 < \varepsilon\}\,dv\right] du.
	\]
	As \(\varepsilon = \lambda b\), the inner integral above equals
	\[
	\lambda^K \int_{\mathbb{R}^K} e^{-\|x\|^2/2}\,\mathbb{I}\{\|x\|_2 < b\}\,dx=\lambda^K (2\pi)^{K/2}\Phi_K(b),
	\]
	where $\Phi_K(b)$ is the Gaussian probability contained in a sphere around the center of radius $||b||$, i.e., if $Z\stackrel{d}{=} N_K(0,I)$ and $\Phi_K(b)=\mathbb{P}(||Z||_2\leq b)=\mathbb{P}(\chi^2_K \leq b^2)$.  Thus, we have
	\[
	I_\lambda^{(3)}(h) =c_3 \lambda^K\,\bar{\Delta}(h)\,\Phi_K(b),
	\]
	where $c_3$ is independent of $\lambda$ and $b$.  
	
	Next, we consider the approximation error in $I^{(3)}$ from $I^{(2)}$. By definition,
	\[
	I_\lambda^{(2)}(h) = \iint_{\mathbb{R}^K \times \mathbb{R}^K} f(u)\,f(v)\,\mathcal{K}_\lambda(u,v)\,||\dot{h}(u)||_2^2\,\mathbb{I}\{\|u - v\|_2 < \varepsilon\}\,du\,dv.
	\]
	It follows that
	\begin{align*}
	\big|I_\lambda^{(2)}(h) - I_\lambda^{(3)}(h)\big| &\leq \int_{\mathbb{R}^K}   f(u) ||\dot{h}(u)||_2^2 \int_{\mathbb{R}^K}\,\mathcal{K}_\lambda(u,v)\,|f(v)-f(u)| \mathbb{I}\{\|u - v\|_2< \varepsilon\}\,dv\,du,
	\\
	&\leq  \int_{\mathbb{R}^K } f(u)||\dot{h}(u)||_2^2\, \int_{\mathbb{R}^K}\,\mathcal{K}_\lambda(u,v)\ L_f\,\varepsilon\, dv\;\; du,
	\end{align*}
	where the second inequality follows by applying the fact that $f$ is Lipschitz with constant $L_f$. Thus,  we arrive at: 
	\begin{align}\label{eq:2.6}
	|I_\lambda^{(2)}(h) - I_\lambda^{(3)}(h)| \leq L_f\,\varepsilon\, \lambda^K \Delta^{(2)}(h).
	\end{align}
	
	Combining, \eqref{eq:2.1}, \eqref{eq:2.5} and \eqref{eq:2.6}, we get $|D_\lambda(h) - I_\lambda^{(3)}(h) |$ bounded above by
	\[
	B(h;\lambda,b)=\Delta^{(2)}(h)\, e^{-b^2/2}
	+ c_1\, \varepsilon^{K+1}\, \dot \Delta^{(1)}(h)
	+ L_f\, \varepsilon \lambda^K\, \Delta^{(2)}(h).
	\]
	It follows that
	\[
	D_\lambda(h) \geq I_\lambda^{(3)}(h)-B(h;\lambda,b) = \lambda^K\, \bar{\Delta}(h) \Phi_K(b)-B(h;\lambda,b)
	\]
	which implies that
	\[
	\Phi_K(b) \bar{\Delta}(h)
	\leq \lambda^{-K}  [D_\lambda(h)
	+ B(h;\lambda,b)].
	\]
	This completes the proof.

	\subsection{Proof of Lemma \ref{lem.a2}}\label{sec:prf_lem.a2}
	
	We first decompose $[\dot \Delta^{(p)}(h)]^p$ based on the set
	\(\{\|x\|_2 \leq M\}\) and its complement, and then upper bound it by
	\begin{align*}
	&(2L)^{p-1} M^{p-1} \mathbb{E}\left\{\|\dot{h}(X)\|_2\right\}
	+  \mathbb{E}\left\{\|\dot{h}(X)\|_2^p \mathbf{1}_{\{\|X\|_2 > M\}}\right\}\\
	&= (2L)^{p-1} M^{p-1} [\dot \Delta^{(1)}(h)] + L^p \mathbb{E}\left[\|X\|_2^p \mathbf{1}_{\{\|X\|_2 > M\}}\right].
	\end{align*}
	The second term above is upper bounded by a (constant) multiple of 
	$M^{p-1} e^{-\lambda_f M}$, where \(\lambda_f\) is the decay rate of \(f\). Thus, we have
	\begin{align}\label{eq:temp.A1}
	\{\dot \Delta^{(p)}(h)\}^p \lesssim  M^{p-1} [\dot \Delta^{(1)}(h) + e^{-\lambda_f M}].
	\end{align}
	Substituting $M=\lambda_f^{-1}p^{-1} \log n$ and $p=2$, we have 
	$$\Delta^{(2)}(h)=\{\dot \Delta^{(2)}(h)\}^2 \lesssim  (\log n)\, [\{\dot \Delta^{(1)}(h)\} + n^{-1}].$$
	
	From \eqref{eq:temp.A1}, we have: 
	\[
	\dot \Delta^{(p)}(h) \lesssim  M^{1-1/p} [\dot \Delta^{(1)}(h) + e^{-\lambda_f M}]^{1/p} \lesssim  M^{1-1/p} [\{\dot \Delta^{(1)}(h)\}^{1/p} + e^{-\lambda_f M/p}].
	\]
	In particular, let $M=\lambda_f^{-1}p^{-1} \log n$. We have 
	\[
	\dot \Delta^{(p)}(h) \lesssim  (\log n) [\{\dot \Delta^{(1)}(h)\}^{1/p} + n^{-1}].
	\]

	\subsection{Proof of Lemma \ref{lemma.Jn}}\label{sec:prf_Jn}
	We upper bound $\hat{\mathcal{J}}_n$ using the following decomposition:
	\begin{align}\label{eq:temp-A1}
	\hat{\mathcal{J}}_n \leq  (1-n^{-1}) \sup_{h \in \fclass} |D_{\lambda}(h)-\hat D_{\lambda,n}(h)|+n^{-1} \sup_{h \in \fclass} \big|n^{-1}\sum_{i=1}^n \kappa_\lambda[h](X^i,X^i)-\mathbb E_f ||h(X)||^2\big|.
	\end{align}
	By lemma~\ref{lemma.g1}, the first term above is bounded by $O_p(n^{-1/2}\log n )$. For the second term note that: 
	\[
	\kappa_\lambda[h](u,u ) = \mathcal{K}_\lambda(u, u) ||{h}(u)||_2^2 + 2 {h}^T(u)\nabla_1 \mathcal{K}_\lambda(u, u)  + \text{trace}(\nabla_{1,2} \mathcal{K}_\lambda(u, u)),
	\]
	where the Gaussian kernel with identity covariance is
	$\mathcal{K}_\lambda(u,v) = \exp( -(2\lambda^2)^{-1}{\|u - v\|_2^2})$. Note that we have excluded the normalizing constant throughout the paper. It follows that
	\[
	\mathcal{K}_\lambda(u, u) = 1, \quad \nabla_1 \mathcal{K}_\lambda(u,u) = 0, \quad \text{trace}(\nabla_{1,2} \mathcal{K}_\lambda(u, u)) = \lambda^{-2}. 
	\]
	Therefore we have
	\[
	\kappa_\lambda\big[ {h}\big](X^i, X^i) = ||{h}(X^i)||_2^2 + \lambda^{-2} \text{ for all } i=1,\ldots,n~.
	\]
	It follows that the second term in \eqref{eq:temp-A1} is bounded above by
	$$
	n^{-1} \sup_{h \in \fclass} \bigg|n^{-1}\sum_{i=1}^n ||{h}(X^i)||_2^2-\mathbb E_f ||h(X)||^2\bigg| + n^{-1} \lambda^{-2}.
	$$
	By lemma~\ref{lemma.g1}, we conclude that the first term above is in the order of $O_p(n^{-3/2} \log n)$ and the second term is in the order of $O(n^{-1/2})$, when $n^{1/4}\lambda \to \infty$ as $n \to \infty$. This completes the proof.

	\subsection{Proof of Lemma \ref{lem.a3}}\label{sec:prf_lem.a3}
	We decompose $\dot \Delta^{(1)}(h) = \int \| \dot{h}(x) \|_2 f(x)\, dx$ into
	\begin{align}\label{eq:temp.A2}
	\int \| \dot{h}(x) \|_2 f(x)\, \mathbf{1}\{\|x\|_2 \leq M\}\, dx + \mathbb{E}\left[\|\dot{h}(x)\|_2 \mathbf{1}\{\|x\|_2 > M\}\right].
	\end{align}
	Using Cauchy-Schwarz inequality, the first term is upper bounded by
	\[
	\left( \int \|\dot{h}(x)\|_2^2 f^2(x) \mathbf{1}\{\|x\|_2 \leq M\}\, dx \right)^{1/2}
	\left( \int \mathbf{1}\{\|x\|_2 \leq M\}\, dx \right)^{1/2}
	\]
	which is again bounded above by
	\[
	\left( \int \|\dot{h}(x)\|_2^2 f^2(x) \, dx \right)^{1/2}
	\left(C_K\, M^{K/2}\right)
	\lesssim \left[ \bar{\Delta}(h) \right]^{1/2} M^{K/2}.
	\]
	The second term in \eqref{eq:temp.A2} is bounded above by
	\[
	(2L) \int \|x\|_2 f(x)\, \mathbf{1}\{\|x\|_2 > M\}\, dx \leq e^{-\lambda_f M},
	\]
	where \(\lambda_f\) is the decay rate of \(f\).
	
	Combining the above two bounds, we get, for all \(h \in \fclass\),
	\[
	\dot \Delta^{(1)}(h) \lesssim M^{K/2} \left[\bar{\Delta}(h)\right]^{1/2} + e^{-\lambda_f M}.
	\]
	Choosing \(M = \lambda_f^{-1} \log n\), we have,
	\[
	\dot \Delta^{(1)}(h) \lesssim \log^{K/2} n \left[\bar{\Delta}(h)\right]^{1/2} + {n}^{-1}.
	\]

	\subsection{Proof of Lemma \ref{lem.a1}}\label{sec:prf_lem.a1}
	
	For any $M >0$, consider the following decomposition: 
	\begin{equation}\label{eqn:ph}
	\Delta^{(p)}(h)=\mathbb E ||\dot h(U)||_p^p 1\{||U||_p\leq M\} + \mathbb E ||\dot h(U)||_p^p 1\{||U||_p> M\}.
	\end{equation}
	
	As $\mathcal{W}_0\in \fclass$, all $\dot{h}$ are also Lipschitz with Lipschitz constant less than $2L$. It follows that the first term on the right hand side of \eqref{eqn:ph} is bounded by $(2L)^p M^p+KB^p$. Using the Lipschitz property again, we see that the second term on the right hand side of \eqref{eqn:ph} is less than 
	$$
	KB^p+(2L)^p\mathbb E_{U\sim f}(||U||_p^p). 
	$$
	As $f$ is sub-exponential all its moments are finite. It follows that there exists a constant $\mathfrak{m}_p$ (the raw moment) such that:  
	\begin{align*}
	\Delta^{(p)}(h)\leq 2^pL^p (M^p+\mathfrak{m}_p)+ 2 KB^p,
	\end{align*}
	proving the desired result. 
}

\section{Proof of Theorem \ref{thm:bayesrisk}}
\label{sec:proof-thm2}
%----------------------------------------------------------------------
%----------------------------------------------------------------------
We first state two lemmata that are needed for proving Theorem \ref{thm:bayesrisk}. Denote $c_0,c_1,\ldots$ some generic positive constants which may vary in different statements.

\begin{lemma}
	\label{lem:1}
	If Assumption \ref{assump:4} holds, then with probability tending to $1$,
	$C_1/\log n\le \tau \le C_2\log n$ and $|\mu|\le C_3\log n$
	for some positive constants $C_1, C_2$ and $C_3$. 
\end{lemma}
%\begin{lemma}
%\label{lem:2}
%Model \eqref{eq:normmodel}, Assumption \ref{assump:4} and Lemma \ref{lem:1} imply that with probability tending to $1$ as $n\to\infty$,
%$$f(y,s^2)\ge \dfrac{c_0 (s^2)^{(m-3)/2}a_n}{n^2(\log n)^{m/2}},
%$$
%where $a_n=\exp\{-\sqrt{(m-1)\log n}\}$.
%\end{lemma}
\begin{lemma}
	\label{lem:3}
	Consider Model \eqref{eq:normmodel}. Suppose Assumption \ref{assump:4} holds. Then with probability tending to $1$, 
	$\mathbb{E}(\tau_i|y_i,s_i^2) \ge c_0/\log n$.
\end{lemma}

Lemmata \ref{lem:1} and \ref{lem:3} are proved in Sections \ref{prf:lem1} and \ref{prf:lem3}, respectively. We now prove Theorem \ref{thm:bayesrisk}.

To establish the first part of Theorem \ref{thm:bayesrisk}, note that
\begin{eqnarray*}
	\dfrac{1}{n}\|\bm \delta^{\sf ds}_n(\lambda)-\bm \delta^{\pi}_{(1)}\|_2^2&=& \dfrac{1}{nm^2}\sum_{i=1}^{n}\Big|\dfrac{w_{1,i}}{\hat{\tau}_i^{\pi}}-\dfrac{\hat{w}^{(1)}_{\lambda,n}(i)}{\tau_{i,n}^{\sf ds}(\lambda)}\Big|^2\nonumber\\
	&\le&\dfrac{2}{nm^2}\sum_{i=1}^{n}\dfrac{1}{[\tau_{i,n}^{\sf ds}(\lambda)]^2}\Big|w_{1,i}-\hat{w}_{\lambda,n}^{(1)}(i)\Big|^2+\dfrac{2}{nm^2}\sum_{i=1}^{n}\Big|w_{1,i}\Big|^2~\Big|\dfrac{1}{\tau_{i,n}^{\sf ds}(\lambda)}-\dfrac{1}{\hat{\tau}_i^{\pi}}\Big|^2
	\\ & \coloneqq & T_1+T_2. 
\end{eqnarray*} 
Consider the first term $T_1$. From the discussion in Section \ref{Subsec:bandwidth}, there is a positive constant $c_0$ such that $\tau_{i,n}^{\sf ds}(\lambda)>c_0>0$ for all $i=1,\ldots,n$. It follows that for some constant $c_1>0$ depending on the fixed $m$,
\begin{eqnarray}
\label{eq:prf_thm3_2}
T_1&\le& \dfrac{c_1}{n}\Big\|\bm w^{(1)}_0-\hat{\bm w}^{(1)}_{\lambda,n}\Big\|_2^2\le\dfrac{c_1}{n}\Big\|\mathcal W_0-\hat{\mathcal W}_n (\lambda)\Big\|_F^2,
\end{eqnarray}
where $\hat{\bm w}^{(1)}_{\lambda,n}$ and $\bm w^{(1)}_0$ are, respectively, the first column of $\hat{\mathcal W}_n(\lambda)$ and $\mathcal{W}_0$.
From Theorem \ref{thm:w} the last term on the right hand side of the inequality in equation \eqref{eq:prf_thm3_2} is $O_p\{n^{-\gamma}\}$ for some constant $\gamma<1/8$.

Next consider the second term  $T_2$. We have
\begin{eqnarray}
\label{eq:prf_thm3_3}
T_2 \le \dfrac{c_2}{n}\sum_{i=1}^{n}\Big|\dfrac{w_{1,i}}{\hat{\tau}_i^{\pi}}\Big|^2\Big|w_{2,i}-\hat{w}^{(2)}_{\lambda,n}(i)\Big|^2.
\end{eqnarray}
We will use Lemma \ref{lem:3} to bound the terms $|w_{1,i}/\hat{\tau}_i^{\pi}|$ in equation \eqref{eq:prf_thm3_3}. First note that since $\bm X^i$ is sub-exponential, $|w_{1,i}|\le c_1\log n$ with high probability from the discussion in the proof of Theorem \ref{thm:w}. Next, Model \eqref{eq:normmodel}, Assumption \ref{assump:4} and Lemma \ref{lem:1} imply that with probability tending to 1, 
$|Y_i|\le c_2\log n$ and 
$
(m-1)n^{-1}\le(m-1)S_i^2\tau_i\le (m-1)+2\sqrt{(m-1)\log n}+2\log n
$
[cf. Lemma 1 of \cite{laurent2000adaptive}]. So conditional on these events, we have, from Lemma \ref{lem:3}, $|w_{1,i}/\hat{\tau}_i^{\pi}|\le c_3(\log n)^2$. Thus,
\begin{eqnarray}
\label{eq:prf_thm3_4}
T_2&\le& \dfrac{c_4(\log n)^4}{n}\sum_{i=1}^{n}\Big|w_{2,i}-\hat{w}^{(2)}_{\lambda,n}(i)\Big|^2,\nonumber
\end{eqnarray}
which is $O_p\{n^{-\gamma}\}$ from Theorem \ref{thm:w}. Thus, $n^{-1}\|\bm \delta^{\sf ds}_n(\lambda)-\bm \delta^{\pi}_{(1)}\|_2^2$ is $O_p\{n^{-\gamma}\}$.

\bigskip

Now we will prove the second part of Theorem \ref{thm:bayesrisk}. Observe that $|l^{(1)}_n(\bm \mu,\bm \delta^{\pi}_{(1)};\bm \tau)-l^{(1)}_n\{\bm \mu,{\bm\delta}^{\sf ds}_n({\lambda});\bm \tau\}|$ equals
%$$
%\dfrac{1}{n}\Big|\big\|\bm \mu-\bm \delta^{\pi}_{(1)}\big\|_2-\big\|\bm \mu-\bm\delta^{\sf ds}(\lambda)\big\|_2\Big|~~\Big|\big\|\bm \mu-\bm \delta^{\pi}_{(1)}\big\|_2+\big\|\bm \mu-\bm\delta^{\sf ds}(\lambda)\big\|_2\Big|
%$$ 
$$
\Big|\sqrt{l^{(1)}_n(\bm \mu,\bm \delta^{\pi}_{(1)};\bm \tau)}-\sqrt{l^{(1)}_n\{\bm \mu,{\bm\delta}^{\sf ds}_n({\lambda});\bm \tau\}}\Big|~~\Big|\sqrt{l^{(1)}_n(\bm \mu,\bm \delta^{\pi}_{(1)};\bm \tau)}+\sqrt{l^{(1)}_n\{\bm \mu,{\bm\delta}^{\sf ds}_n({\lambda});\bm \tau\}}\Big|
$$ 
and Triangle inequality implies
\begin{eqnarray}
\Big|\sqrt{l^{(1)}_n(\bm \mu,\bm \delta^{\pi}_{(1)};\bm \tau)}-\sqrt{l^{(1)}_n\{\bm \mu,{\bm\delta}^{\sf ds}_n({\lambda});\bm \tau\}}\Big|&\le& \Bigl|\sqrt{\dfrac{1}{n}\sum_{i=1}^{n}\tau_i\bigl\{\delta_{i,(1)}^{\pi}-\delta^{\sf ds}_{i,n}(\lambda)\bigr\}^2}\Bigr|\nonumber\\
\label{eq:prf_thm_3_4}
&\le& c_0\sqrt{\dfrac{\log n}{n}}\big\|\bm{\delta}^{\sf ds}_n(\lambda)-\bm{\delta}^{\pi}_{(1)}\big\|_2,%\dfrac{1}{\sqrt{n}}\big\|\bm{\delta}^{\sf ds}(\lambda)-\bm{\delta}^{*}\big\|_2.
\end{eqnarray}
where the last inequality in Equation \eqref{eq:prf_thm_3_4} follows from Lemma \ref{lem:1}. Thus, from the first part of Theorem \ref{thm:bayesrisk} and Lemma \ref{lem:1}, the quantity on the right hand side of the inequality in equation \eqref{eq:prf_thm_3_4} is $O_p\{(\log n)^{1/2}n^{-\gamma/2}\}$. Thus, it follows from equation \eqref{eq:prf_thm_3_4} that $$\sqrt{l^{(1)}_n\{\bm \mu,{\bm\delta}^{\sf ds}_n(\lambda);\bm \tau\}}\le\sqrt{l^{(1)}_n(\bm \mu,{\bm\delta}^{\pi}_{(1)};\bm \tau)}+O_p\{(\log n)^{1/2}n^{-\gamma/2}\}, \; \mbox{and}$$  
$$\big|l^{(1)}_n(\bm \mu,\bm \delta^{\pi}_{(1)};\bm \tau)-l^{(1)}_n\{\bm \mu,{\bm\delta}^{\sf ds}_n({\lambda});\bm \tau\}\big|\le 4\sqrt{l^{(1)}_n(\bm \mu,{\bm\delta}^{\pi}_{(1)};\bm \tau)}~\big|\sqrt{l^{(1)}_n(\bm \mu,\bm \delta^{\pi}_{(1)};\bm \tau)}-\sqrt{l^{(1)}_n\{\bm \mu,{\bm\delta}^{\sf ds}_n({\lambda});\bm \tau\}}\big|\bigl\{1+o_p(1)\bigr\}.$$
%Now Assumption \ref{assump:4}, together with Lemmata \ref{lem:1} and \ref{lem:3} imply ${l_n(\bm \mu,\bm \delta^{\pi}_{(1)})}$ is $O_p(\log^6 n)$. 
Now $\bm \delta^{\pi}_{(1)}$ is the Bayes estimator of $\bm \mu$ under the weighted squared error loss and so its risk $\mathbb{E}l^{(1)}_n(\bm \mu,\bm \delta^{\pi}_{(1)};\bm \tau)<\infty$. This implies $l^{(1)}_n(\bm \mu,\bm \delta^{\pi}_{(1)};\bm \tau)$ is $O_p(1)$. Thus, from equation \eqref{eq:prf_thm_3_4}, the first part of Theorem \ref{thm:bayesrisk} and the display above, we have the desired result.\hfill$\blacksquare$

\subsection{Proof of Lemma \ref{lem:1}}
\label{prf:lem1}
The proof of Lemma \ref{lem:1} follows directly from Assumption \ref{assump:4} and Markov's inequality. For example, fix a $\nu>0$ and note that, for $r=\epsilon_2^{-\nu}>1$, 
$$\mathbb{P}\Big(\tau\le \dfrac{\epsilon_2^{1+\nu}}{\log n}\Big)\le \dfrac{\mathbb{E}_H\Bigl\{\exp(\epsilon_2/\tau)\Bigr\}}{n^{r}}.
$$ \hfill$\blacksquare$

\subsection{Proof of Lemma \ref{lem:3}}
\label{prf:lem3}
Recall that 
$f(y_i, s_i^2) = \int_{\mathbb{R}^{+}}\int_{\mathbb{R}}f_1(y_i|\mu,\tau)f_2(s_i^2|\tau)g(\mu|\tau)h(\tau)\mathrm{d}\mu\mathrm{d}\tau,
$ 
where $g(\cdot|\tau)$ and $h(\cdot)$ are, respectively, the density functions associated with the distribution functions $G_{\mu}(\cdot|\tau)$ and $H_{\tau}(\cdot)$ in Model \eqref{eq:normmodel}, $f_1$ is the density of a Gaussian random variable with mean $\mu$ and variance $1/(m\tau)$ and $f_2$ is the density of $S^2$ where $(m-1)S^2\tau\sim \mathcal{X}^2_{m-1}$.

We will first analyze the behavior of $f(y_i,s_i^2)$. Gaussian concentration implies that with high probability $\{(Y_i-\mu_i)^2m\tau_i\}\le 2\log n$ and so, conditional on this event,
\begin{equation}
\label{eq:prf_lem2_1}
f_1(y_i|\mu,\tau)\ge c_0\dfrac{\sqrt{\tau}}{n}.
\end{equation}
Moreover, using the Chi-square concentration in Lemma 1 of \cite{laurent2000adaptive}, $(m-1)S_i^2\tau_i\le (m-1)+2\sqrt{(m-1)\log n}+2\log n$ and $S_i^2\tau_i\ge n^{-1}$  with high probability. It follows that
\begin{equation}
\label{eq:prf_lem2_2}
f_2(s_i^2|\tau)\ge c_1\tau\dfrac{a_n}{n^{(m-1)/2}},
\end{equation}
conditional on this event, where $a_n=\exp\{-\sqrt{(m-1)\log n}\}$. Using equations \eqref{eq:prf_lem2_1} and \eqref{eq:prf_lem2_2}, we have
\begin{equation}
\label{eq:prf_lem2_3}
f(y_i,s_i^2)\ge c_2\dfrac{a_n}{n^{(m+1)/2}} \int_{\mathbb{R}^{+}}\tau^{3/2}h(\tau)\mathrm{d}\tau.
\end{equation}
Now, use Assumption \ref{assump:4} and Lemma \ref{lem:1} on the quantity $\int_{\mathbb{R}^{+}}\tau^{3/2}h(\tau)\mathrm{d}\tau$ in equation \eqref{eq:prf_lem2_3} to conclude that 
\begin{equation}
\label{eq:prf_lem2_4}
\int_{\mathbb{R}^{+}}\tau^{3/2}h(\tau)\mathrm{d}\tau\ge \dfrac{c_3}{(\log n)^{3/2}}\mathbb{P}\Big(\tau\ge C_2/\log n\Big). 
\end{equation}
So, with equations \eqref{eq:prf_lem2_4}, \eqref{eq:prf_lem2_3} and Lemma \ref{lem:1}, we have with high probability,
\begin{equation}
\label{eq:prf_lem2_5}
f(y_i,s_i^2)\ge c_4\dfrac{a_n}{n^{(m+1)/2}(\log n)^{3/2}}.
\end{equation}
%The first statement of Lemma \ref{lem:3} thus follows from equations \eqref{eq:prf_lem2_5} and \eqref{eq:prf_lem3_10}.

Now we proceed to prove the statement of Lemma \ref{lem:3}. Fix $\nu>0$ such that $\epsilon^{-\nu}>2m$ in Lemma \ref{lem:1} and let $C_1=\epsilon^{1+\nu}$. First note that from Assumption \ref{assump:4}, $\mathbb{P}(\tau\le C_1/\log n)\le c_0/n^{2m}$. Now, Markov's inequality implies,
$$\mathbb{E}(\tau|y_i,s_i^2) \ge \dfrac{C_1}{\log n}\Bigl\{1-\mathbb{P}\Big[\tau\le C_1(\log n)^{-1}\Big|y_i,s_i^2\Big]\Bigr\}.
$$
Moreover,
\begin{eqnarray}
\label{eq:prf_lem4_1}
\mathbb{P}\Big(\tau\le \dfrac{C_1}{\log n}\Big|y_i,s_i^2\Big) = \{f(y_i,s_i^2)\}^{-1} \int_{0}^{C_2/\log n}h(\tau)\Bigl\{\int_{\mathbb{R}}g(\mu|\tau)f_1(y_i|\mu,\tau)f_2(s_i^2|\tau)\mathrm{d}\mu\Bigr\}\mathrm{d}\tau\nonumber.
\end{eqnarray}
Now $f_1(y_i|\mu,\tau)\le c_0\sqrt{\tau}$ and $f_2(s_i^2|\tau)\le c_1\tau$ where $c_0,~c_1>0$ are constants. So for some positive constant $c_2$, 
\begin{eqnarray}
\label{eq:prf_lem4_2}
\mathbb{P}(\tau\le C_2/\log n|y_i,s_i^2) &\le& \dfrac{c_2}{f(y_i,s_i^2)}\int_{0}^{C_2/\log n}\tau^{3/2}h(\tau)\Bigl\{\int_{\mathbb{R}}g(\mu|\tau)\mathrm{d}\mu\Bigr\}\mathrm{d}\tau\nonumber\\
&\le& \dfrac{c_3}{f(y_i,s_i^2)(\log n)^{3/2}}\int_{0}^{C_2/\log n}h(\tau)\mathrm{d}\tau=\dfrac{c_3}{f(y_i,s_i^2)(\log n)^{3/2}}\mathbb{P}(\tau\le C_2/\log n).\nonumber
\end{eqnarray}
Thus, from the above display, Assumption \ref{assump:4} and Lemma \ref{lem:1},
\begin{eqnarray}
\mathbb{E}(\tau|y_i,s_i^2) \ge \dfrac{C_2}{\log n}\Bigl\{1-c_3\dfrac{n^{-2m}}{f(y_i,s_i^2)(\log n)^{3/2}}\Bigr\}.\nonumber
\end{eqnarray}
Finally, equation \eqref{eq:prf_lem2_5} and the above display prove the statement of Lemma \ref{lem:3}.\hfill$\blacksquare$

\section{Proof of Other Results in the Main Text}\label{proof-other-main}

%%%%%%%%%%%%%%%%%%%%%%%%%%%%%%%%%%%%%%%%%%%%%%%%%%%%%%%%%%%%%%%%%

\subsection{Proof of Proposition \ref{thm:spq}}

Consider the sample criterion $\hat{\mathbb{M}}_{\lambda,n}(\tilde{\mathcal W})$ (Equation \eqref{eq:samplecrit}) and population criterion ${\mathbb{M}}_\lambda(\tilde{\mathcal W})$ (Equation \eqref{eq:sqp}). We have
\begin{equation}
\label{eq:prf_thm1_1}
\begin{split}
\Big|\hat{\mathbb{M}}_{\lambda,n}(\tilde{\mathcal W})-{\mathbb{M}}_\lambda(\tilde{\mathcal W})\Big|&\le\Big|\dfrac{1}{n^2}\sum_{i=1}^{n}\sum_{j=1}^{n}\kappa_\lambda[\tilde{\bm w}(\bm X^i),\tilde{\bm w}(\bm X^j)](\bm X^i,\bm X^j){I}(i\ne j)-\mathbb{M}_\lambda(\tilde{\mathcal W})\Big|+\\
&\Big|\dfrac{1}{n^2}\sum_{i=1}^{n}\kappa_\lambda[\tilde{\bm w}(\bm X^i),\tilde{\bm w}(\bm X^i)](\bm X^i,\bm X^i)\Big|\\%+\dfrac{\lambda(n)}{n^2}\|\tilde{\mathcal W}\|_F^2\\
&\coloneqq I_1+I_2.
\end{split}
\end{equation}
Define $\bar{\mathbb{M}}_{\lambda,n}(\tilde{\mathcal W})=[n(n-1)]^{-1}\sum_{i=1}^{n}\sum_{j=1}^{n}\kappa_\lambda[\tilde{\bm w}(\bm X^i),\tilde{\bm w}(\bm X^j)](\bm X^i,\bm X^j){I}(i\ne j)$. Then 
$$I_1\le |\bar{\mathbb{M}}_{\lambda,n}(\tilde{\mathcal W})-{\mathbb{M}}_\lambda(\tilde{\mathcal W})|+n^{-1}|\bar{\mathbb{M}}_{\lambda,n}(\tilde{\mathcal W})|.
$$By Assumption \ref{assump:1}, $n^{-1}|\bar{\mathbb{M}}_{\lambda,n}(\tilde{\mathcal W})|$ is $O_p(n^{-1})$. {Now, note that $\bar{\mathbb{M}}_{\lambda,n}(\tilde{\mathcal W})$ is an unbiased estimator of $\mathbb{M}_\lambda(\tilde{\mathcal{W}})$ and is a U-statistic with a symmetric kernel function $\kappa_\lambda[\tilde{\bm w}(\bm X^i),\tilde{\bm w}(\bm X^j)](\bm X^i,\bm X^j)$. From Assumption \ref{assump:1}, $\kappa_\lambda[\tilde{\bm w}(\bm X^i),\tilde{\bm w}(\bm X^j)](\bm X^i,\bm X^j)$ has finite second moments. Moreover, from theorem 4.1 of \cite{liu2016kernelized}, $\bar{\mathbb{M}}_{\lambda,n}(\tilde{\mathcal W})$ is a non-degenerate U-statistic whenever $f\ne \tilde{f}$. 
	%To see that, define $\sigma^2_\kappa=\text{Var}_{\bm X\sim f}\{\mathbb{E}_{\bm X'\sim f}\kappa[\tilde{\bm w}(\bm X),\tilde{\bm w}(\bm X')](\bm X,\bm X')\}$ and assume $\sigma^2_\kappa=0$ when $f\ne f'$. Then we must have $\mathbb{E}_{\bm X'\sim f}\kappa[\tilde{\bm w}(\bm X),\tilde{\bm w}(\bm X')](\bm X,\bm X')=c$, where $c$ is a constant. Furthermore,
	%$$ c=\mathbb{E}_{\bm X\sim \tilde{f}}\Bigl\{(\mathbb{E}_{\bm X'\sim f}\kappa[\tilde{\bm w}(\bm X),\tilde{\bm w}(\bm X')](\bm X,\bm X')\Bigr\}=\mathbb{E}_{\bm X'\sim %{f}}\Bigl\{\mathbb{E}_{\bm X\sim \tilde{f}}\kappa[\tilde{\bm w}(\bm X),\tilde{\bm w}(\bm X')](\bm X,\bm X')\Bigr\}
	%.$$ But
}  Thus, from the CLT for U-statistics (\cite{serfling2009approximation} section 5.5), $|\bar{\mathbb{M}}_{\lambda,n}(\tilde{\mathcal W})-{\mathbb{M}}_\lambda(\tilde{\mathcal W})|$ is $O_p(n^{-1/2})$. Moreover, from Assumption \ref{assump:1}, $I_2$ is $O_p(n^{-1})$. Proposition \ref{thm:spq} is thus proved by combining these results.\hfill$\blacksquare$

\subsection{Proof of Equations \eqref{eq:generalized-Tweedie} and \eqref{eq:generalized-Tweedie-var}}
\label{sec:prf_eqs}
The proof follows by first recalling that under the hierarchical model of Equation \eqref{eq:normmodel}, 
$$f_{m_i}(y_i,s_i^2|\mu_i,\tau_i) \propto \exp\Bigl\{-\dfrac{\tau_i}{2}\big[m_iy_i^2+(m_i-1)s_i^2\big]+m_i\tau_i\mu_i y_i-\dfrac{m_i}{2}\tau_i\mu_i^2+\dfrac{m_i-3}{2}\log s_i^2\Bigr\}.
$$
Therefore from Bayes theorem,
$$f_{m_i}(\mu_i,\tau_i|y_i,s_i^2) = \dfrac{f_{m_i}(y_i,s_i^2|\mu_i,\tau_i)}{f_{m_i}(y_i,s_i^2)}g(\mu_i|\tau_i)h(\tau_i)\propto\exp\Bigl\{\bm \eta_i^T\bm T(\mu_i,\tau_i)-A(\bm \eta_i)\Bigr\}g(\mu_i|\tau_i)h(\tau_i).
$$
Here $\bm \eta_i = (m_iy_i,-m_iy_i^2-(m_i-1)s_i^2)\coloneqq (\eta_{1i},\eta_{2i})$, $\bm T(\mu_i,\tau_i) = (\tau_i\mu_i,\tau_i/2)$ and 
\begin{eqnarray}
A(\bm \eta_i) &=& -0.5(m_i-3)\log\gamma(\eta_{1i},\eta_{2i})+\log f_{m_i}\left\{m_i^{-1}\eta_{1i},\gamma(\eta_{1i},\eta_{2i})\right\},\nonumber\\
\gamma(\eta_{1i},\eta_{2i})&=&\dfrac{-\eta_{2i}-m_i^{-1}\eta_{1i}^2}{m_i-1},\nonumber
\end{eqnarray}
with $f_{m}(y,s^2)=\int\int f_m(y,s^2|\mu,\tau)g_\mu(\mu|\tau)h_\tau(\tau)\mathrm{d}\mu\mathrm{d}\tau$ being the marginal density function of $(Y,S^2)$.
So the posterior distribution of $(\mu_i,\tau_i)$ belongs to a 2-parameter exponential family and using the properties of exponential family distributions we have, dropping subscript $i$,
\begin{eqnarray}
\hat{\tau}^{\pi}\coloneqq \hat{\tau}^{\pi}(y,s^2,m)=\mathbb{E}(\tau|y,s^2,m) = 2\dfrac{\partial A(\bm \eta)}{\partial \eta_2}=\dfrac{m-3}{(m-1)s^2}-\dfrac{2}{m-1}w_{2}({y},s^2;m).\nonumber
\end{eqnarray}
Furthermore, with $\zeta=\tau\mu$,
\begin{eqnarray}
\hat{\zeta}^{\pi}\coloneqq\hat{\zeta}^{\pi}(y,s^2,m)&=&\mathbb{E}(\zeta|{y},s^2,m)=\dfrac{\partial A(\bm \eta)}{\partial \eta_1}=\dfrac{(m-3)y}{(m-1)s^2}+m^{-1}w_1(y,s^2;m)-\dfrac{2}{m-1}yw_2(y,s^2;m)\nonumber\\
&=&{y}\mathbb{E}(\tau|{y},s^2,m)+m^{-1} w_{1}({y},s^2;m).\nonumber
\end{eqnarray}\hfill$\blacksquare$

\section{Proof of Propositions in the appendix}\label{proof-pro-ec}

\subsection{Proof of Proposition \ref{prop:twf}}
%\subsubsection{Proof of Lemma \ref{lem:0}}
\label{sec:prf_lem0}
Recall from Definition \ref{def:twfs2} that
$$
\delta_i^{\sf TF}\coloneqq\delta^{\sf TF}(y_i,s_i^2,m_i) =y_i+\dfrac{s_i^2}{m_i} w_1(y_i,s_i^2;m_i),
$$
where $w_1({y},s^2;m)\coloneqq \dfrac{\partial}{\partial{y}}\log f_m({y},s^2)$. Since the $n$ study units are independent, we will focus on unit $i$. 

We note that the inequality in the left hand side of Proposition \ref{prop:twf} follows from the fact that $\delta_{i,(0)}^{\pi}$ is the Bayes estimator of $\mu_i$ that uniquely minimizes the expected squared error loss under Model \eqref{eq:normmodel}. To prove the inequality in the right hand side of Proposition \ref{prop:twf}, we proceed as follows.

Denote $w_{1,i}\coloneqq w_1(y_i,s_i^2;m_i)$. We have,
\begin{eqnarray}
\label{eq:prf_lem0_1}
\mathbb{E}(\mu_i-\delta_i^{\sf TF})^2&=&\dfrac{1}{m_i}\mathbb{E}(1/\tau_i)+\dfrac{1}{m_i^2}\mathbb{E}(S_i^2w_{1,i})^2+\dfrac{2}{m_i}\mathbb{E}(Y_i-\mu_i)S_i^2w_{1,i}\nonumber\\
&=&\mathbb{E}(\mu_i-Y_i)^2+\dfrac{1}{m_i^2}\mathbb{E}(S_i^2w_{1,i})^2+\dfrac{2}{m_i^2}\mathbb{E}\dfrac{S_i^2}{\tau_i}w_{1,i}^{'},
\end{eqnarray}
where $w_{1,i}^{'}\coloneqq w_1^{'}(y_i,s_i^2;m_i)$ and $w_1^{'}(y,s^2;m)=\dfrac{\partial}{\partial y}w_1(y,s^2;m)$. The equality in Equation \eqref{eq:prf_lem0_1} follows from integration by parts and the fact that $Y_i|\mu_i,\tau_i\sim N(\mu_i,1/(m_i\tau_i))$. Consider the term $T_1\coloneqq \dfrac{1}{m_i^2}\mathbb{E}(S_i^2w_{1,i})^2+\dfrac{2}{m_i^2}\mathbb{E}\Big(\dfrac{S_i^2}{\tau_i}w_{1,i}^{'}\Big)$ and note that
\begin{eqnarray}
\label{eq:prf_lem0_2}
T_1&=&\dfrac{1}{m_i^2}\mathbb{E}\Big(S_i^2w_{1,i}\Big)^2+\dfrac{1}{m_i^2}\mathbb{E}\Big[(S_i^2)^2w_{1,i}^{'}\Big]+\dfrac{2}{m_i^2}\mathbb{E}\Big(\dfrac{S_i^2}{\tau_i}w_{1,i}^{'}\Big)-\dfrac{1}{m_i^2}\mathbb{E}\Big[(S_i^2)^2w_{1,i}^{'}\Big]\nonumber\\
&=&\dfrac{1}{m_i^2}\mathbb{E}\Big[\dfrac{2S_i^2}{\tau_i}-(S_i^2)^2\Big]w_{1,i}^{'}.
\end{eqnarray}
The equality in Equation \eqref{eq:prf_lem0_2} follows because, dropping subscript $i$,
$$\mathbb{E}\Big[(S^2)^2(w_{1}^2+w_{1}^{'})\Big]=\mathbb{E}\Big[(S^2)^2\dfrac{f_{m,(1)}^{''}(Y,S^2)}{f_m(Y,S^2)}\Big]=0,
$$ where $f_{m,(1)}^{''}(y,s^2)$ is the second order partial derivative of $f_m(y,s^2)$ with respect to $y$.
Now, we can re-write Equation \eqref{eq:prf_lem0_2} as
\begin{eqnarray}
\label{eq:prf_lem0_3}
T_1 =\dfrac{1}{m_i^2}\mathbb{E}_{\mu_i,\tau_i}\mathbb{E}_{Y_i,S_i^2|\mu_i,\tau_i}\Bigl\{\Big[\dfrac{2S_i^2}{\tau_i}-(S_i^2)^2\Big]w_{1,i}^{'}\Bigr\}
=\dfrac{1}{m_i^2}\mathbb{E}_{\mu_i,\tau_i}\Big(T_2\Big)+\dfrac{1}{m_i^2}\mathbb{E}_{\mu_i,\tau_i}\Big(T_3\Big),
\end{eqnarray}
where $\mathbb{E}_{\mu,\tau}$ is the expectation with respect to the joint distribution of $(\mu,\tau)$, $\mathbb{E}_{Y,S^2|\mu,\tau}$ is the expectation with respect to the joint distribution of $(Y,S^2)$ conditional on $(\mu,\tau)$ and
\begin{eqnarray}
T_2&=&\mathbb{E}_{Y_i,S_i^2|\mu_i,\tau_i}\Bigl\{\Big[\dfrac{2S_i^2}{\tau_i}-(S_i^2)^2\Big]w_{1,i}^{'}\Big|S_i^2<\dfrac{2}{\tau_i}\Bigr\}\mathbb{P}\Big(S_i^2<\dfrac{2}{\tau_i}\Big|\tau_i\Big),\nonumber\\
T_3&=&\mathbb{E}_{Y_i,S_i^2|\mu_i,\tau_i}\Bigl\{\Big[\dfrac{2S_i^2}{\tau_i}-(S_i^2)^2\Big]w_{1,i}^{'}\Big|S_i^2>\dfrac{2}{\tau_i}\Bigr\}\mathbb{P}\Big(S_i^2>\dfrac{2}{\tau_i}\Big|\tau_i\Big).\nonumber
\end{eqnarray}
Denote $p\coloneqq \mathbb{P}(S_i^2>{2}/{\tau_i}~|~\tau_i)$ and let $c_i$ be the partial derivative of $w_1(y,s^2;m)$ with respect to $y$ and evaluated at $(y_i,2/\tau_i,m_i)$. Now, using equations \eqref{eq:prf_lem0_2} and \eqref{eq:prf_lem0_3} in Equation \eqref{eq:prf_lem0_1}, we get
\begin{eqnarray}
\label{eq:prf_lem0_4}
\mathbb{E}(\mu_i-\delta_i^{\sf TF})^2=\mathbb{E}(\mu_i-Y_i)^2+\dfrac{1}{m_i^2}\mathbb{E}_{\mu_i,\tau_i}(T_2+T_3).
\end{eqnarray}
We will show that $T_2+T_3< 0$ which will be enough to prove the statement of Proposition \ref{prop:twf} using Equation \eqref{eq:prf_lem0_4}.

We first state a few results that are straightforward consequences of Model \eqref{eq:normmodel} and the assumptions of Proposition \ref{prop:twf}. We have,
\begin{enumerate}
	\item Under Model \ref{eq:normmodel}, $(m_i-1)S_i^2\tau_i\sim \chi^2_{m_i-1}$.
	\item Additionally, $$\mathbb{E}_{Y_i,S_i^2|\mu_i,\tau_i}\Big[\dfrac{2S_i^2}{\tau_i}-(S_i^2)^2\Big]=\dfrac{m_i-3}{(m_i-1)\tau_i^2}>0,$$ since $m_i>3$ in the statement of Proposition \ref{prop:twf}.
	\item Since $f_m(y,s^2)$ is a log-concave density, $w_{1,i}^{'}\le 0$ and so $T_2\le 0$ while $T_3\ge 0$.
\end{enumerate}
Assume, without loss of generality, $w_{1,i}^{'}<0$. Since $\mathbb{E}_{Y_i,S_i^2|\mu_i,\tau_i}[{2S_i^2}/{\tau_i}-(S_i^2)^2]>0$, we have
\begin{eqnarray}
\label{eq:prf_lem0_5}
(1-p)\mathbb{E}_{Y_i,S_i^2|\mu_i,\tau_i}\Big[\dfrac{2S_i^2}{\tau_i}-(S_i^2)^2\Big|S_i^2<\dfrac{2}{\tau_i}\Big]>-p\mathbb{E}_{Y_i,S_i^2|\mu_i,\tau_i}\Big[\dfrac{2S_i^2}{\tau_i}-(S_i^2)^2\Big|S_i^2>\dfrac{2}{\tau_i}\Big]>0.
\end{eqnarray}
Furthermore, as $w_{1,i}^{'}< 0$ and $w_{1,i}^{'}$ is a non-decreasing function of $s_i^2$,
\begin{eqnarray}
\label{eq:prf_lem0_6}
T_2&\le& (1-p)\mathbb{E}_{Y_i,S_i^2|\mu_i,\tau_i}\Bigl\{\Big[\dfrac{2S_i^2}{\tau_i}-(S_i^2)^2\Big]c_i\Big|S_i^2<\dfrac{2}{\tau_i}\Bigr\},
\end{eqnarray}
Note that $c_i< 0$, and as defined earlier, it is the partial derivative of $w_1(y,s^2;m)$ with respect to $y$ and evaluated at $(y_i,2/\tau_i,m_i)$.
%$c\coloneqq \dfrac{\partial}{\partial y}w_{1}(y,2/\tau;m)< 0$. 
Therefore, using equations \eqref{eq:prf_lem0_5} and \eqref{eq:prf_lem0_6},
\begin{eqnarray}
\label{eq:prf_lem0_7}
T_2\le(1-p)\mathbb{E}_{Y_i,S_i^2|\mu_i,\tau_i}\Bigl\{\Big[\dfrac{2S_i^2}{\tau_i}-(S_i^2)^2\Big]c_i\Big|S_i^2<\dfrac{2}{\tau_i}\Bigr\}< -p\mathbb{E}_{Y_i,S_i^2|\mu_i,\tau_i}\Bigl\{\Big[\dfrac{2S_i^2}{\tau_i}-(S_i^2)^2\Big]c_i\Big|S_i^2>\dfrac{2}{\tau_i}\Bigr\}< 0.
\end{eqnarray}
Now, we consider the term $T_3$. Recall that 
$$T_3=p\mathbb{E}_{Y_i,S_i^2|\mu_i,\tau_i}\Bigl\{\Big[\dfrac{2S_i^2}{\tau_i}-(S_i^2)^2\Big]w_{1,i}^{'}\Big|S_i^2>\dfrac{2}{\tau_i}\Bigr\}> 0.$$ Since $w_{1,i}^{'}< 0$ and $w_{1,i}{'}$ is a non-decreasing function of $s_i^2$,
\begin{eqnarray}
\label{eq:prf_lem0_8}
0<	T_3\le -p\mathbb{E}_{Y_i,S_i^2|\mu_i,\tau_i}\Bigl\{\Big[\dfrac{2S_i^2}{\tau_i}-(S_i^2)^2\Big]|c_i|\Big|S_i^2>\dfrac{2}{\tau_i}\Bigr\}.
\end{eqnarray}
So, using equations \eqref{eq:prf_lem0_7} and \eqref{eq:prf_lem0_8}
\begin{eqnarray}
T_2+T_3< -p\mathbb{E}_{Y_i,S_i^2|\mu_i,\tau_i}\Bigl\{\Big[\dfrac{2S_i^2}{\tau_i}-(S_i^2)^2\Big]\Big(c_i+|c_i|\Big)\Big|S_i^2>\dfrac{2}{\tau_i}\Bigr\}= 0,\nonumber
\end{eqnarray}
Hence the desired result follows from the display above and Equation \eqref{eq:prf_lem0_4}.\hfill$\blacksquare$

%%%%%%%%%%%%%%%%%%%%%%%%%%%%%%%%%%%%%%%%%%%%%%%%%%%%%%%%%%%%%%%%%%
\subsection{Proof of Proposition \ref{prop:oraclerisks}}
We will first collect a few notations that will be used throughout the proof. Denote $w_{1,i}\coloneqq w_1(y_i,s_i^2;m_i),~w_{2,i}\coloneqq w_2(y_i,s_i^2;m_i)$ and $\gamma_i\coloneqq \gamma(y_i,s_i^2,m_i)$. Let $w_1^{'}(y,s^2;m)=\dfrac{\partial}{\partial y}w_1(y,s^2;m)$ and denote $w_{1,i}^{'}\coloneqq w_1^{'}(y_i,s_i^2;m_i)$. Similarly, $\nu(y,s^2;m)=\dfrac{1}{f_m(y,s^2)}\dfrac{\partial^2}{\partial y^2}f_m(y,s^2)$ and denote $\nu_i\coloneqq \nu(y_i,s_i^2;m_i)$. Finally, let $\gamma^{'}(y,s^2,m)=\dfrac{\partial}{\partial y}\gamma(y,s^2,m)$ and denote $\gamma_i^{'}\coloneqq\gamma^{'}(y_i,s_i^2,m_i)$.

The proof of Proposition \ref{prop:oraclerisks} will use the following two lemmata.
\begin{lemma}
	\label{lem:22}
	Suppose $f_m(y,s^2)$ is a log concave density and Assumption \ref{assump_5} holds.
	Then under Model \eqref{eq:normmodel} and $m_i> 5$, we have, 
	$$\mathbb{E}\Bigl\{\Big[\dfrac{2S_i^2}{\tau_i}\Big(1-\gamma_i\Big)-\Big(S_i^2\Big)^2\Big(1-\gamma_i^2\Big)\Big]w_{1,i}^{'}\Bigr\}\ge 0.$$
\end{lemma}
\begin{lemma}
	\label{lem:33}
	Under Assumption \ref{assump_6} and Model \eqref{eq:normmodel}, we have, 
	$$ \dfrac{1}{m_i^2}\mathbb{E}\Big[S_i^2\Big(S_i^2\gamma_i^2 \nu_{i}+2\dfrac{w_{1,i}\gamma_i^{'}}{\tau_i}\Big)\Big]\le 0.$$
\end{lemma}
Lemmata \ref{lem:22} and \ref{lem:33} are proved in Sections \ref{sec:prf_lem22} and \ref{sec:prf_lem33} respectively. We now prove Proposition \ref{prop:oraclerisks}. Recall from Definition \ref{def:oraclenest} that the oracle NEST estimator for $\mu_i$ is 
$$\delta_{i,(1)}^{\pi}= y_i+\dfrac{s_i^2}{m_i}\gamma(y_i,s_i^2,m_i)w_1(y_i,s_i^2,m_i),$$ where
$$\gamma(y_i,s_i^2,m_i) = \dfrac{m_i-1}{m_i-3-2s_i^2w_2(y_i,s_i^2;m_i)}.$$
Since the $n$ study units are independent, we will focus on unit $i$. 

Under the squared error loss, $\delta_{i,(0)}^{\pi}$ is the Bayes estimator of $\mu_i$ in Model \eqref{eq:normmodel}. This establishes the first inequality on the left hand side of Proposition \ref{prop:oraclerisks}. Together with Proposition \ref{prop:twf}, we only need to show $r_0(\bm \delta^{\pi}_{(1)},\mathcal{G})\le r_0(\bm \delta^{\sf TF},\mathcal{G})$. First note that
\begin{eqnarray}
\label{eq:prf_prop11_0}
r_0(\bm \delta^{\sf TF},\mathcal{G})-r_0(\bm \delta^{\pi}_{(1)},\mathcal{G})=\dfrac{1}{m_i^2}\mathbb{E}\Big[(S_i^2)^2\Big(1-\gamma_i^2\Big)w_{1,i}^2\Big]+\dfrac{2}{m_i}\mathbb{E}\Big[(Y_i-\mu_i)S_i^2\Big(1-\gamma_i\Big)w_{1,i}\Big]\nonumber\\
=\dfrac{1}{m_i^2}\mathbb{E}\Big[(S_i^2)^2\Big(1-\gamma_i^2\Big)w_{1,i}^2\Big]+\dfrac{2}{m_i^2}\mathbb{E}\Bigl\{\dfrac{S_i^2}{\tau_i}\Big[(1-\gamma_i)w_{1,i}^{'}-w_{1,i}\gamma_i^{'}\Big]\Bigr\}.
\end{eqnarray}
%Here $w_{1,i}^{'}$ is the partial derivative of $w_1(y,s^2;m)$ with respect to $y$ and evaluated at $(y_i,s_i^2,m_i)$. Similarly, $\gamma_i^{'}$ is the partial derivative of $\gamma(y,s^2,m)$ with respect to $y$ and evaluated at $(y_i,s_i^2,m_i)$.  %$w_{1}^{'}=\dfrac{\partial}{\partial{y}}w_1({y},s^2;m),~\gamma^{'}=\dfrac{\partial}{\partial{y}}\gamma({y},s^2;m)$ 
The equality in equation \eqref{eq:prf_prop11_0} follows from integration by parts and the fact that $Y_i\sim N\left(\mu_i,\frac 1{m_i\tau_i}\right)$. We can re-write Equation \eqref{eq:prf_prop11_0} as,
\begin{eqnarray}
\label{eq:prf_prop11_1}
\begin{split}
\dfrac{1}{m_i^2}\mathbb{E}\Big[(S_i^2)^2\Big(1-\gamma_i^2\Big)w_{1,i}^2\Big]+\dfrac{1}{m_i^2}\mathbb{E}\Big[(S_i^2)^2\Big(1-\gamma_i^2\Big)w_{1,i}^{'}\Big]-\dfrac{2}{m_i^2}\mathbb{E}\Big(\dfrac{S_i^2}{\tau_i}w_{1,i}\gamma_i^{'}\Big)\\+\dfrac{1}{m_i^2}\mathbb{E}\Bigl\{\Big[\dfrac{2S_i^2}{\tau_i}\Big(1-\gamma_i\Big)-\Big(S_i^2\Big)^2\Big(1-\gamma_i^2\Big)\Big]w_{1,i}^{'}\Bigr\}.
\end{split}
\end{eqnarray}
From Lemma \ref{lem:22}, the last term in Equation \eqref{eq:prf_prop11_1} is non-negative. Let us consider the first three terms in Equation \eqref{eq:prf_prop11_1} and denote them by,
\begin{equation*}
T\coloneqq \dfrac{1}{m_i^2}\mathbb{E}\Big[(S_i^2)^2\Big(1-\gamma_i^2\Big)w_{1,i}^2\Big]+\dfrac{1}{m_i^2}\mathbb{E}\Big[(S_i^2)^2\Big(1-\gamma_i^2\Big)w_{1,i}^{'}\Big]-\dfrac{2}{m_i^2}\mathbb{E}\Big(\dfrac{S_i^2}{\tau_i}w_{1,i}\gamma_i^{'}\Big).
\end{equation*}
As shown in the proof of Proposition \ref{prop:twf}, $\mathbb{E}[(S^2)^2(w_{1}^2+w_{1}^{'})]=0.$
The above display involving the term $T$ can be written as
\begin{equation}
\label{eq:prf_prop11_2}
T = -\dfrac{1}{m_i^2}\mathbb{E}\Big[(S_i^2)^2\gamma_i^2 \nu_{i}\Big]-\dfrac{2}{m_i^2}\mathbb{E}\Big(\dfrac{S_i^2}{\tau_i}w_{1,i}\gamma_i^{'}\Big).
\end{equation}
From Lemma \ref{lem:33}, the term $T$ in Equation \eqref{eq:prf_prop11_2} is non-negative. This establishes the inequality on the right hand side of $r_0(\bm\delta^{\pi}_{(1)},\mathcal{G})$ in Proposition \ref{prop:oraclerisks} and completes the proof.\hfill$\blacksquare$

%%%%%%%%%%%%%%%%%%%%%%%%%%%%%%%%%%%%%%%%%%%%%%%%%%%%%%%%%%%%%

\subsection{Proof of Lemma \ref{lem:22}}
\label{sec:prf_lem22}

Denote
$$ Z_i\coloneqq \dfrac{2S_i^2}{\tau_i}(1-\gamma_i)-(S_i^2)^2(1-\gamma_i^2)=\dfrac{Q_i}{\tau_i},
$$ where $Q_i =2S_i^2(1-\gamma_i)-(S_i^2)^2(1-\gamma_i^2)\tau_i$.  Since $f_m(y,s^2)$ is a log-concave density, $w_{1,i}^{'}\le 0$. %Suppose, without loss of generality, $w_{1,i}^{'}<0$. 
We will show that $Q_i\le 0$ which will be sufficient to prove $Z_iw_{1,i}^{'}\ge 0$ and hence the statement of Lemma \ref{lem:22}.

Dropping subscript $i$,
\begin{equation*}
\label{eq:prf_lem22_0}
\mathbb{E}(Q) = \mathbb{E}_{Y,S^2}\Big[2S^2(1-\gamma)-(S^2)^2(1-\gamma^2)\hat{\tau}^{\pi}\Big]=\mathbb{E}_{Y,S^2}[R(Y,S^2)],
\end{equation*}
where $\hat{\tau}^{\pi}=\mathbb{E}(\tau|y,s^2,m)=(s^2\gamma)^{-1}$ from Equation \eqref{eq:generalized-Tweedie-var} and $\mathbb{E}_{Y,S^2}$ is expectation with respect to the joint marginal distribution of $(Y,S^2)$. Suppose, if possible, $Q_i> 0$ for all $(\tau_i,y_i,s^2_i)\in \mathbb{R}^{+}\times\mathbb{R}\times\mathbb{R}^{+}$. We will show that $\mathbb{E}(Q)\le 0$ which will present a contradiction to $Q> 0$ for all $(\tau,y,s^2)\in \mathbb{R}^{+}\times\mathbb{R}\times\mathbb{R}^{+}$.
Fix a $y\in\mathbb{R}$ and consider the following cases.
\\[0.5ex]
\noindent\textbf{Case 1} -- Suppose $0<\gamma(y,s^2,m)\le 1/(2c)$ where $c\ge 1$ is a constant. Then, we have $s^2\hat{\tau}^{\pi}\ge 2c$ and consequently $s^2(1+\gamma)\hat{\tau}^{\pi}> 2$. So $R(y,s^2)<0$. Now, from Assumption \ref{assump_5}, $\gamma$ is a continuous and non-increasing function of $s^2$. Therefore, there exists $c_{1}(y)\in\mathbb{R}^{+}$, depending on $y$, such that $s^2>c_{1}(y)$ whenever $0<\gamma(y,s^2,m)\le 1/(2c)$. Thus, $R(y,s^2)<0$ for $s^2>c_{1}(y)$.
\\[0.5ex]
\textbf{Case 2} -- Next, suppose $1/2<\gamma(y,s^2,m)\le (m-1)/(2m-6)$. Then $(2m-6)/(m-1)\le s^2\hat{\tau}^{\pi}<2$ and $s^2(1+\gamma)\hat{\tau}^{\pi}\ge (3m-7)/(m-1)$. Since $m> 5$, $(3m-7)/(m-1)> 2$ and so $s^2(1+\gamma)\hat{\tau}^{\pi}>2$. Thus $R(y,s^2)< 0$ and using Assumption \ref{assump_5}, $R(y,s^2)<0$ for $c_2(y)< s^2\le c_1(y)$, where $c_2(y)$ is such that $s^2> c_2(y)$ whenever $\gamma(y,s^2,m)\le (m-1)/(2m-6)$. 

The remaining four cases proceed in a similar manner as follows:
\\[0.5ex]
\textbf{Case 3} -- Suppose $(m-1)/(2m-6)<\gamma(y,s^2,m)\le (2m-6)/(3m-11) $. Then $(3m-11)/(2m-6)\le s^2\hat{\tau}^{\pi}$. So $s^2(1+\gamma)\hat{\tau}^{\pi}\ge (5m-17)/(2m-6)$. Since $(5m-17)/(2m-6)>2$ if $m>5$, we have $s^2(1+\gamma)\hat{\tau}^{\pi}> 2$ and so $R(y,s^2)< 0$ for $c_3(y)< s^2\le c_2(y)$. Similarly, we can show that $R(y,s^2)\le 0$ for $c_4(y)< s^2\le c_3(y)$ where $s^2\in(c_4(y),c_3(y)]$ whenever $(2m-6)/(3m-11)<\gamma(y,s^2,m)\le 1$.
\\[0.5ex]
\textbf{Case 4} -- Now suppose, $1< \gamma(y,s^2,m)\le (2m-5)/(2m-6)$. Then $(2m-6)/(2m-5)\le s^2\hat{\tau}^{\pi}< 1$ and $s^2(1+\gamma)\hat{\tau}^{\pi}<2$. Note that here $\gamma> 1$ as opposed to $\gamma\le 1$ in the earlier cases. Therefore, $s^2(1+\gamma)\hat{\tau}^{\pi}< 2$ implies $R(y,s^2)< 0$ for $c_5(y)< s^2\le c_4(y)$. %The upper limit of the interval for $s^2(1+\gamma)\hat{\tau}^{\pi}$ is $(4m-11)/(2m-6)$ which, for $m> 7$, is approximately 2 and converges to $2$ for a moderately large $m$.  
\\[0.5ex]
\textbf{Case 5} -- Similarly, if $\gamma$ is in the intervals $((2m-5)/(2m-6),~(m-2)/(m-3)],~((m-2)/(m-3),~(m-1)/(m-3)]$ and $((m-1)/(m-3),~(m+1)/(m-3)]$ then we have $\gamma\ge 1$ and $s^2(1+\gamma)\hat{\tau}^{\pi}< 2$. Thus, on each of the corresponding intervals for $s^2$, $R(y,s^2)< 0$. 
\\[0.5ex]
\textbf{Case 6} -- Denote $r_1=1$ and $r_t=2r_{t-1}+3$ for $t=2,3,\ldots$. Suppose $(m+r_{t-1})/(m-3)<\gamma(y,s^2,m)\le (m+r_t)/(m-3)$. Then for each of these intervals indexed by $t\ge 2$, we have $\gamma\ge 1$, $s^2(1+\gamma)\hat{\tau}^{\pi}< 2$ and so $R(y,s^2)< 0$ on the corresponding intervals for $s^2$.

So from these six cases, $R(y,s^2)\le 0$ for all $s^2>0$ and consequently $\mathbb{E}_{Y,S^2}[R(Y,S^2)]\le 0$. Therefore, $\mathbb{E}(Q)\le 0$ which contradicts that $Q> 0$ for all $(\tau,y,s^2)\in \mathbb{R}^{+}\times\mathbb{R}\times\mathbb{R}^{+}$. Now suppose that for some $\Omega\subset \mathbb{R}^{+}\times\mathbb{R}\times\mathbb{R}^{+}$,  $Q>0$ whenever $(\tau,y,s^2)\in\Omega$.  However,  Assumption \ref{assump_5} and the aforementioned six cases imply that $\mathbb{E}(Q|\Omega)\le 0$. Thus, $Q\le 0$ for all $(\tau,y,s^2)\in \mathbb{R}^{+}\times\mathbb{R}\times\mathbb{R}^{+}$.
So, we have $Zw_1^{'}\ge 0$ and this completes the proof of Lemma \ref{lem:22}.\hfill$\blacksquare$

\subsection{Proof of Lemma \ref{lem:33} }
\label{sec:prf_lem33}
Dropping subscript $i$, denote,
$$T\coloneqq -\dfrac{1}{m^2}\mathbb{E}\Big[(S^2\gamma(Y,S^2,m))^2 \nu(Y,S^2,m)\Big]-\dfrac{2}{m^2}\mathbb{E}\Big[\dfrac{S^2}{\tau}w_{1}(Y,S^2,m)\gamma^{'}(Y,S^2,m)\Big].$$
We will show that $T\ge 0$.

Let $\gamma\coloneqq\gamma(y,s^2,m),\nu\coloneqq\nu(y,s^2,m),w_1\coloneqq w_{1}(y,s^2,m)$ and $\gamma^{'}\coloneqq \gamma^{'}(y,s^2,m)$. First note that using standard integration by parts, we have
\begin{equation*}
-\dfrac{1}{m^2}\mathbb{E}\Big[(S^2\gamma)^2 \nu\Big] = \dfrac{2}{m^2}\mathbb{E}\Big[(S^2)^2\gamma w_1\gamma^{'}\Big].
\end{equation*}
So, we can write
\begin{equation*}
T=	\dfrac{2}{m^2}\mathbb{E}\Bigl\{S^2w_1\gamma^{'}\Big[S^2\gamma-\dfrac{1}{\tau}\Big]\Bigr\}.
\end{equation*}
Furthermore, from Definition \ref{def:oraclenest},
$$\gamma^{'} = \dfrac{2}{m-1}\gamma^2s^2w_2^{'},
$$ where $w_2^{'}\coloneqq \dfrac{\partial}{\partial y}w_2(y,s^2,m)$.
So, we have
\begin{eqnarray}
T=	\dfrac{4}{m^2(m-1)}\mathbb{E}\Bigl\{(S^2\gamma)^2w_1w_2^{'}\Big[S^2\gamma-\dfrac{1}{\tau}\Big]\Bigr\}.\nonumber
\end{eqnarray}
Now conditional on $(Y,S^2)$, note that from Jensen's inequality,
$$ s^2\gamma = \dfrac{1}{\mathbb E(\tau\mid y,s^2)}\le \mathbb E\Big(\dfrac{1}{\tau}\Big| y,s^2\Big).
$$
Furthermore, since $w_1w_2^{'}\le 0$ from Assumption \ref{assump_6}, we have
$$ T = \dfrac{4}{m^2(m-1)}\mathbb{E}_{Y,S^2}\Bigl\{(S^2\gamma)^2w_1w_2^{'}\Big[S^2\gamma-\mathbb E\Big(\dfrac{1}{\tau}\Big| Y,S^2\Big)\Big]\Bigr\}\ge 0.
$$
This completes the proof of Lemma \ref{lem:33}.\hfill$\blacksquare$

\section{Additional Numerical Experiments}
\label{sec:add_numexp}
\subsection{Compound estimation of Normal means under squared error loss}
\label{sec:nomrsimuMeans_se}
%%%%%%%%%%%%%%%%%%%%%%%%%%%%%%%%
We focus on the hierarchical Model of Equation \eqref{eq:normmodel} and compare six approaches for estimating $\bm \mu$ under the squared error loss when the variances $\sigma_i=1/\tau_i$ are assumed to be unknown. These approaches can be categorized into three types: the first consists of the NEST method (NEST Orc. $\lambda$), which estimates $\lambda$ by minimizing the true loss, the proposed data-driven NEST method and Tweedie's formula (TF) that uses sample variances. For both NEST and TF, $\lambda$ is chosen using the modified cross-validation approach described in Section \ref{Subsec:bandwidth} with ${\vartheta}_n(\lambda;\mathcal U,\mathcal V)$ defined as $$
{\vartheta}_n(\lambda;\mathcal U,\mathcal V)=\dfrac{1}{n}\sum_{i=1}^{n}\left\{\bar{V}_i-\delta_i^{\sf ds}(\bar{U}_i;\mathcal{U},\lambda)\right\}^2,
$$ to reflect the squared error loss. The second are linear shrinkage methods: the group linear estimator (Grp Linear) of \citet{Weinsteinetal18} %; the semi-parametric monotonically constrained SURE estimator that shrinks towards the grand mean (XKB.SG) from \citet{Xieetal12}; and 
and the semi-parametric linear shrinkage rule (Jing.SM) from \citet{jing2016sure}. Finally, the third type is the g-modelling approach of \citet{gu2017empirical, gu2017unobserved}. For Grp. Linear we use code provided by \citet{Weinsteinetal18} while for Jing.SM we write our own routine in R. We continue to rely on the function WGLVmix in the R package REBayes \citep{koenker2017rebayes} for NPMLE. We note that amongst the six methods considered here, NEST and NEST Orc. $\lambda$ are designed to estimate the means under the weighted squared error loss while the remaining four approaches target the squared error loss. Furthermore, with the exception of Grp. Linear, all other methods considered here estimate $\bm \mu$ when the variances are unknown. Grp. Linear, on the other hand, assumes full knowledge of the unknown variances for shrinkage estimation of the means and here we use sample variances for its implementation. 

\begin{figure}[!t]
	\centering
	\includegraphics[width=1\linewidth]{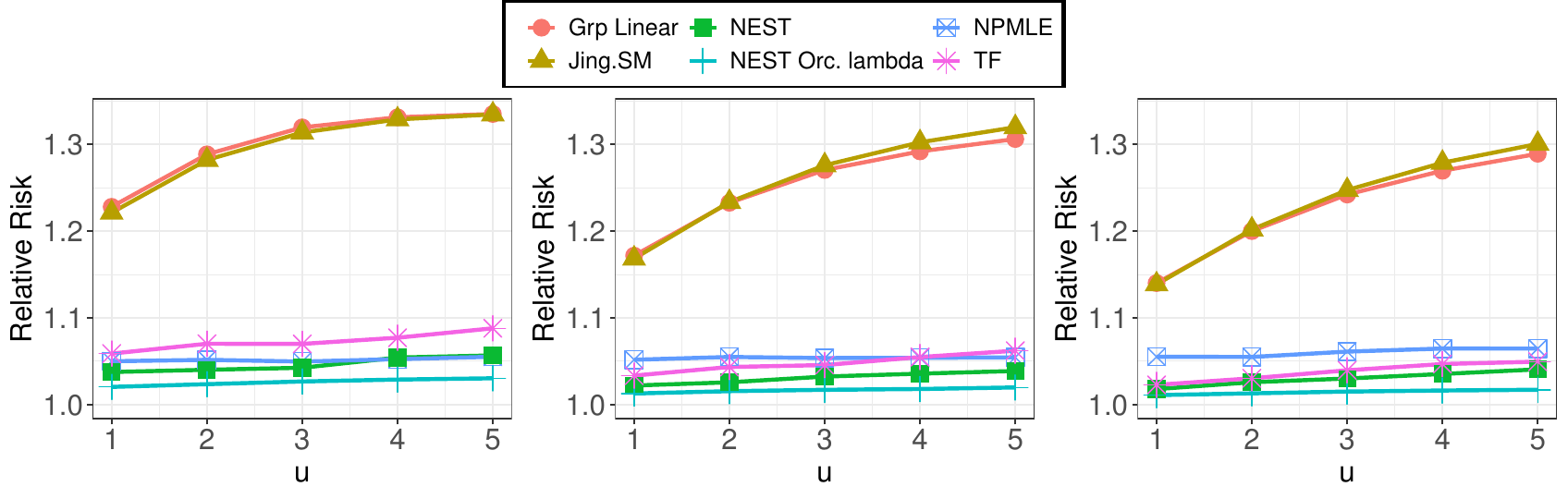}
	\caption{Comparison of relative risks when $(\mu_i, \sigma_i^2)$ are independent. Here $\mu_i\stackrel{i.i.d}{\sim} 0.7~N(0,.1)+0.15~N(1,3)+0.15~N(-1,3)$ and $\sigma_i^2\stackrel{i.i.d}{\sim}U(0.5,u)$. Plots show $m=10,15,20$ left to right.}
	\label{fig:simInd}
\end{figure}

\begin{figure}[!t]
	\centering
	\includegraphics[width=1\linewidth]{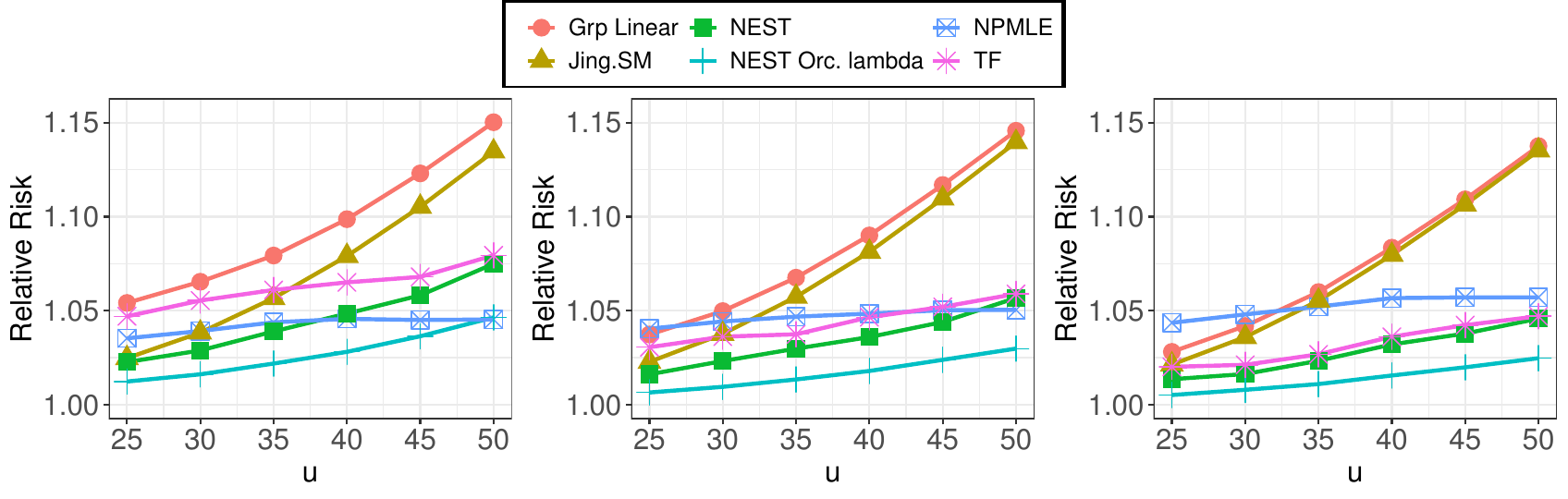}
	\caption{Comparison of relative risks for correlated $(\mu_i, \tau_i)$. Here $\tau_i\stackrel{i.i.d}{\sim}0.5\Gamma(20,\texttt{rate}=20)+0.5\Gamma(20,\texttt{rate}=u)$ and $\mu_i|\tau_i\stackrel{ind.}{\sim} 0.5~N(0.5/\tau_i,0.5^2)+0.5~N(-0.5/\tau_i,0.5^2)$. Plots show $m=10,15,20$ left to right.}
	\label{fig:simCorr}
\end{figure}

\begin{figure}[!h]
	\centering
	\includegraphics[width=1\linewidth]{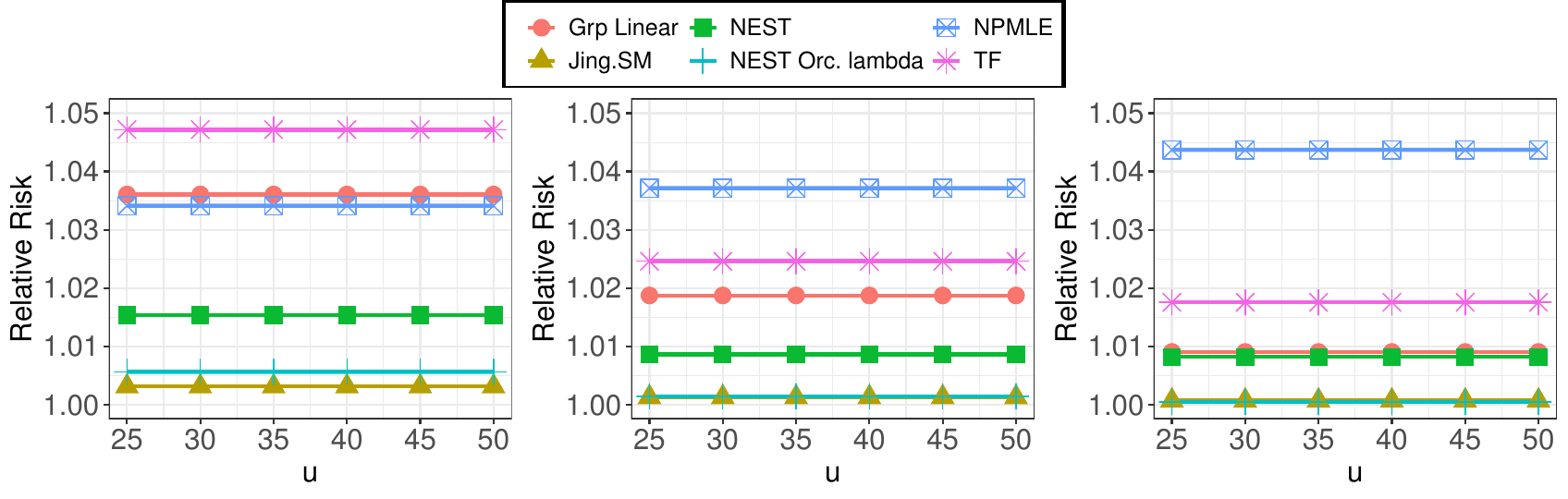}
	\caption{Comparison of relative risks when $(\mu_i, \tau_i)$ have conjugate priors. Here $\mu_i|\tau_i\stackrel{ind.}{\sim} N(0,0.5/\tau_i)$ and $\tau_i\stackrel{i.i.d}{\sim}\Gamma(20,\texttt{rate}=u)$. Plots show $m=10,15,20$ left to right.}
	\label{fig:conjugate}
\end{figure}

\begin{figure}[!t]
	\centering
	\includegraphics[width=1\linewidth]{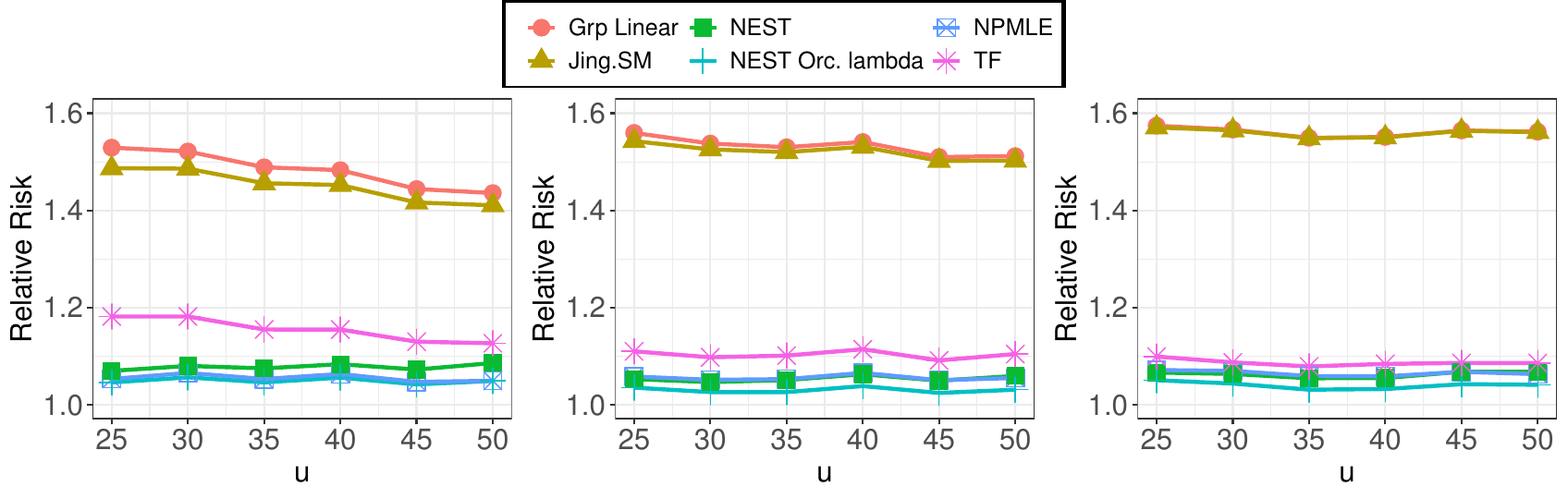}
	\caption{Comparison of relative risks when $\bm \mu$ is sparse. Here $\tau_i\stackrel{i.i.d}{\sim} 0.5~\Gamma(20,\texttt{rate}=20)+0.5~\Gamma(20,\texttt{rate}=u)$ and $\mu_i|\tau_i\stackrel{ind.}{\sim} 0.7~N(0,0.01)+0.15~N(0.5/\tau_i,1)+0.15~N(-0.5/\tau_i,1)$. Plots show $m=10,15,20$ left to right.}
	\label{fig:simSparse}
\end{figure}

\begin{figure}[!t]
	\centering
	\includegraphics[width=1\linewidth]{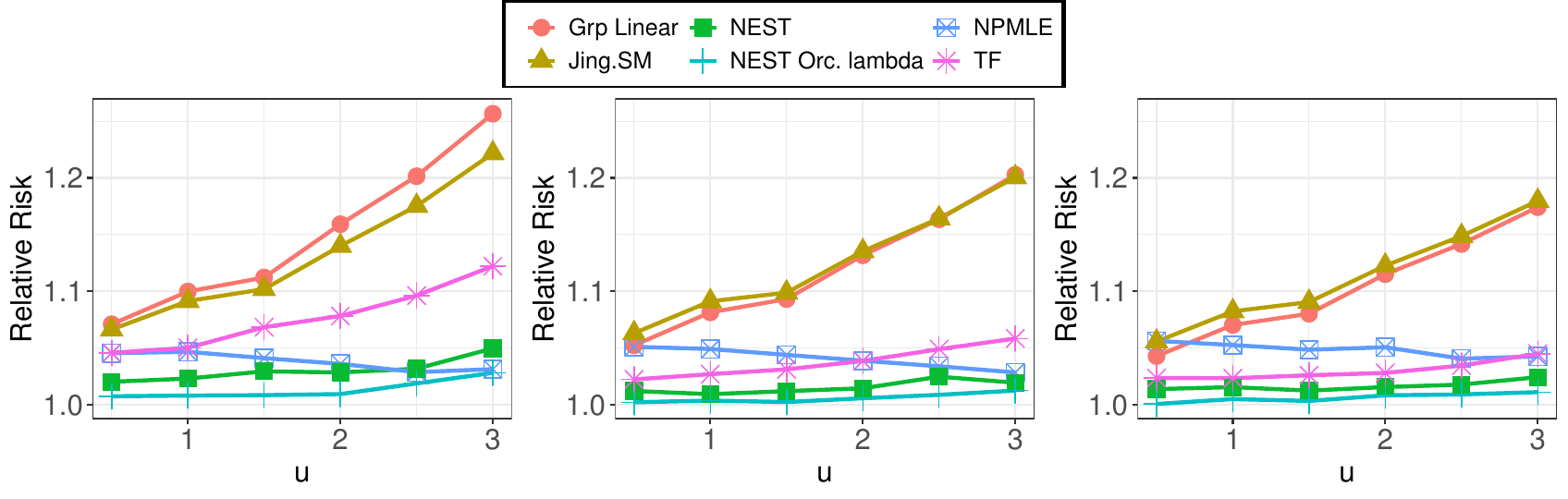}
	\caption{Comparison of relative risk for non-normal data. Here $Y_{ij}|\mu_i,\sigma_i\stackrel{i.i.d}{\sim} U(\mu_i-\sqrt{3\sigma_i^2},\mu_i+\sqrt{3\sigma_i^2})$, $\sigma_i^2$ are sampled independently from $N(u,1)$ truncated below at $0.1$ and $\mu_i|\sigma_i^2\stackrel{ind.}{\sim} 0.8~N(0.25\sigma_i^2,0.25)+0.2~N(\sigma_i^2,1)$. Plots show $m=10,15,20$ left to right.}
	\label{fig:simMisspec1}
\end{figure}

The aforementioned six approaches are evaluated on five different simulation settings, with the goal of assessing the relative performance of the competing estimators as the heterogeneity in the variances $\sigma_i^2$ is varied while keeping the sample sizes $m_i$ fixed at $m$. The five simulation settings can be categorized into three types: a setting where mean and variances are independent; three settings where mean and variance are correlated; and a setting that represents departure from the Normal data-generating model. For each setting we set $n=1{,}000$ and compute the average squared error risk for each competing estimator of $\bm\mu$ across $50$ Monte Carlo repetitions.  Figures \ref{fig:simInd} to \ref{fig:simMisspec1} plot the relative risk which is the ratio of the average squared error risk for any competing estimator to that of oracle Bayes estimator $\bm \delta^{\pi}_{(0)}$ of $\bm \mu$ (Equation \eqref{eq:bayes}) so that a ratio bigger than 1 represents a poorer risk performance of the competing estimator relative to the Bayes oracle.

The first setting, Figure \ref{fig:simInd}, corresponds to the independent case. Here, for each $i=1,\ldots,n$, $\mu_i\stackrel{i.i.d}{\sim} 0.7~N(0,0.1)+0.15~N(1,3)+0.15~N(-1,3)$ and $\sigma_i^2\stackrel{i.i.d}{\sim}U(0.5,u)$ where we let $u$ vary across five levels, $\{1, 2,3,4,5\}$.  
The three plots in Figure \ref{fig:simInd} show the relative risks as $u$ varies for $m = 10, 15$ and $20$ (left to right). We see that for $m = 10$, the competing methods split into two levels of performance. The group with the lowest relative risks consists of NPMLE, TF and NEST while the two linear shrinkage methods exhibit substantially higher relative risks. Moreover, we also see that as heterogeneity increases with increasing $u$, the gap between the two groups' relative risks increases, indicating that NPMLE, TF and the proposed NEST method are particularly useful for compound estimation of normal means when the variances are unknown and heterogeneous, and the sample size for estimating those variances are themselves small. As $m$ increases, the performance of the two linear shrinkage methods and TF improve which is expected as there are now more replicates per unit of study to construct a relatively reliable estimate of the unknown variances. However, the performance of NEST improves too and particularly at $m=20$ (Figure \ref{fig:simInd} right), NPMLE exhibits a slightly higher relative risk than NEST and TF.

The second setting, Figure \ref{fig:simCorr}, corresponds to the correlated case. The precisions $\tau_i = 1/\sigma_i^2$ are generated independently from a gamma mixture, with an even chance of drawing  $\Gamma(20, {\sf rate} = 20)$ or  $\Gamma(20, {\sf rate} =u)$ and given $\tau_i$, the means $\mu_i$ are independently $0.5~N(0.5/\tau_i, 0.5^2)+0.5~N(-0.5/\tau_i, 0.5^2)$. In this setting, the magnitude of the variances increase with $u$ and the means grow with the variances. We note from Figure \ref{fig:simCorr} that Grp. Linear and Jing.SM exhibit improved performance particularly for small values of $u$. As $u$ increases, TF, NPMLE and NEST perform well although their relative risk profiles are substantially away from $1$ as $u$ increases. For TF and NEST this behavior is expected given the statements of Propositions \ref{prop:twf} and \ref{prop:oraclerisks}. %Additionally, Jing.SM has a lower relative risk in comparison to  Grp. Linear while the relative risk of TF is slightly higher than that of NPMLE and NEST. 
The improved performance of Jing.SM in this setting is potentially related to the observation that when $u$ is small, the rate mixture of Gamma distributions on $\tau_i$ can be well approximated by a single Gamma distribution and that coincides with the parametric prior that \cite{jing2016sure} use on the precision to derive their empirical Bayes estimator for the means.

{In the third setting, Figure \ref{fig:conjugate}, $(\mu_i,\tau_i)$ continue to be correlated and have a conjugate prior distribution under Model \eqref{eq:normmodel}. The precisions $\tau_i$ are drawn from $\Gamma(20, {\sf rate} =u)$ and conditional on $\tau_i$, $\mu_i$ are independently $N(0,0.5/\tau_i)$. Under this data generating scheme, the posterior mean of $\mu_i$ is $my_i/(m+2)$ which is independent of $u$. This is the reason that the relative risks of the competing estimators in Figure \ref{fig:conjugate} do not vary with the heterogeneity in the variances. Compared to the first two settings, we see that the linear shrinkage estimators have a relatively better performance and Jing.SM dominates all other shrinkage estimators. This is expected because in this setting the posterior mean of $\mu_i$ is indeed a linear function of the sample mean $y_i$. For $m=10$ and $15$, we notice that the relative risk of NEST is marginally better than the competing estimators while at $m=20$ Grp Linear and NEST have similar risk performance. } 

Figure \ref{fig:simSparse} presents the fourth setting where the precisions $\tau_i$ are drawn from the gamma mixture of Setting 2 and $\mu_i|\tau_i\stackrel{ind.}{\sim} 0.7~N(0,0.01)+0.15~N(0.5/\tau_i,1)+0.15~N(-0.5/\tau_i,1)$. We see a similar pattern to that in Figure \ref{fig:simInd} at $m=10$. For $m=15$ and $20$, we notice that the relative risks of the linear shrinkage methods are now higher than their levels at $m=10$. This is not surprising for in this setting, while the risk performance of all methods have improved with larger sample sizes, NPMLE, TF and NEST exhibit a bigger improvement in risk than those of Grp Linear and Jing.SM.

The fifth setting, Figure \ref{fig:simMisspec1}, corresponds to the setting where the data $Y_{ij}|(\mu_i, \sigma^2_i)$ are not normally distributed. 
Here the proposed NEST method demonstrates robustness to departures from the Normal model particularly in comparison to TF, Grp. Linear and Jing.SM. 
%Proposition 7 in \cite{barp2019minimum} guarantees that, in general, the influence function of minimum KSD estimators, such as the NEST estimator, is bounded under data corruption and the behavior of the NEST estimator in Settings 5 and 6 is potentially an example of such a robustness property.% of minimum KSD estimators. 

Overall, the results of the preceding five simulation settings corroborate the statement of Proposition \ref{prop:oraclerisks} and reveal that when the variances are unknown, the NEST estimation framework enjoys a relatively better risk performance for estimating the means under the squared error loss than the linear shrinkage methods and Tweedie's formula that rely on sample variances. 

\subsection{Numerical Experiments with unequal sample sizes $m_i$}
\label{sec:mi}
\begin{figure}[!t]
	\centering
	\begin{subfigure}[b]{1\textwidth}
		\includegraphics[width=1\linewidth]{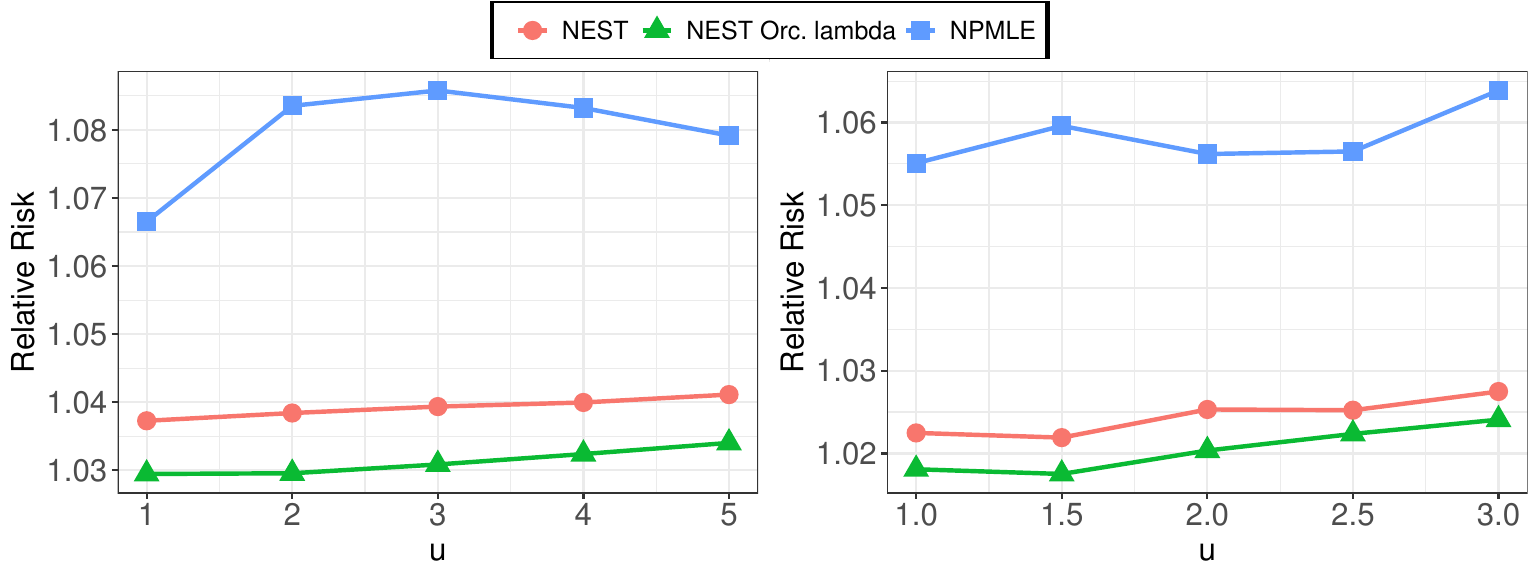}
		\caption{Simulation settings 4 (left) and 7 (right) from Section \ref{sec:simulations} where estimation is conducted under the weighted squared error loss.}
		\label{fig:sims123_mi}
	\end{subfigure}
	\begin{subfigure}[b]{1\textwidth}
		\includegraphics[width=1\linewidth]{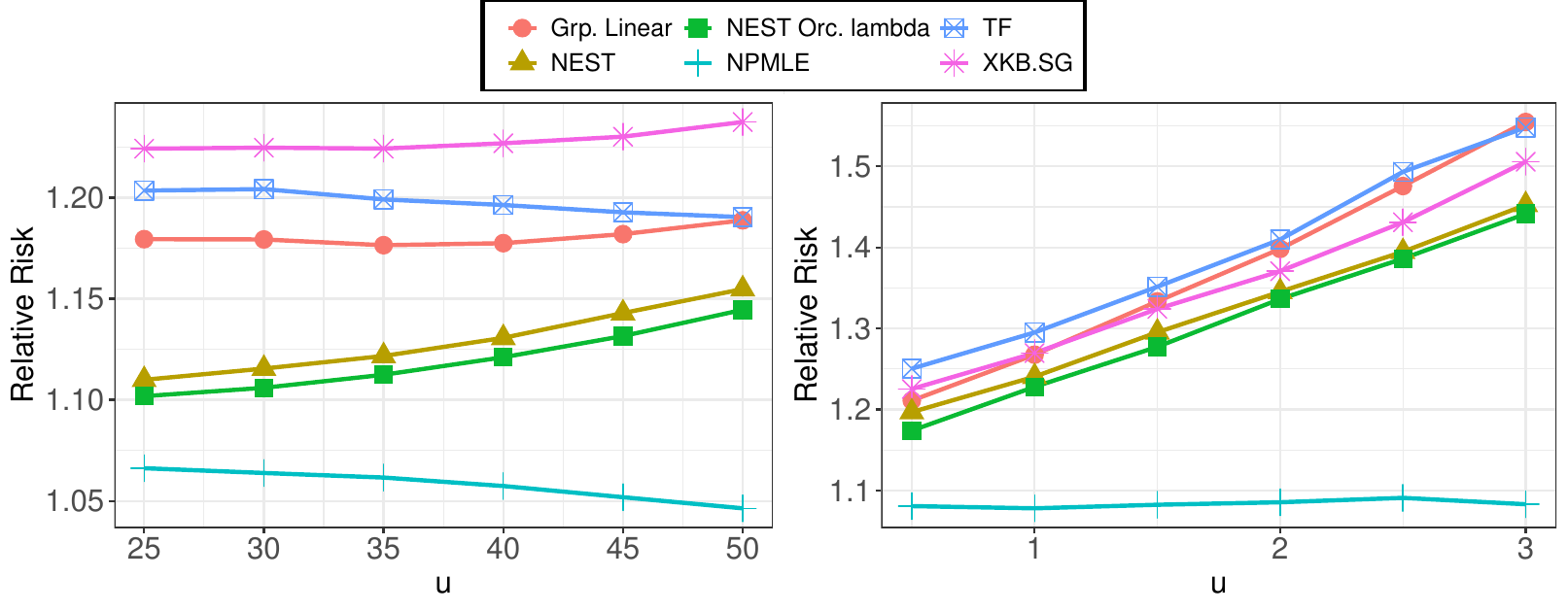}
		\caption{Simulation settings 2 (left) and 5 (right) from Section \ref{sec:nomrsimuMeans_se} where estimation is conducted under the squared error loss.}
		\label{fig:sims456_mi}
	\end{subfigure}
	\caption{Compound estimation of means under small and unequal sample sizes $\bm m$. Here $\bm m$ is a fixed vector of size $n$ with elements sampled randomly from $(4,5,6)$ with replacement.}
\end{figure}
In this section, we present the risk performance of the competing approaches of sections \ref{sec:simulations} and \ref{sec:nomrsimuMeans_se} when the sample sizes $m_i$ are small and differ across the $n=1000$ units of study. \red{We note that a small $m_i$ represents a challenging setting for estimating $(\mu_i,\tau_i)$ for any empirical Bayes method, with the extreme case of $m_i=1$, where they are not statistically identifiable.} 

We use the following four simulation settings: Settings 4 and 7 from Section \ref{sec:simulations} and Settings  2 and 5 from Section \ref{sec:nomrsimuMeans_se}. However, we change how $\bm m=(m_1,\ldots,m_n)$ are generated in these settings. Figures \ref{fig:sims123_mi} and \ref{fig:sims456_mi} present the relative risks of the competing estimators across these four scenarios. Here $\bm m$ is generated as a fixed vector of size $n$ with elements sampled randomly from $\{4,5,6\}$ with replacement. The case where $m\in\{4,5\}$ represents a particularly challenging scenario for NEST which is based on the following observation: in Equation \eqref{eq:generalized-Tweedie-var} $(m_i-3)\{(m_i-1)S_i^2\}^{-1}$ is an unbiased estimator of $\tau_i$ which follows from the fact that $\{(m_i-1)S_i^2\tau_i\}^{-1}$ has an inverse Chi-square distribution with $m_i-1$ degrees of freedom. However the variance of this distribution does not exist unless $m_i>5$. 

Figure \ref{fig:sims123_mi} represents Settings 4 (left) and 7 (right) from Section \ref{sec:simulations}. We note that NEST continues to dominate NPMLE under the weighted squared error loss even when $m_i$ are small. However, at such small sample sizes the risk of the NEST estimator is relatively larger than that of $\bm \delta_{(1)}^{\pi}$. 

Figure \ref{fig:sims456_mi} exhibits Settings 2 (left) and 5 (right) from Section \ref{sec:nomrsimuMeans_se} where the estimation is conducted under the squared error loss. Here, we use the semi-parametric monotonically constrained SURE estimator that shrinks towards the grand mean, XKB.SG, from \cite{Xieetal12} in place of Jing.SM from Section \ref{sec:nomrsimuMeans_se} as the latter was originally designed for the case $m_i=m$. While the extension of Jing.SM to unequal $m_i$ is straightforward, we do not pursue that direction in this article and, instead, use XKB.SG in its place. 

In Figure \ref{fig:sims456_mi} we note that NPMLE dominates NEST and NEST exhibits a relatively better risk performance than TF, Grp. Linear and XKB.SG. However, in Setting 5 (right panel of Figure \ref{fig:sims456_mi}), where the data $Y_{ij}|(\mu_i,\sigma_i^2)$ are not normally distributed, the performance of NEST is substantially poorer than NPMLE when $u$ is large. The two main reasons for this behavior are related to (1) Proposition \ref{prop:oraclerisks} where the oracle NEST estimator $\bm \delta_{(1)}^{\pi}$ is not, in general, the optimal estimator of the means under the squared error loss, and (2) when the sample size $m_i$ is less than 6 the variance of $\{(m_i-1)S_i^2\tau_i\}^{-1}$ under Model \eqref{eq:normmodel} does not exist. 
\subsection{Compound Estimation of Ratios}
\label{num:simuRatios}
In this section we demonstrate the use of the NEST estimation framework for compound estimation of $n$ ratios $\theta_i=\sqrt{m_i}\mu_i/\sigma_i$ which represent a popular financial metric for assessing mutual fund performance (see Section \ref{sec:mfdata} for a related real data application involving compound estimation of  mutual fund Sharpe ratios.).
\begin{figure}[!ht]
	\centering
	\includegraphics[width=1\linewidth]{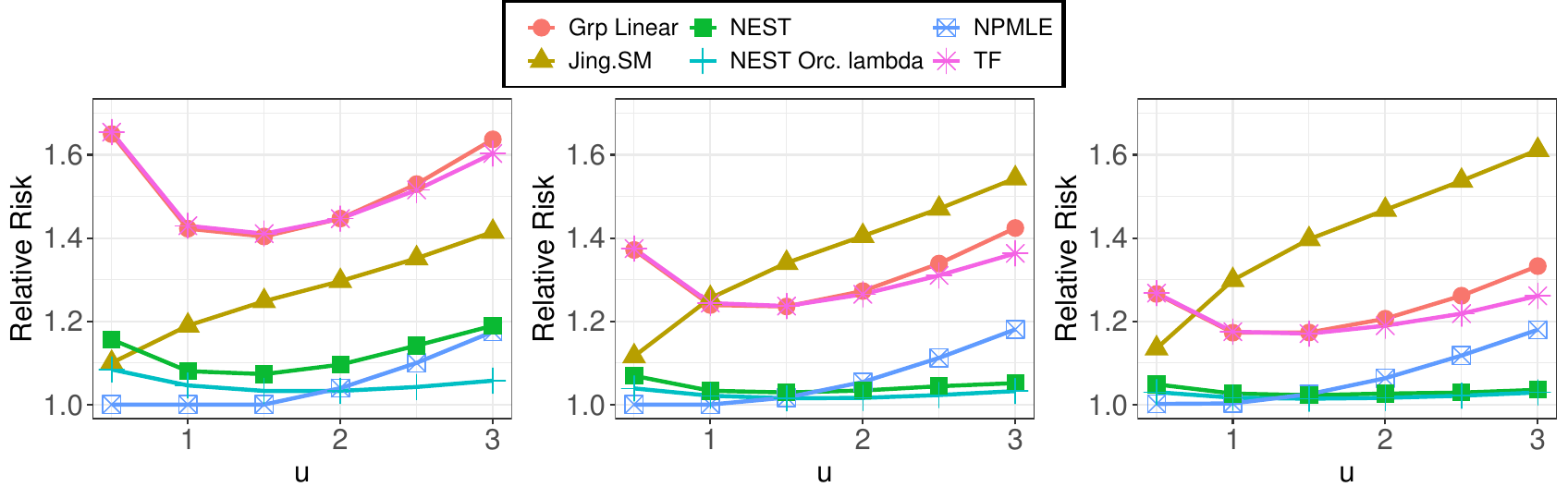}
	\caption{Comparison of relative risk for estimating $\bm \theta$. Here $\sigma_i^2\stackrel{i.i.d}{\sim}U(0.1,u)$ and $\mu_i|\sigma_i\stackrel{ind.}{\sim} 0.5~N(0.5\sigma_i^2,0.5^2)+0.5~N(-0.5\sigma_i^2,0.5^2)$. Plots show $m=10,15,20$ left to right.}
	\label{fig:simRatio1}
\end{figure}
We evaluate the performance of the same six methods under the squared error loss with $n$ fixed at $1000$ and $m_i=m$ for $i=1,\ldots,n$. We consider two simulation settings and plot the risk performance of the competing estimators of $\bm\theta=(\theta_1,\ldots,\theta_n)$ relative to the optimal Bayes estimator that estimates $\bm \theta$ using the vector of posterior means $\mathbb{E}(\theta_i|y_i,s_i^2)$ for $i=1,\ldots,n$. Note that amongst the six methods considered here, NPMLE is the only method that is designed to estimate these posterior means. The other methods estimate $\theta_i$ by separately estimating $\mu_i$ and $\sigma_i$, and then take their ratio to construct an estimate. Grp. Linear and TF, in particular, rely on the sample standard deviation for estimating $\sigma_i$ while NEST and Jing.SM employ their respective empirical Bayes estimators of $\sigma_i^2$. 

Setting 1 is presented in Figure \ref{fig:simRatio1}. The data $Y_{ij}$ are generated independently from $N(\mu_i, \sigma^2_i)$, the variances $\sigma_i^2$ are simulated uniformly between $0.1$ and $u$, and the means are independently drawn from a mixture model with half chance  $N(-\sigma_i^2/2, 0.5^2)$ and the other half $N(\sigma_i^2/2, 0.5^2)$. We continue to see that NEST has a lower relative risk than Grp. Linear and TF, both of which use sample variances. NEST also dominates Jing.SM for almost all values of $u$ while NPMLE dominates NEST for small values of $u$. As $u$ increases the heterogeneity in the data grows and we see that the relative risks of Grp. Linear and Tweedie's Formula across all $m$ first decrease and then increase. The shift in the behavior of these estimators is related to the observation that as $u$ increases, the centers of the mixture model that generates $\mu_i$, are on average, further away from one another. This makes estimating the numerator of the ratio easier for all methods up until a point. As heterogeneity increases further, the risks of these methods that use the sample standard deviation in the denominator of $\theta_i$ are relatively worse than the risk of NEST and NPMLE. 
\begin{figure}[!ht]
	\centering
	\includegraphics[width=1\linewidth, height=2.2in]{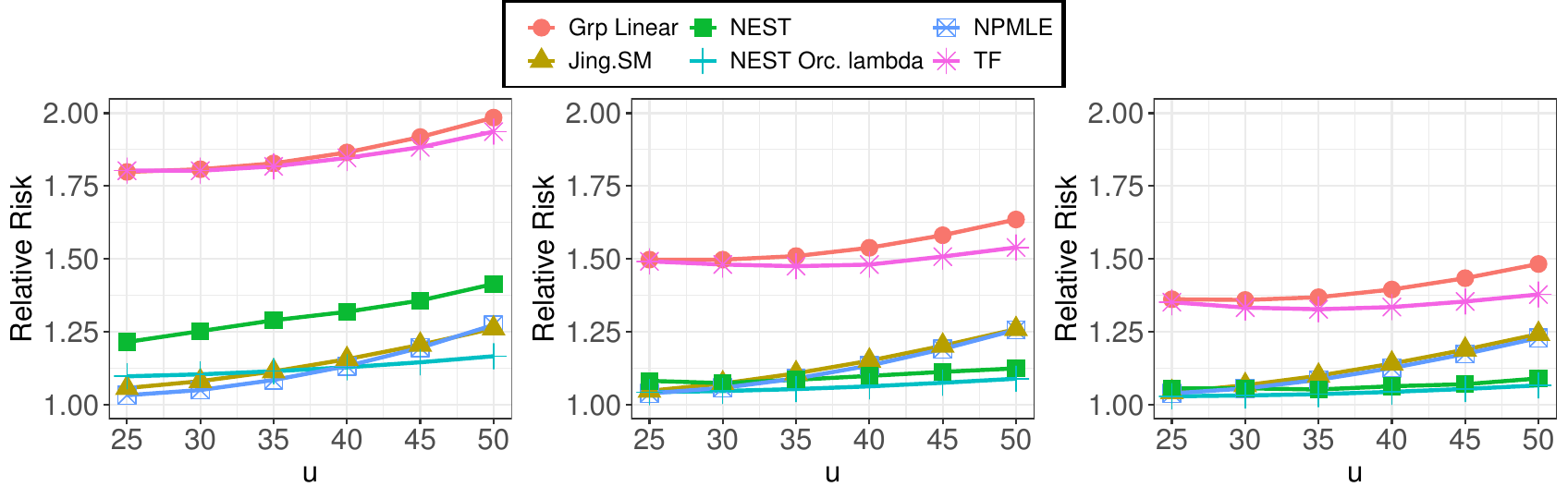}
	\caption{Comparison of relative risk for estimating $\bm \theta$. Here $\tau_i\stackrel{i.i.d}{\sim}0.5\Gamma(20,\texttt{rate}=20)+0.5\Gamma(20,\texttt{rate}=u)$ and $\mu_i|\tau_i\stackrel{ind.}{\sim} 0.5~N(0.5/\tau_i,0.5^2)+0.5~N(-0.5/\tau_i,0.5^2)$. Plots show $m=10,15,20$ left to right.}
	\label{fig:simRatio2}
\end{figure}

Setting 2 is presented in Figure \ref{fig:simRatio2} where the means are generated according to Setting 1 but the precisions are independently drawn from a mixture model with half chance  $\Gamma({\sf shape}=20, {\sf rate}=20)$ and the other half $\Gamma({\sf shape}=20, {\sf rate}=u)$. We note that while NEST dominates TF and Grp. Linear, NPMLE and Jing.SM dominate NEST when the heterogeneity is relatively smaller and when $m=10$. The improved performance of Jing.SM in this setting is potentially related to the observation that when $u$ is small, the rate mixture of Gamma distributions on $\tau_i$ can be well approximated by a single Gamma distribution and that coincides with the parametric prior that \cite{jing2016sure} use on the precision to derive their empirical Bayes estimator for the variances. In Figures \ref{fig:simRatio1} and \ref{fig:simRatio2} we note that for $m=10$, the relative risk of NEST is substantially higher than NEST Orc. $\lambda$. This is not unexpected because NEST Orc. $\lambda$ estimates $\lambda$ by minimizing the true squared error loss involving $\bm \theta$ while the data-driven NEST estimator relies on the modified cross-validation approach described in Section \ref{Subsec:bandwidth} with ${\vartheta}_n(\lambda;\mathcal U,\mathcal V)$ defined as $$
{\vartheta}_n(\lambda;\mathcal U,\mathcal V)=\dfrac{1}{n}\sum_{i=1}^{n}\left\{\bar{V}_i-\delta_i^{\sf ds}(\bar{U}_i;\mathcal{U},\lambda)\right\}^2,
$$ to choose $\lambda$.
%%%%%%%%%%%%%%%%%%%%%%%%%%%%%%%%%%%%%%%%%%%%%%%%%%
\section{Real Data Analyses}
\label{sec:realdata}
\subsection{Baseball Data}
\label{sec:baseball}
We analyze the monthly data on the number of ``at bats" and ``hits" for all U.S Major League baseball players over the regular seasons from 2002 until 2011. In this analysis we focus on both pitchers and non-pitchers using an approach similar to that of \cite{gu2017empirical}. The data are available from the R package \texttt{REBayes} and have been aggregated into half seasons to produce an unbalanced panel. It includes observations on 932 players who have at least ten at bats in any half season and appear in no fewer
than five half-seasons (note that there are a total of 20 half-seasons that a player can appear in).

Following \cite{brown2008season}, let the transformed batting average $Y_{ij}$ for player $i(=1,\ldots,n)$ at time $j(=1,\ldots,m_i)$ be denoted by $Y_{ij}=\text{arcsin}\left(\sqrt{\dfrac{H_{ij}+0.25}{N_{ij}+0.5}}\right)$ where $H_{ij}$ denotes the number of ``hits" and $N_{ij}$ denotes the number of ``at bats" at time $j$ for player $i$. We assume that $Y_{ij}\sim N(\mu_i,v_{ij}^2/\tau_i)$ where $\mu_i=\text{arcsin}(\sqrt{p_i})$, $p_i$ being player $i$'s batting success probability, and $v_{ij}^2=1/(4N_{ij})$. Here $1/\tau_i$ are player specific scale parameters as described in \cite{gu2017empirical}. Under this setup, the sufficient statistics are 
\begin{eqnarray*}
	\hat{\mu}_i & = &  \left(\sum_{j=1}^{m_i}1/v_{ij}^2\right)^{-1}\sum_{j=1}^{m_i}Y_{ij}/v_{ij}^2\sim N(\mu_i,v_i^2/\tau_i)\text{ with }v_i^2=\left(4\sum_{j=1}^{m_i}N_{ij}\right)^{-1}, \\
	S_i^2 & = & \dfrac{1}{m_i-1}\sum_{j=1}^{m_i}(Y_{ij}-\hat{\mu}_i)^2/v_{ij}^2 \text{ with }(m_i-1)S_i^2\tau_i\sim \mathcal{X}^2_{m_i-1}.
\end{eqnarray*}

In this analysis, the goal is to use the 2002-2011 data to predict the batting averages of the players in 2012. Players are divided into three categories: all, non-pitchers, and pitchers. We consider the following seven estimators of $\mu_i$: two non-parametric maximum likelihood based estimators, denoted NPMLE-Indep and NPMLE-Dep which assume, respectively, independent and dependent priors on $(\mu_i,1/\tau_i)$, the sufficient statistics $\hat{\bm \mu}=(\hat{\mu}_1,\ldots,\hat{\mu}_n)$ of $\bm\mu$, the grand mean across all players in the 2011 season  $\bar{Y}^{\sf 2011}=n^{-1}\sum_{i=1}^{n}Y_{i,2011}$, the proposed NEST estimator and its counterpart with an oracle choice for the tuning parameter $\lambda$ (NEST orc $\lambda$), and the naive estimator that uses 2011 batting averages $\bm Y_{2011}=(Y_{1,2011},\ldots,Y_{n,2011})$. To assess how well these methods predict 2012 batting averages $\bm Y_{2012}=(Y_{1,2012},\ldots,Y_{n,2012})$, we consider three criteria for evaluating any estimate $\delta_i$ of $\mu_i$: total squared error from \cite{brown2008season} and defined as $TSE(\bm\delta) = \sum_{i=1}^{n}\Bigl\{(Y_{i,2012}-\delta_i)^2-(4N_{i,2012})^{-1}\Bigr\}$, normalized squared error from \cite{gu2017empirical} and defined as $NSE(\bm\delta) = \sum_{i=1}^{n}\Bigl\{4N_{i,2012}(Y_{i,2012}-\delta_i)^2\Bigr\}$, and total squared error on a probability scale from \cite{jiang2010empirical} which is defined as $TSEp(\hat{\bm p}) = \sum_{i=1}^{n}\Bigl\{(p_{i,2012}-\hat{p}_i)^2-p_{i,2012}(1-p_{i,2012})(4N_{i,2012})^{-1}\Bigr\}$. Here $\hat{p}_i=\sin^2(\delta_i)$ and $p_{i,2012}=\sin^2(Y_{i,2012})$. 
% Table generated by Excel2LaTeX from sheet 'Sheet2'
% Table generated by Excel2LaTeX from sheet 'Sheet2'
\begin{table}[!t]
	\centering
	\caption{Performance of the competing estimators relative to the performance of the naive estimator $\bm Y_{2011}$. Here R-$TSE(\bm\delta)=TSE(\bm\delta)/TSE(\bm Y_{2011})$ with similar definitions for R-NSE and R-TSEp. The smallest two relative errors are bolded}
	\scalebox{0.8}{
		\begin{tabular}{rlcccccc}
			\toprule
			&       & NPMLE-Indep & NPMLE-Dep & $\hat{\bm \mu}$    & $\bar{Y}^{\sf 2011}$  & NEST    & NEST orc $\lambda$ \\
			\midrule
			\multicolumn{1}{c}{All} &       &       &       &       &       &       &  \\
			\multicolumn{1}{l}{$n$ for estimation: 932} & R-TSE & 0.463 & 0.491 & 0.352 & 1.958 & \textbf {0.350} &  \textbf{0.350} \\
			\multicolumn{1}{l}{$n$ for prediction: 370} & R-NSE & \textbf{0.668} & 0.678 & 0.676 & 1.495 & \textbf{0.670} & \textbf{0.670} \\
			& R-TSEp & 0.582 & 0.610 & 0.501 & 1.783 & \textbf{0.498} & \textbf{0.498} \\
			\midrule
			\multicolumn{1}{c}{Nonpitchers} &       &       &       &       &       &       &  \\
			\multicolumn{1}{l}{$n$ for estimation: 792} & R-TSE & 0.535 & 0.559 & {0.503} & 0.551 & \textbf{0.488} & \textbf{0.484} \\
			\multicolumn{1}{l}{$n$ for prediction: 325} & R-NSE & \textbf{0.656} & {0.665} & 0.679 & 0.973 & 0.667 & \textbf{0.659} \\
			& R-TSEp & 0.651 & 0.677 & {0.619} & 0.682 & \textbf{0.604} & \textbf{0.601} \\
			\midrule
			\multicolumn{1}{c}{Pitchers} &       &       &       &       &       &       &  \\
			\multicolumn{1}{l}{$n$ for estimation: 140} & R-TSE & 0.659 & 0.662 & 0.629 & 0.804 & \textbf{0.628} & \textbf{0.620} \\
			\multicolumn{1}{l}{$n$ for prediction: 45} & R-NSE & 0.659 & 0.663 & \textbf{0.649} & 0.769 & \textbf{0.649} & \textbf{0.638} \\
			& R-TSEp & \textbf{0.124} & 0.124 & 0.133 & 0.337 &  0.133 & \textbf{0.122} \\
			\bottomrule
	\end{tabular}}
	\label{tab:baseball}%
\end{table}%

In Table \ref{tab:baseball}, we report the performance of the competing estimators relative to the performance of the naive estimator $\bm Y_{2011}$ wherein R-$TSE(\bm\delta)=TSE(\bm\delta)/TSE(\bm Y_{2011})$ with similar definitions for R-$NSE$ and R-$TSE_p$. Thus, a smaller value of R-$TSE$, R-$NSE$ or R-$TSE_p$ indicates a relatively better prediction error. Across ``All" and ``Nonpitchers", NEST exhibits the best relative risk for two of the three performance metrics. It is interesting to note that the sufficient statistics $\hat{\bm \mu}$ are quite competitive in this example while NPMLE with independent priors dominate the one with dependent priors across ``All", ``Nonpitchers" and ``Pitchers". The compound estimation problem for ``Pitchers" is an example of a setting where $n$ is relatively small and NEST demonstrates a better risk performance than NPMLE for total squared error and normalized squared error losses. 
\subsection{Mutual Fund Sharpe Ratios}
\label{sec:mfdata}
In this section we analyze a dataset on $n_1=5{,}000$ monthly mutual fund returns spanning 12 months from January 2014 to December 2014. This data are sourced from the Wharton research data services \citep*{wrds}. The goal in this analysis is to use Sharpe ratios constructed using the data on the first $m_1=6$ months, January 2014 - June 2014, to predict the corresponding Sharpe ratios for the next 6 months. Formally, let $Y_{ij}$ denote the excess return of fund $i (=1,\ldots, n_1)$ in month $j(=1,\ldots,m_1)$ over the return on the 3 month treasury yield. Denote $\bar{Y}_i=m_{1}^{-1}\sum_{j=1}^{m_{1}}Y_{ij}$, $S_i^2=(m_{1}-1)^{-1}\sum_{j=1}^{m_{1}}(Y_{ij}-\bar{Y}_i)^2$ and $\delta_i^{\sf naive}=\bar{Y}_i/\sqrt{S_i^2}$ to be, respectively, the sample mean, the sample variance and the observed Sharpe ratio of the monthly excess returns. %In our analysis we include funds that have at least $4$ months worth of returns available during January 2014 - June 2014 and so $m_{i,1}\in[4,6]$. 
Of the $5{,}000$ funds available during these first 6 months, there are $n_2=4{,}958$ funds that appear in the next 6 months, July 2014 - December 2014, and have at least $3$ months of returns available during this period. For our prediction, we consider these $n_2$ funds to assess the performance of various estimators for predicting $\theta_i = \mu_i/\sigma_i$ where $\mu_i$ and $\sigma_i$ are the sample mean and sample standard deviation of the excess returns of the $n_2$ funds during the next 6 months.
% Table generated by Excel2LaTeX from sheet 'Sheet1'

We consider the following estimators of $\bm \theta=(\theta_1,\ldots,\theta_n)$: NEST, NEST orc. $\lambda$, Tweedie's formula (TF), Grp Linear, Jing.SM and NPMLE from Section \ref{sec:nomrsimuMeans_se}. Additionally, we consider the SURE estimators XKB.SG and XKB.G from \citet{Xieetal12}. Note that for predicting $\theta_i$, TF, Grp Linear, XKB.SG and XKB.G rely on the sample variances $S_i^2$. %and the SURE estimator (XKB.SG) from \citet{Xieetal12} %that shrinks towards the grand mean.
To evaluate the performance of these estimators for predicting $\bm \theta$, we consider the following three criteria with $m_{i,2}\in[3,6]$: Total Squared Error : $TSE(\bm\delta) = \sum_{i=1}^{n_2}(\theta_{i}-\delta_i)^2$; weighted Squared Error : $WSE(\bm\delta) = \sum_{i=1}^{n_2}m_{i,2}(\theta_{i}-\delta_i)^2$; and weighted Absolute Error :  $WAE(\bm \delta) = \sum_{i=1}^{n_2}m_{i,2}|1-\delta_i/\theta_i|$. In Table \ref{tab:mfdata}, we present the performance of the competing estimators relative to the performance of the naive estimator $\bm \delta^{\sf naive}=(\delta_{i}^{\sf naive}:1\le i\le n)$ so that a smaller value of R-TSE, R-WSE or R-WAE indicates a relatively better prediction error.
\begin{table}[!t]
	\centering
	\caption{Performance of the competing estimators relative to the performance of the naive estimator $\bm\delta^{\sf naive}$. Here R-$TSE(\bm\delta)=TSE(\bm\delta)/TSE(\bm\delta^{\sf naive})$ with similar definitions for R-WSE and R-WAE. The smallest two relative errors are bolded}
	\scalebox{0.9}{
		\begin{tabular}{ccccccccccc}
			\toprule
			& $n_1$  & $n_2$ & Grp Lin. & XKB.G  & XKB.SG & NPMLE & Jing.SM &TF & NEST & NEST orc. $\lambda$   \\
			\midrule
			R - TSE & 5000  & 4958  & 0.896 & 0.931 &  0.922 &0.842 & 0.720 &  0.997&\textbf{0.688} & \textbf{0.686} \\
			R - WSE & 5000  & 4958  & 0.893 & 0.930 & 0.920 & 0.837& 0.716 &0.997&\textbf{0.686} & \textbf{0.684} \\
			R - WAE & 5000  & 4958  & 1.067 & 0.974 & 0.983 &\textbf{0.479}&\textbf{0.171} & 0.999&0.782 & {0.780} \\
			\bottomrule
	\end{tabular}}%
	\label{tab:mfdata}%
\end{table}%

Along the performance measures of Total Squared Error and Weighted Squared Error, NEST has the smallest relative risk among all competing estimators considered in this example. With respect to the Weighted Absolute Error, Jing.SM has a substantially smaller relative risk than NEST while Grp Linear appears to be doing relatively worse and exhibits an R-WAE bigger than $1$. %Note that under this performance measure, the data-driven NEST has a smaller relative risk than the NEST estimator with oracle $\lambda$. This is expected because the former relies on the modified cross-validation approach described in Section \ref{Subsec:bandwidth} to choose $\lambda$ while the latter minimizes the true squared error loss to derive the oracle $\lambda$.  
When compared against the three linear shrinkage methods considered here, NEST and NPMLE demonstrate an overall value in joint shrinkage estimation of the means $\mu_i$ and the variances $\sigma_i^2$ for predicting $\bm \theta$.
%%%%%%%%%%%%%%%%%%%%%%%%%%%%%%%%
\section{Extensions}
\label{sec:extensions}
%%%%%%%%%%%%%%%%%%%%%%%%%%%%%%%%
This section considers the extension of our methodology to several well known members in the two-parameter exponential family. We will focus on examples where the nuisance parameter is known. Our proposed estimation framework is motivated by the double shrinkage idea, but the approach nonetheless handles the case with known nuisance parameters. %We discuss four examples and in each of these examples the key idea is to express the posterior distribution of the parameter of interest as a one-parameter exponential family. Thereafter, 
We discuss four examples, in each of which we derive the Bayes estimator of the natural parameter under the squared error loss. The Bayes estimator in these examples relies on the unknown score function (of the marginal density of the sufficient statistic), which can be consistently estimated using the ideas in Section \ref{sec:opt_criteria}. 
\begin{example}[Location mixture of Gaussians]
	\label{example:1}
	Consider the following hierarchical model 
	\begin{eqnarray}
	\label{eq:gaussianlocation}
	Y_{i}~|~\mu_i,\tau_i \stackrel{ind.}{\sim} N(\mu_i, 1/\tau_i), \quad 
	\mu_i\stackrel{i.i.d}{\sim} G_\mu(\cdot), \quad \mbox{for $i=1,\ldots,n$},
	\end{eqnarray}
	where $\tau_i$ are known and $G_\mu(\cdot)$ is an unspecified prior. Equation \eqref{eq:gaussianlocation} represents the heteroskedastic normal means problem with known variances $1/\tau_i$ [see for example \cite{Weinsteinetal18}]. In this setting, the sufficient statistic for $\mu_i$ is $Y_i$ and the Bayes estimator of $\mu_i$ under the squared error loss is given by 
	$$\mu_i^{\pi}\coloneqq\mathbb{E}(\mu_i|y_i,\tau_i) = y_i+\dfrac{1}{\tau_i}\dfrac{\partial}{\partial y_i}\log f(y_i|\tau_i),    
	$$
	where $f(\cdot|\tau_i)$ is the pdf of the distribution of $Y_i$ given $\tau_i$ marginalizing out $\mu_i$. From Section \ref{sec:opt_criteria} and with $m_i=1,~\bm x_i = (y_i,\tau_i)$, the NEST estimate of $\mu_i$  is given by
	$\delta_{i,n}^{\sf nest}(\lambda) = y_i+\dfrac{1}{\tau_i}\hat{w}^{(1)}_{\lambda,n}(i).$
\end{example}
\begin{example}[Scale mixture of Gamma distributions]
	\label{example:2}
	Consider the following model
	\begin{eqnarray}
	\label{eq:gamma_scale}
	Y_{ij}~|~\alpha_i,1/\beta_i \stackrel{i.i.d}{\sim} \Gamma(\alpha_i, 1/\beta_i), \quad 
	1/\beta_i\stackrel{i.i.d}{\sim} G(\cdot)\nonumber,
	\end{eqnarray}
	where the shape parameters $\alpha_i$ are known and $G(\cdot)$ is an unspecified prior distribution on scale parameters $1/\beta_i$. Here $T_i=\sum_{j=1}^{m}Y_{ij}$ is a sufficient statistic and $T_i|\alpha_i,\beta_i\stackrel{ind.}{\sim}\Gamma(m\alpha_i,1/\beta_i)$. The posterior distribution of $1/\beta_i$ belongs to a one-parameter exponential family with density
	\begin{equation}
	\label{eq:gamma_posterior}
	f(1/\beta_i|T_i,\alpha_i)\propto \exp{\Bigl\{-{T_i}{\beta_i}+(m\alpha_i-1)\log T_i-\log f(T_i|\alpha_i)\Bigr\}},
	\end{equation}
	where $f(\cdot|\alpha_i)$ is the pdf of the distribution of $T_i$ given $\alpha_i$ (marginalizing out $1/\beta_i$). From Equation \eqref{eq:gamma_posterior}, the Bayes estimator of $\beta_i$ under the squared error loss is given by 
	\begin{equation*}
	\label{eq:gamma_scale_bayes}
	\beta_i^{\pi}\coloneqq\mathbb{E}(\beta_i|T_i,\alpha_i) = \dfrac{m\alpha_i-1}{T_i}-\dfrac{\partial}{\partial T_i}\log f(T_i|\alpha_i). 
	\end{equation*}
	With $\bm x_i = (T_i,\alpha_i)$, the NEST estimate of $\beta_i$ is given by
	$\delta_{i,n}^{\sf nest}(\lambda) = \dfrac{m\alpha_i-1}{T_i}-\hat{w}^{(1)}_{\lambda,n}(i).
	$
\end{example}
\begin{example}[Shape mixture of Gamma distributions]
	\label{example:3}
	We consider the following model:
	\begin{eqnarray}
	\label{eq:gamma_shape}
	Y_{ij}~|~\alpha_i,1/\beta_i \stackrel{i.i.d}{\sim} \Gamma(\alpha_i, 1/\beta_i), \quad 
	\alpha_i\stackrel{i.i.d}{\sim} G(\cdot)\nonumber,
	\end{eqnarray}
	where the scale parameters $1/\beta_i$ are known and $G(\cdot)$ is an unspecified prior distribution on the shape parameters $\alpha_i$. Let $Y_i=\sum_{j=1}^{m}Y_{ij}$. Then  $Y_i|\alpha_i,\beta_i\stackrel{ind.}{\sim}\Gamma(m\alpha_i,1/\beta_i)$ and $T_i=\log Y_i$ is a sufficient statistic. Moreover, the posterior distribution of $\alpha_i$ belongs to a one-parameter exponential family with density
	\begin{equation}
	\label{eq:gamma_posterior_2}
	f(\alpha_i|T_i,1/\beta_i)\propto \exp{\Bigl\{(m\alpha_i)T_i-\beta_i{\exp{(T_i)}}-\log f(T_i|1/\beta_i)\Bigr\}},
	\end{equation}
	where $f(\cdot|1/\beta_i)$ is the density of the distribution of $T_i$ given $1/\beta_i$ marginalizing out $\alpha_i$. From Equation \eqref{eq:gamma_posterior_2}, the Bayes estimator of $\alpha_i$ under the squared error loss is given by 
	\begin{equation*}
	\label{eq:gausslocation_bayes}
	\alpha_i^{\pi}\coloneqq\mathbb{E}(\alpha_i|T_i,1/\beta_i) = \dfrac{\beta_i\exp{(T_i)}}{m}+\dfrac{1}{m}\dfrac{\partial}{\partial T_i}\log f(T_i|1/\beta_i). \end{equation*}
	With $\bm x_i = (T_i,1/\beta_i)$, the NEST estimate of $\alpha_i$ is
	$\delta_{i,n}^{\sf nest}(\lambda) = \dfrac{\beta_i\exp{(T_i)}}{m}+\dfrac{1}{m}\hat{w}^{(1)}_{\lambda,n}(i).
	$
\end{example}
\begin{example}[Scale mixture of Weibulls]
	\label{example:4}
	We consider the following model:
	\begin{eqnarray}
	\label{eq:weibull_scale}
	Y_{ij}~|~k_i,\beta_i \stackrel{i.i.d}{\sim} \text{Weibull}(k_i, \beta_i), \quad 
	\beta_i\stackrel{i.i.d}{\sim} G(\cdot).
	\end{eqnarray}
	We have $f(y|~k,\beta)=\beta k y^{k-1}\exp(-\beta y^k)$. In Equation \eqref{eq:weibull_scale} the shape parameters $k_i$ are known, $G(\cdot)$ is an unspecified prior distribution on the scale parameters $\beta_i$,  $T_i=\sum_{j=1}^{m}\{Y_{ij}\}^{k_i}$ is a sufficient statistic, and $T_i|k_i,1/\beta_i\stackrel{ind.}{\sim}\Gamma(m,1/\beta_i)$. From Example 2, the Bayes estimator of $\beta_i$ is 
	\begin{equation*}
	\label{eq:weibulllocation_bayes}
	\beta_i^{\pi}\coloneqq\mathbb{E}(\beta_i|T_i,k_i) = \dfrac{m-1}{T_i}+\dfrac{\partial}{\partial T_i}\log f(T_i|k_i). 
	\end{equation*}
	With $\bm x_i = (T_i,k_i)$, the NEST estimate of $\beta_i$ is given by
	$\delta_i^{\sf nest}(\lambda) = \dfrac{m-1}{T_i}-\hat{w}^{(1)}_{\lambda,n}(i).
	$
\end{example}
The preceding examples present a setting with known nuisance parameter. When both parameters are unknown, extensions of our estimation framework to an arbitrary member of the two-parameter exponential family is difficult. The main reason is that in the Gaussian case the sufficient statistics are independent and their marginal distributions are known. However, for other distributions such as the Gamma and Beta, the joint distribution of the two sufficient statistics is generally unknown. This impedes a full generalization of our approach. We anticipate that an iterative scheme that conducts shrinkage estimation on the primary and nuisance coordinates in turn may be developed by combining the ideas in Examples \ref{example:3} and \ref{example:4} above. We do not pursue those extensions in this article. 
\section{Computational complexity}
\label{sec:complexity}
Here we discuss the computational complexities of NEST and NPMLE, and  provide a comparison of their running time. 

In contrast to the linear shrinkage estimators, such as Group Linear \cite{Weinsteinetal18}, SURE estimators of \cite{Xieetal12} and \cite{jing2016sure}, NEST and NPMLE are similar in the sense that both these approaches rely on solving a convex optimization problem to estimate the means. %Here we discuss the computational complexities of the underlying optimization problems for NEST and NPMLE, and then provide a comparison of their running time as $n$ increases.
\begin{figure}[!t]
	\centering
	\includegraphics[width=0.95\linewidth]{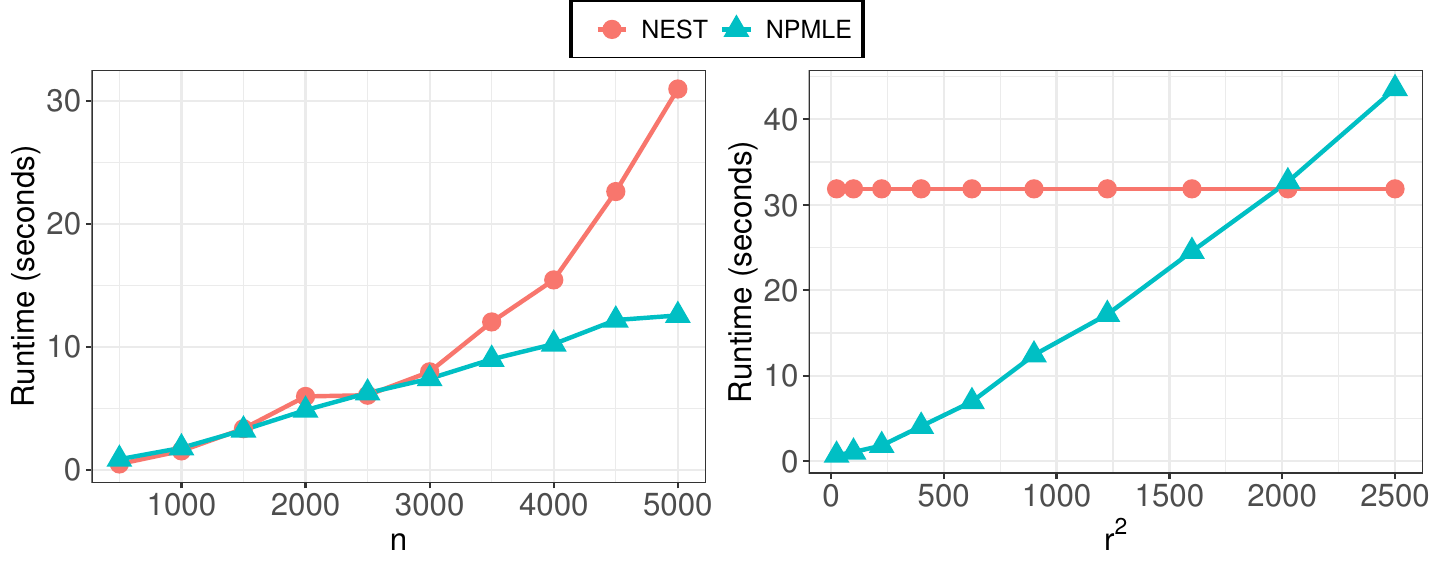}
	\caption{Running time comparison of NEST and NPMLE. Left: $r$ fixed at $30$. Right: $n$ fixed at $5000$.}
	\label{fig:complexity}
\end{figure}
An implementation of the convex optimization problem for estimating the joint prior distribution of the means and the variances via the NPMLE is available in the R package REBayes \citep{koenker2017rebayes}. This package relies on interior point methods for solving the convex problem. See \cite{KoeMiz14} for more details. The main effort in solving this problem depends on computing a Hessian and the computational complexity of that is $O(n^2r^2)$ where $r$ is the number of grid points at which the prior masses for each of the two prior distributions 
will be estimated. Usually $n\gg r$. Recently, \cite{kim2020fast} propose a fast sequential quadratic programming based algorithm for estimating a one dimensional prior distribution using the NPMLE however their algorithm is not available for the case where the prior is two dimensional, such as the scenario when both mean and variances are unknown. For NEST, the convex quadratic optimization problem described in Equation \eqref{eq:quad_opt} depends on the $n\times n$ matrix $\bm K_\lambda$ and uses the interior-point optimizer in MOSEK. The worst case computational complexity for evaluating the hessian of the underlying objective function is $O(n^3)$, which is substantially larger than that of NPMLE. Figure \ref{fig:complexity} provides a comparison of the running time of NEST and NPMLE for increasing $n$ and $r$. In the left panel of Figure \ref{fig:complexity} we fix $r=30$, the default choice implemented in the function {\sf WGLVmix} in REBayes, and vary $n$. In the right panel, $n$ is fixed at $5000$ with $r$ varying. We see that for problems with large $n$ and $r$ fixed, NPMLE is substantially faster than NEST and NEST, in its current implementation, may not be as scalable as NPMLE is to large $n$ problems. Our future research efforts will be geared towards developing faster first-order methods to solve Equation \eqref{eq:quad_opt} rather than relying on the interior-point solver in MOSEK.
\section{Sensitivity to $\alpha$}
\label{sec:alpha}
We consider four scenarios from Section \ref{sec:simulations} and report the impact of $\alpha$ on the estimation of $\lambda$. Specifically, we vary $\alpha$ over $\{0.2,0.5,0.8\}$ and present the risks of $\bm \delta_n^{\sf ds}(\hat{\lambda})$ and $\bm \delta_n^{\sf ds}({\lambda})$ relative to the risk of the oracle Bayes estimator $\bm \delta_{(1)}^{\pi}$ for $m\in\{10,15\}$.

\begin{figure}[!t]
	\centering
	\begin{subfigure}[b]{0.65\textwidth}
		\includegraphics[width=1\linewidth]{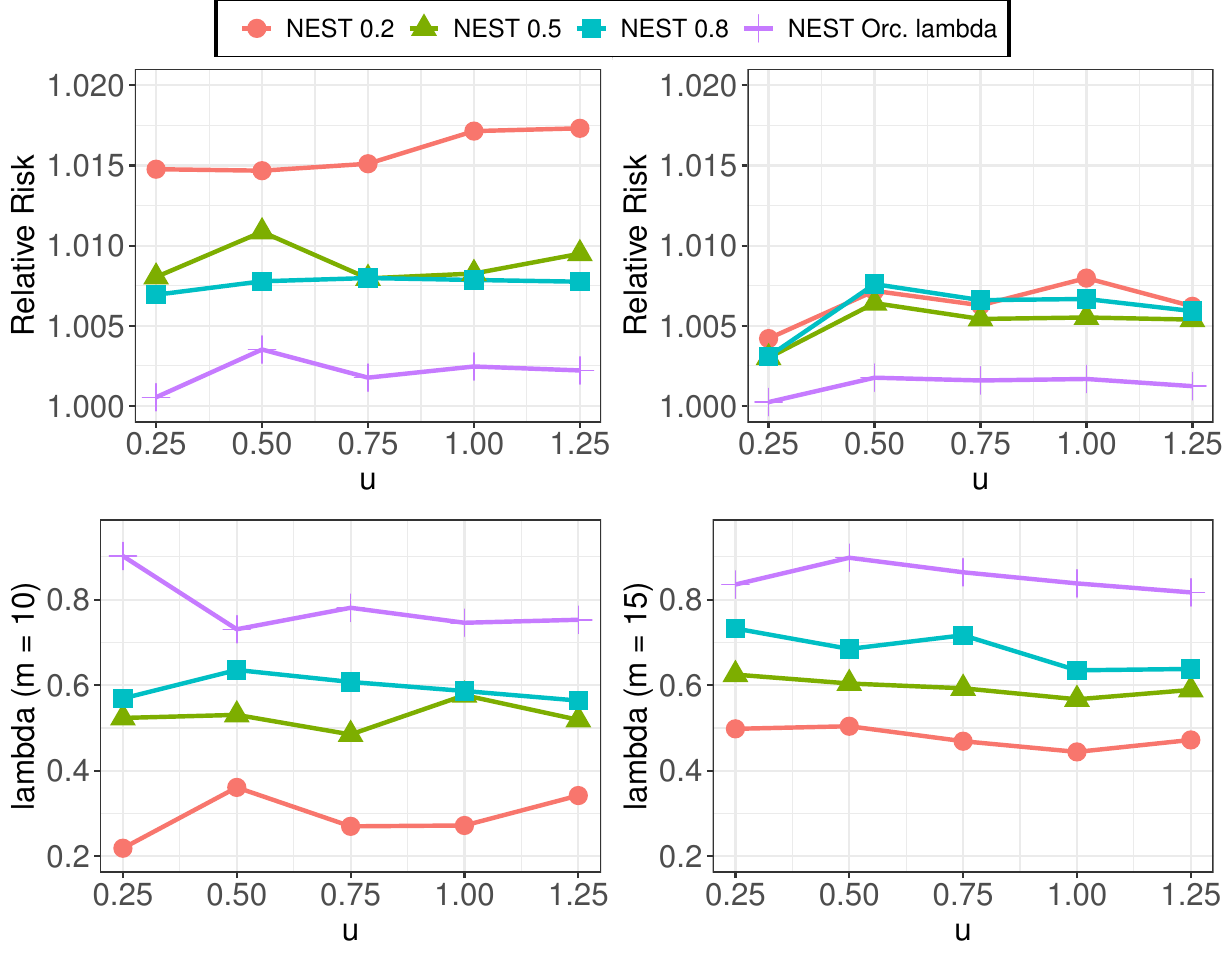}
		\caption{ Setting 2: Here $\sigma_i^2\stackrel{i.i.d}{\sim}N(u,1)$ truncated below $0.1$ and $\mu_i\stackrel{i.i.d}{\sim}~{\sf Laplace}({\sf location}=0,{\sf scale}=1)$. Left $m=10$ and right $m=15$.}
		\label{fig:setting_2} 
	\end{subfigure}
	%\end{figure}
	%\begin{figure}[!t]
	%	\centering
	\begin{subfigure}[b]{0.65\textwidth}
		\includegraphics[width=1\linewidth]{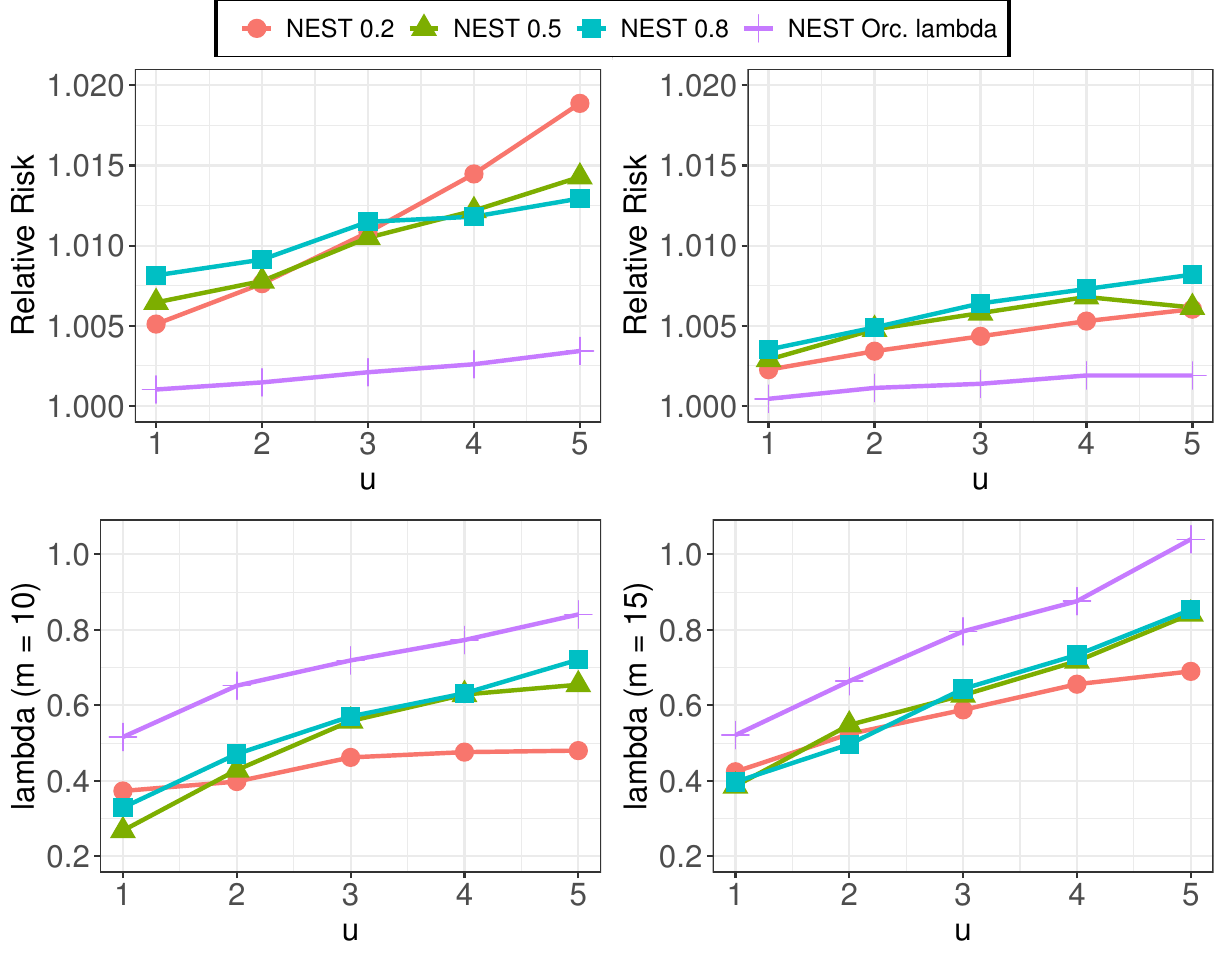}
		\caption{Setting 4: $\sigma_i^2\stackrel{i.i.d}{\sim}U(0.25,u)$ and $\mu_i~|~\sigma_i^2\stackrel{ind.}{\sim}~0.5~N(\sigma_i^2,\sigma_i^2)+0.5~N(-\sigma_i^2,\sigma_i^2)$. Left $m=10$ and right $m=15$.}
		\label{fig:setting_4}
	\end{subfigure}
	\caption{Panels (a) and (b) top left: relative risks of $\bm \delta_n^{\sf ds}(\hat{\lambda})$ for $\alpha\in\{0.2,0.5,0.8\}$. Panels (a) and (b) bottom left: the corresponding estimates of $\lambda$ along with the oracle $\lambda$ obtained by minimizing the true loss. In each panel the first column  corresponds to $m=10$ and the second column pertains to $m=15$.}
	\label{fig:alpha_sense_1}
\end{figure}

\begin{figure}[!t]
	\centering
	\begin{subfigure}[!t]{0.65\textwidth}
		\includegraphics[width=1\linewidth]{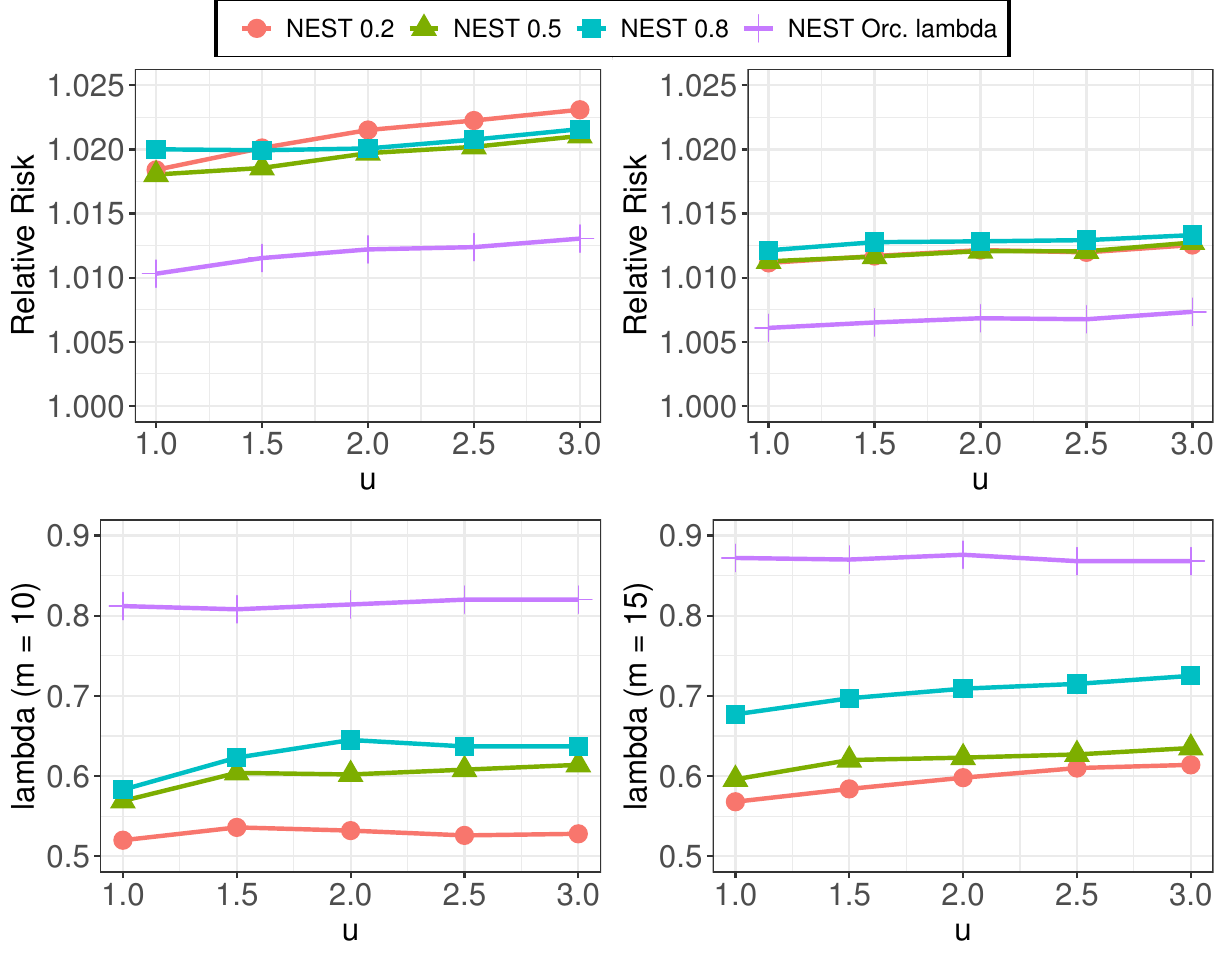}
		\caption{ Setting 6:  $\sigma_i^2\stackrel{i.i.d}{\sim}U(0.1,u)$ and $\mu_i~|~\sigma_i^2\stackrel{ind.}{\sim}~{\sf Laplace}({\sf location}=0,{\sf scale}=\sqrt{\sigma_i^2/2})$. Left $m=10$ and right $m=15$.}
		\label{fig:setting_6} 
	\end{subfigure}
	%\end{figure}
	%\begin{figure}[!t]
	%	\centering
	\begin{subfigure}[!t]{0.65\textwidth}
		\includegraphics[width=1\linewidth]{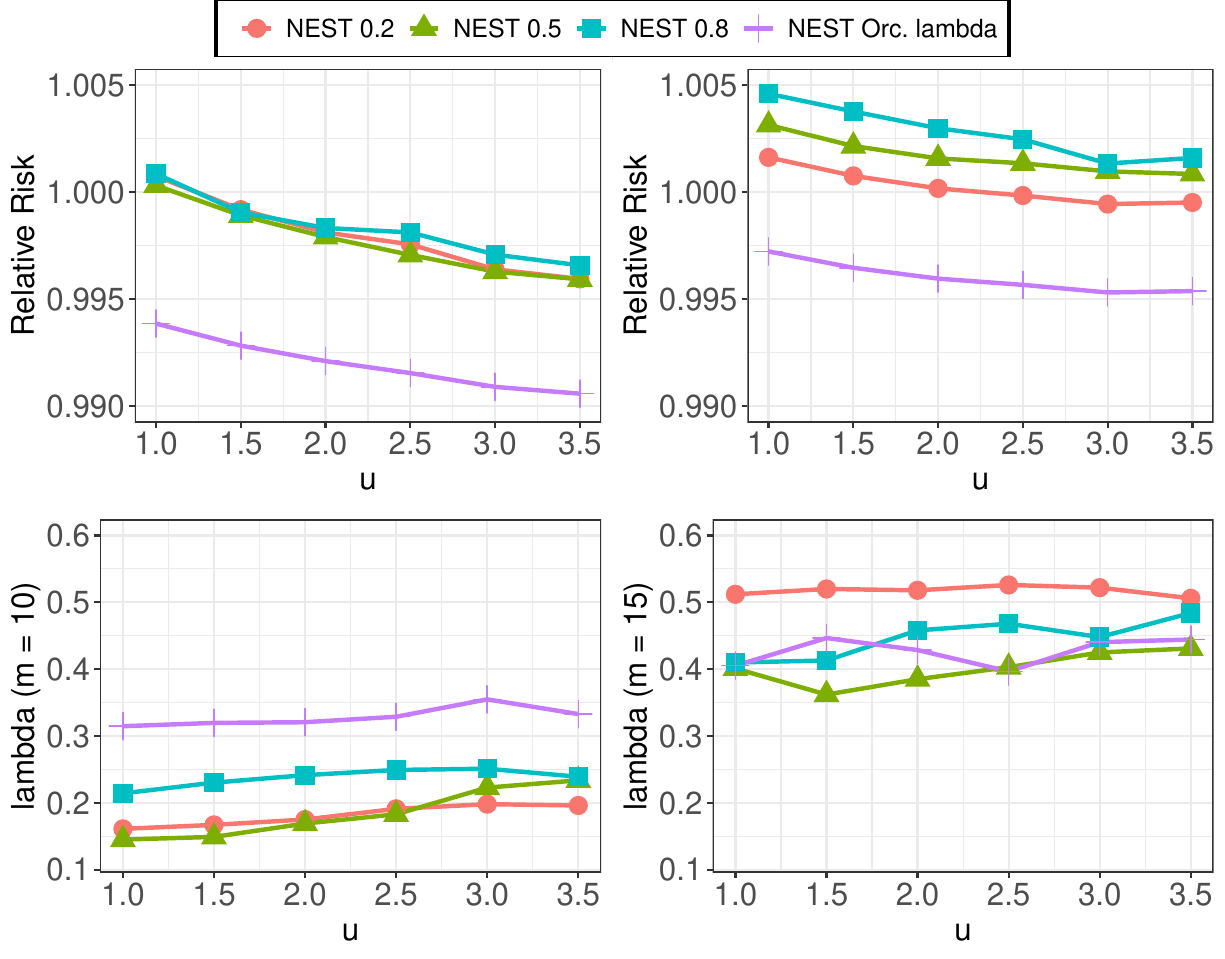}
		\caption{Setting 8:$(\mu_i,~\sigma_i^2)$ generated according to Setting 5 and $Y_{ij}|(\mu_i, \sigma^2_i)\stackrel{i.i.d.}{\sim}{\sf Laplace}({\sf location}=\mu_i,{\sf scale}=\sqrt{\sigma_i^2/2})$. Left $m=10$ and right $m=15$.}
		\label{fig:setting_8}
	\end{subfigure}
	\caption{Panels (a) and (b) top left: relative risks of $\bm \delta_n^{\sf ds}(\hat{\lambda})$ for $\alpha\in\{0.2,0.5,0.8\}$. Panels (a) and (b) bottom left: the corresponding estimates of $\lambda$ along with the oracle $\lambda$ obtained by minimizing the true loss. In each panel the first column corresponds to $m=10$ and the second column pertains to $m=15$.}
	\label{fig:alpha_sense_2}
\end{figure}

The top left plot in figures \ref{fig:setting_2}--\ref{fig:setting_8} represent the relative risks of $\bm \delta_n^{\sf ds}(\hat{\lambda})$ for three different choices of $\alpha$ at $m=10$. The bottom left plot presents the corresponding estimates of $\lambda$ along with the oracle $\lambda$ obtained by minimizing the true loss. The second column in figures \ref{fig:alpha_sense_1} and \ref{fig:alpha_sense_2} pertain to $m=15$. As discussed in Remark \ref{rem:1}, we note that the estimates of $\lambda$ for the various choices of $\alpha$ are substantially closer to the oracle $\lambda$ at $m=15$ than at $m=10$. Furthermore, at $m=10$ the NEST estimator with $\alpha=0.8$ appears to have a marginally improved relative risk than the NEST estimator with $\alpha=0.5$. Overall, we observe the following pattern, namely (i) for large $m$ the different choices of $\alpha$ have a relatively smaller impact on the overall performance of the NEST estimator, and (ii) at $m=10$,  the NEST estimator with $\alpha=0.8$ has a marginally improved relative risk than the NEST estimator with $\alpha=0.5$.
%%%%%%%%%%%%%%%%%
\end{document}